\documentclass[aps,pra,twocolumn,showpacs,nofootinbib,longbibliography,notitlepage,floatfix]{revtex4-2}
\usepackage[T1]{fontenc}
\usepackage[main=english,french]{babel}
\babeltags{en=english}
\makeatletter
\@ifundefined{l@en}{\@namedef{l@en}{\l@english}}{}
\@ifundefined{captionsen}{\@namedef{captionsen}{\captionsenglish}}{}
\@ifundefined{daten}{\@namedef{daten}{\dateenglish}}{}
\makeatother
\usepackage{etex}
\usepackage[normalem]{ulem}
\usepackage{pgfplots}
\pgfplotsset{compat=1.18}
\usepackage{xcolor}
\usepackage{amsmath,amssymb,amsthm}
\usepackage{float}
\usepackage[colorlinks=true,citecolor=blue,urlcolor=blue]{hyperref}
\usepackage{cleveref}
\usepackage{times,txfonts}
\usepackage{braket}
\usepackage{color}
\usepackage{physics}
\usepackage{natbib}
\usepackage{amsmath,blkarray}
\usepackage{mathtools}
\usepackage{latexsym}
\usepackage{tabularx, booktabs}
\usepackage{graphics,epstopdf}
\usepackage{graphicx}
\usepackage{graphicx}
\usepackage{amsfonts}
\usepackage{cleveref}
\usepackage{comment}
\usepackage{color,soul}
\newcommand{\X}{\sigma_x}
\newcommand{\Z}{\sigma_z}

\newcommand{\be}{\begin{equation}}
\newcommand{\ee}{\end{equation}}
\newcommand{\ba}{\begin{eqnarray}}
\newcommand{\ea}{\end{eqnarray}}
\newcommand{\eac}[2]{\expval{\acomm{#1}{#2}}}

\newcommand{\I}{\mathbb{I}}

\newcommand{\half}{\frac{1}{2}}

\newcommand{\ps}{\ket{\psi}}
\usepackage{mathrsfs}
\newcommand{\na}{\nonumber}

\newcommand{\tcr}{\textcolor{red}}

\begin{document}
	\title{Robust self-testing based on Gisin's arbitrary-input  Bell inequality} 
\author{Rajdeep Paul}\email{rajdeeppaul22@gmail.com}\affiliation{Department of Physics, Indian Institute of Technology Hyderabad, Telangana-502284, India }
         \author{A. K. Pan}
	\email{akp@phy.iith.ac.in}
	 \affiliation{Department of Physics, Indian Institute of Technology Hyderabad, Telangana-502284, India }
\begin{abstract}
    Self-testing refers to the strongest device-independent (DI) certification method that validates the nature of a quantum system and devices solely based on the observed statistics.  We demonstrate the self-testing of state and measurements based on the Gisin Bell inequality (GBI) featuring arbitrary inputs for both parties. We introduce a systematic and elegant sum-of-squares (SOS) approach that enables the dimension-independent derivation of the optimal quantum violation of GBI. We derive the state and the interrelation between the local observables directly from the optimization condition. Since the practical experimental scenario involves inevitable noise and imperfection, we present a comprehensive strategy for robust self-testing. 
\end{abstract}
 \maketitle
 \section*{Introduction}
 
Self-testing establishes a powerful certification paradigm that infers the properties of physical systems exclusively from observed input–output statistics. The quantum violation of a Bell inequality \cite{Bell1964, Clauser1969, Brunner2014, Braunstein1990} facilitates such DI self-testing for both the source and the measurement devices. In contrast to classical physics, quantum states can be entangled, and measurements can be incompatible. These distinct features give rise to striking phenomena: the outcomes of incompatible measurements performed locally on the subsystem of an entangled state can display correlations that exceed those permitted by any classical model.  Such a violation of the classical bound rules out the adequacy of local realist models, highlighting the intrinsically non-classical nature of quantum correlations. The optimal quantum violation of a Bell inequality enables the certification of untrusted sources and black-box devices solely from observed input–output statistics, making no assumptions on the inner working of the devices or on the dimension of the quantum system.

Since the first self-testing scheme introduced by Mayers and Yao \cite{mayers1998,mayers2004s}, numerous such DI protocols have been introduced. These include DI self-testing schemes for arbitrary pure, non-maximally entangled two-qubit states \cite{Acin2012, Coladangelo2017, Rai2021, Wooltron2022, Rai2022, Bamps2015}, as well as parallel self-testing of several maximally entangled two-qubit pairs \cite{McKague2016, Wu2016}. The methodology has been generalized to multipartite systems, yielding self-tests for tripartite W states \cite{Wu2014}, $N$-partite GHZ states \cite{Panwar2023, Singh2025}, maximally entangled two-qudit states \cite{Sarkar2021}, and multipartite graph states \cite{McKague2014}. 

Beyond states, self-testing has been demonstrated for Pauli measurements \cite{Bowles2018, Pan2021, Bowles2018/2}, quantum instruments \cite{Wagner2020}, quantum memories \cite{SekatskiPrl2018, SekatskiPrl2023}, and unsharp measurements \cite{Roy2023, Paul2024,Gomez2016}. In network nonlocality settings, results include self-testing of all entangled states \cite{Supic2023}, commuting observables \cite{Munshi2023PRA}, and even the formalism of complex quantum theory itself \cite{Renou2021}. 
Relaxing the assumption of dimension independence has led to substantial advances in semi-DI protocols, particularly in prepare-and-measure scenarios \cite{Tavakoli2018,mohan2019, Abhyoudai2023, Miklin2020, supic2020input, Mukherjee2021, Singh2025prepare, Paulunitary}. Experimental demonstrations of self-testing for states and measurements have also been realized \cite{Smania2020, Gomez2018,zhang2018, Hu2018, Hu2023}. For an extensive overview of these developments, we refer a review article \cite{Supic2020rev}.

In realistic experimental conditions, unavoidable noise in quantum systems prevents the attainment of the maximal quantum violation of Bell inequalities. Consequently, self-testing methods must remain reliable for noisy states and imperfect devices. Robust self-testing of bipartite entangled states and local measurements, with a focus on noise resilience, has been investigated in Refs.~\cite{McKague2012,Kaniewski2016, McKague2014, Bamps2015, Supic2016, Wu2014, Bowles2018/2,Paul2026}, and later generalized to three-qubit W states \cite{Wu2014} and more general multipartite states \cite{Singh2025, Panwar2023, Sarkar2022, Zhang2019}. More recently, experimental self-testing has also been extended to the quantum network \cite{Agresti2021}.

In this work, we introduce a DI self-testing protocol based on the GBI\cite{Gisin1999}, featuring an arbitrary  $n$ number of measurement settings on each side. In bipartite Bell tests with only two measurement settings, DI self-testing generally exploits the Jordan lemma \selectlanguage{french}\cite{Jordanlemma}\selectlanguage{english}, which ensures a basis where the two dichotomic observables decompose into $2\times 2$ blocks. However, the Jordan lemma does not generalize to scenarios with more than two measurements or outcomes, and hence extending the DI self-testing protocol to more than two input scenarios becomes nontrivial. We overcome this limitation by developing a systematic and streamlined SOS approach that yields a dimension-independent derivation of the optimal quantum violation of GBI, and enables the self-testing of the quantum state and observables from the optimal quantum value.

To make our method more suitable for experimental implementation, we develop a swap-circuit approach that explicitly realizes the local isometry required to self-test both the underlying state and the observables. Within this framework, local unitaries acting on ancillary systems of known dimension are implemented to extract a target state from an uncharacterized device, thereby mapping the properties of the unknown system onto a trusted one and enabling DI certification. In addition, we present a systematic procedure to quantify the robustness of these self-testing in the presence of  noise and experimental imperfections, demonstrating the feasibility of the protocol in practical scenarios.

The structure of the paper is as follows. In Section~\ref{SECI}, we introduce the GBI and present the construction of the SOS approach. In Section~\ref{SOSsec}, we give a dimension-independent analytical derivation of the optimal quantum violation of the GBI for an arbitrary number $n$ of measurement settings per party. From the SOS optimization conditions, we further obtain the unique underlying state and measurements. We then illustrate the derivation of the optimal quantum violation of the GBI, together with the corresponding state and observables, for general $n$ we obtain a unified formulation, and for $n=3,4$ we present detailed derivations; additional explicit cases for $n=5,6,11$ are provided in Appendix~\ref{SOS6,7,11}. Section~\ref{sec3} presents the explicit construction of the swap circuit and demonstrates self-testing of the maximally entangled bipartite state and associated observables via a local isometry. The robustness analysis of this self-testing scheme, based on the swap circuit, is given in Section~\ref{Sec4}. Finally, Section~\ref{Sec5} concludes with a summary of our results and a discussion of future research directions.
 \section{GBI for arbitrary \texorpdfstring{$n$}{n} input per party}\label{SECI}
 The general form of arbitrary $n$ input GBI \cite{Gisin1999} is given by 
\begin{equation}
\label{gisinineq}
\mathcal{G}_n = \sum\limits_{i=1}^n \qty(\sum\limits_{j=1}^{n-i+1}{A}_i{B}_j-\sum\limits_{j=n-i+2}^n A_i{B}_j)\leq \Big\lfloor\frac{n^2+1}{2}\Big\rfloor
\end{equation}
where $\lfloor \hspace{1mm}\rfloor$ is the floor function. The optimal quantum violation of Eq.~ (\ref{gisinineq}) was derived in \cite{Gisin1999}, for the following set of observables and state is given by  
  \ba
  A_i=-\sin\alpha_i\X+\cos\alpha_i\Z, \quad B_j=\sin\beta_j\X-\cos\beta_j\Z
  \ea
  where $\alpha_i = i\frac{\pi}{n}$, $\beta_j=\frac{3-n-2j}{2}\frac{\pi}{n}$. The required maximum entangled two-qubit state is given by $\ket{\phi^+}=\frac{1}{\sqrt{2}}\qty(\ket{00}+\ket{11})$\\
  
We present an analytical method to determine the optimal quantum violation of the GBI using SOS approach, devoid of assuming the dimension of the quantum system. This allows us to demonstrate that the optimal quantum violation self-tests the observables and the entangled state.
\section{SOS approach: An elegant optimization technique}\label{SOSsec}
We construct an elegant SOS method to derive the maximal quantum violation of the GBI in Eq. (\ref{gisinineq}) without imposing any restriction on the system’s dimension. To this end, we introduce a positive semidefinite operator $\Gamma_n$ such that the Bell functional satisfies $\expval{\mathcal{G}_n} = \eta_n - \expval{\Gamma_n}$. As $\expval{\Gamma_n} \ge 0$, the maximal quantum value of $\mathcal{G}_n$ is reached when $\expval{\Gamma_n} = 0$, which yields $\expval{\mathcal{G}_n}^{\mathrm{opt}}_Q = \eta_n$. We define the operator $\Gamma_n$ as
\begin{eqnarray}\label{gcn}
    \Gamma_{n} = \sum\limits_{i=1}^n\frac{\mu_{n,i}}{2}\mathcal{K}_{n,i}^\dagger\mathcal{K}_{n,i}
\end{eqnarray}  
where 
\begin{eqnarray}
\label{kni}\mathcal{K}_{n,i}=\openone_d\otimes\mathcal{B}_i -A_i \otimes \openone_d, \forall i\in[n]
\end{eqnarray}
and
\begin{eqnarray}  \mathcal{B}_i=\frac{1}{\mu_{n,i}}\qty(\sum\limits_{j=1}^{n-i+1}{B}_j-\sum\limits_{j=n-i+2}^n {B}_j)
\end{eqnarray}
And $\mu_{n,i}$ are defined as suitable norms given by $\mu_{n,i}=||\sum\limits\limits_{j=1}^{n-i+1}{B}_j-\sum\limits\limits_{j=n-i+2}^n {B}_j||_{\rho_{AB}}$. Substituting $\mathcal{K}_{n,i}$ from Eq. (\ref{kni}) in Eq. (\ref{gcn}), we get $\expval{\Gamma_n}=\sum\limits\limits_{i=1}^n\mu_{n,i} -\expval{\mathcal{G}_n}$, which implies that
\begin{eqnarray}
&&\expval{\mathcal{G}_n}^{opt}_Q=\max\qty[\sum\limits\limits_{i=1}^n\mu_{n,i}]
\end{eqnarray}

We use  the following inequality   \ba \label{cnv} \sum\limits\limits_{i=1}^{n}f_i\leq  \sqrt{n\sum\limits\limits_{i=1}^{n}f_i^2}, \quad\forall f_i\geq 0, \ea 
where the equality holds if  $f_i=f_{i'}, \forall i\neq i'\in[n]$ . Hence, we can write 
 \begin{eqnarray}\label{gnbo}
 \mathcal{
G}_{n}&\leq&\sqrt{n\qty[\sum\limits\limits_{i=1}^n (\mu_{n,i})^2]}
 \end{eqnarray}
Here, the expectation value $\langle\{B_{j},B_{j'}\}\rangle=\Tr[\{B_{j},B_{j'}\}\rho_{AB}]$. 

We derive the relation between local observables and the state required to achieve the optimal quantum value. Taking into account the optimization criterion, we can find the constraint $\expval{\Gamma_n}=0,\forall\rho_{AB}$, which leads to
\ba \label{cn} \sum \limits_i \Tr[\mathcal{K}_{n,i}^\dagger\mathcal{K}_{n,i} \ \rho_{AB}] = 0, \quad  \forall i\in[n]\ea.

Since each $\mathcal{K}_{n,i}^\dagger\mathcal{K}_{n,i}$ is a positive Hermitian operator, Eq.~(\ref{cn}) implies that $\Tr[\mathcal{K}_{n,i} \ \rho_{AB}]=0$ for all $i\in[n]$. Consequently, we obtain
\begin{eqnarray}\label{selfn}
&&\Tr[(A_i\otimes \mathcal{B}_i) \ \rho_{AB}]=1\Rightarrow A_i\otimes \mathcal{B}_i \ps_{AB}=\ps_{AB},\forall i\in[n]\quad
\end{eqnarray}

We prove that $\ps_{AB}$ is the maximally entangled state. Consider a bipartite pure state $\ket{\psi}_{AB} \in \mathcal{H}_A \otimes \mathcal{H}_B$ of dimension $d \times d$. If there exists an irreducible set of operators $\{A_i\}$ acting on $\mathcal{H}_A$ and $\{\mathcal{B}_i\}$ acting on $\mathcal{H}_B$ such that the state satisfies the eigenvalue equation,
\begin{equation}
    (A_i \otimes \mathcal{B}_i) \ket{\psi}_{AB} = \ket{\psi}_{AB} \quad \forall i\in[n]\quad,
\end{equation}

Then in density matrix form, we can write
\begin{eqnarray}
    \rho_{AB}=(A_i \otimes \mathcal{B}_i) \rho_{AB} (A_i \otimes \mathcal{B}_i)^\dagger &=&  (A_i \otimes \mathcal{B}_i) \rho_{AB} (A_i^\dagger \otimes \mathcal{B}_i^\dagger)\quad
\end{eqnarray}

To find the local state of the subsystem $A$, we take a partial trace over the subsystem $\mathcal{B}$ so that  
\begin{equation}
    \rho_A = \Tr_\mathcal{B} [\rho_{AB}] = \Tr_\mathcal{B} \left[ (A_i \otimes \mathcal{B}_i) \rho_{AB} (A_i^\dagger \otimes \mathcal{B}_i^\dagger) \right]
\end{equation}

By the linearity of the partial trace, Alice's local operators can be factored as
\begin{equation}
\label{rhoa}
    \rho_A = A_i \left( \Tr_\mathcal{B}[ (\openone_d \otimes \mathcal{B}_i) \rho_{AB} (\openone_d \otimes \mathcal{B}_i^\dagger)] \right) A_i^\dagger
\end{equation}

Using the cyclic property of the trace within the Hilbert space $\mathcal{H}_B$:
\begin{equation}
   \Tr_\mathcal{B} [ (\openone_d \otimes \mathcal{B}_i) \rho_{AB} (\openone_d \otimes \mathcal{B}_i^\dagger) ] = \Tr_\mathcal{B} [ (\openone_d \otimes \mathcal{B}_i^\dagger \mathcal{B}_i) \rho_{AB} ]
\end{equation}
Since $\mathcal{B}_i$ is unitary ($\mathcal{B}_i^\dagger \mathcal{B}_i = \openone_d$), the term inside the parenthesis reduces to $ \rho_A$. Substituting this back into the Eq. (\ref{rhoa}), we get
\begin{equation}
    \rho_A = A_i \rho_A A_i^\dagger \implies \rho_A A_i = A_i \rho_A
\end{equation}

This reveals that the reduced density matrix commutes with all operators in the set, i.e., 
\begin{equation}
    [\rho_A, A_i] = 0, \quad \forall i\in[n]\quad
\end{equation}

By Schur's Lemma \cite{schurlemma}, any operator that commutes with an irreducible set of operators must be a scalar multiple of the identity operator. Therefore, $\rho_A = c \openone_d$. Finally, the normalization condition gives $c = 1/d$. Thus, $\rho_A = \openone_d/d$. Reversing the role of Alice and Bob's observables (presented in the latter part of the paper in Eq.~(\ref{Bobsc1})), one finds $\rho_B = {\openone_d}/{d} $, which concludes that $\ket{\psi}_{AB}$ is a maximally entangled state.

 
\subsection{Derivation of optimal quantum violation of GBI for \texorpdfstring{$n=3$}{n=3}}
Alice and Bob, each measures three dichotomic observables, denoted by $A_i$ for Alice and $B_j$ for Bob, with $i,j \in [3]$. The corresponding GBI is obtained by substituting $n=3$ into Eq.~(\ref{gisinineq}), yielding
\begin{eqnarray}
    \mathcal{G}_3 &=&    A_1\qty(B_1+B_2+B_3)+A_2\qty(B_1+B_2-B_3)\nonumber\\
    &&+A_3\qty(B_1-B_2-B_3) \leq 5
\end{eqnarray}
In order to find the optimal quantum violation of the inequality, we construct the positive semi-definite operator $\Gamma_3$, where
\begin{equation}\label{gamma3}
    \Gamma_{3} =\sum\limits\limits_{i=1}^3\frac{\mu_{3,i}}{2}\mathcal{K}_{3,i}\mathcal{K}_{3,i}^\dagger
\end{equation}
Suitably, we construct the operators $\mathcal{K}_{3,i}$'s in the following form,
\begin{eqnarray}\label{l3}
    &&\mathcal{K}_{3,i}\ket{\psi} = \openone_d\otimes\mathcal{B}_i\ket{\psi}-A_i\otimes \openone_d\ket{\psi},\quad \forall i\in[3]
\end{eqnarray}
where $\mathcal{B}_i=\frac{1}{\mu_{3,i}}\qty(\sum\limits\limits_{j=1}^{4-i}{B}_j-\sum\limits\limits_{j=5-i}^3 {B}_j)$ and $\mu_{3,i}=||\sum\limits\limits_{j=1}^{4-i}{B}_j-\sum\limits\limits_{j=5-i}^3 {B}_j||_{\rho_{AB}}$. Substituting $\mathcal{K}_{3,i}$ from Eq.~(\ref{l3}) in Eq.~(\ref{gamma3}), we get
\begin{eqnarray}
    \mathcal{G}_3=-\Tr[\Gamma_3 \ \rho_{AB}]+\sum\limits_{i=1}^3 \mu_{3,i}
\end{eqnarray}
Hence, the optimal value of following the SOS approach as outlined in Sec.~\ref{SOSsec}, we get that the optimal value of $(\mathcal{G}_{3})_Q$ is obtained if $\Tr[ \Gamma_3\ \rho_{AB}]=0$, which implies
\begin{equation}\label{g31}
   \qty(\mathcal{G}_3)_Q^{\ opt}= \text{max}\sum\limits_{i=1}^3 \mu_{3,i}
\end{equation}
where
\begin{eqnarray}
    &&\mu_{3,1}=\sqrt{3\openone_d+\langle\{B_1,B_2\}\rangle+\langle\{B_2,B_3\}\rangle+\langle\{B_1,B_3\}\rangle}\nonumber\\
    &&\mu_{3,2}=\sqrt{3\openone_d+\langle\{B_1,B_2\}\rangle-\langle\{B_2,B_3\}\rangle-\langle\{B_1,B_3\}\rangle}\nonumber\\
    &&\mu_{3,3}=\sqrt{3\openone_d-\langle\{B_1,B_2\}\rangle+\langle\{B_2,B_3\}\rangle-\langle\{B_1,B_3\}\rangle}\nonumber
\end{eqnarray}
Using the inequality 
  \ba\sum\limits\limits_{i=1}^{n}f_i\leq  \sqrt{n\sum\limits\limits_{i=1}^{n}f_i^2}, \quad\forall f_i\geq 0, \ea 
where the equality holds if  $f_i=f_{i'}, \forall i\neq i'\in[n]$, we can write Eq.~(\ref{g31}) as

\begin{eqnarray}
     \mathcal{G}_3&\leq& \sqrt{3\qty(9+\eac{B_1}{B_2}+\eac{B_2}{B_3}-\eac{B_1}{B_3})}\nonumber\\
     &=& \sqrt{3\qty(9+\eac{B_2}{B_1+B_3}-\eac{B_1}{B_3})}
\end{eqnarray}
Assuming an observable relation $B_2=\frac{B_1+B_3}{\nu^1_{3}}$ we write
\begin{equation}
    (\mathcal{G}_3)_Q \leq\sqrt{3\qty(9+2\sqrt{2+\eac{B_1}{B_3}}-\eac{B_1}{B_3})}
\end{equation}
Apparently, the maximum value is attained only when $\eac{B_1}{B_3}=-1$, which gives $\nu^1_3=\sqrt{2+\eac{B_1}{B_3}}=1$ and the relation between Bob's observables $B_1-B_2+B_3=0$. Subsequently, we also find $\eac{B_1}{B_2}=1$ and $\eac{B_2}{B_3}=1$. Finally, $\mu_{3,1}=\mu_{3,2}=\mu_{3,3}=\csc{\frac{\pi}{6}}$, thus the maximal bound turns out to be 

\begin{eqnarray}
    \qty(\mathcal{G}_3)_Q^{\ opt} = 6=3\csc{\frac{\pi}{6}}
\end{eqnarray}
Here Bob's observables follow the following relation.
\begin{eqnarray} 
     \langle\{B_j,B_{j+x}\}\rangle&=&2\cos\frac{\pi x}{3},\quad \forall j\in[2], \ x\in[3-j]
     \end{eqnarray}
Similarly, from the appendix.~(\ref{Bobo}) we get the Alice's observables also follow the following property
\begin{eqnarray}
    \eac{A_i}{A_{i+x}}=2\cos{\frac{\pi x}{3}},\quad \forall x\in[3-i]
\end{eqnarray}
Again, $\Tr[\Gamma_3\,\rho_{AB}] = 0$ yields $\Tr[\mathcal{K}_{n,i}^\dagger \mathcal{K}_{n,i}\,\rho_{AB}] = 0$, which in turn implies
\begin{eqnarray}
    A_i\otimes \mathcal{B}_i \ps_{AB}&=&\ps_{AB},\quad \forall i\in[3]\label{a3b3}
\end{eqnarray}
Again at the optimal quantum value given in Appendix~(\ref{Bobo}), we obtain
 \begin{eqnarray}\label{selfn13}
     \mathcal{A}_j\otimes B_j \ket{\psi}_{AB}=\ket{\psi}_{AB},\quad \forall j\in[3]
 \end{eqnarray}
 where    $\mathcal{A}_1 = \frac{A_1+A_2+A_3}{2}, \mathcal{A}_2 = \frac{A_1+A_2-A_3}{2}, \mathcal{A}_3 = \frac{A_1-A_2-A_3}{2}$.
 
From the optimization condition of Eq.~(\ref{selfn13}), we see that $\rho_{AB}$ must be a common eigenstate of the operators $\qty(\mathcal{A}_j\otimes B_j),  \forall j\in[n]$. As these operators produce the maximum eigenvalue with the normalized state $\rho_{AB}$, this implies that $\rho_{AB}=|\psi\rangle_{AB}\langle\psi|_{AB}$ needs to be a maximally entangled pure state. We can then write the state $\rho_{AB}\in \mathbb{C}^d\times \mathbb{C}^d$ in this form \cite{Paul2024}
\begin{eqnarray}\label{rho odd}
\rho_{AB} = \frac{1}{d^2} \qty[\openone_d\otimes\openone_d + \sum_{f=1}^{d^2-1} C_{f} \otimes C_{f}]
\end{eqnarray}
Since $\rho_{AB}$ is a pure state, for each $f\neq f'  \in[d^2-1]$, the terms $C_{f}\otimes C_{f}$ and $C_{f'}\otimes C_{f'}$ in Eq.~(\ref{rho odd}) has to be  mutually commuting and $\Tr[C_{f}]=0$. To explicitly derive
$C_{f} \otimes C_{f}$, we use  the following conditions i) $\Tr[(\mathcal{A}_j\otimes B_j) \ \rho_{AB}]=1$, ii) $\Tr[(C_{f} \otimes C_{f})\rho_{AB}]=1$, and 
    iii) $\langle[C_{f}\otimes C_{f}, C_{f'}\otimes C_{f'}]_{f\neq f'}\rangle_{\rho_{AB}}=0$.
Using these conditions, we obtain explicit expressions for any two \(C_{f} \otimes C_{f}\) say,  \(C_1 \otimes C_1\) and \(C_2 \otimes C_2\), in such a way that only these two terms will contribute for \(\Tr[(\mathcal{A}_i \otimes B_i) \rho_{AB}] = 1\), $\forall i \in [n]$. Detailed derivations are provided in the Appendix~\ref{cseodd}. Hence, by putting $n=3$ in Appendix~\ref{cseodd} we get
\begin{eqnarray}
 C_1 \otimes C_1 &=& \mathcal{A}_2 \otimes B_2,
C_2 \otimes C_2=\frac{(\mathcal{A}_1-\mathcal{A}_3) \otimes (B_1-B_3)}{(2-\langle\{B_1,B_3\}\rangle)}\quad
\end{eqnarray} and 
$C_3 \otimes C_3 =(C_1 \otimes C_1)\cdot (C_2 \otimes C_2)$.

In a two-qubit system, a set of choices of observables is given by  \ba A_1=\frac{-\sqrt{3}\sigma_x+\sigma_z}{2},\  A_2=\frac{-\sqrt{3}\sigma_x-\sigma_z}{2},\  A_3=-\sigma_z\nonumber\\
B_1=\frac{-\sqrt{3}\sigma_x-\sigma_z}{2},\   B_2=\frac{-\sqrt{3}\sigma_x+\sigma_z}{2}, \ B_3=\sigma_z\nonumber\ea 
And consequently, we get 
\begin{eqnarray}\label{cop3}
   &&C_1 \otimes C_1 = \mathcal{A}_2 \otimes B_2=\frac{-\sqrt{3}\sigma_x+\sigma_z}{2}\otimes\frac{-\sqrt{3}\sigma_x+\sigma_z}{2}\nonumber\\
   &&C_2 \otimes C_2=\frac{(\mathcal{A}_1-\mathcal{A}_3) \otimes (B_1-B_3)}{(2-\langle\{B_1,B_3\}\rangle)}\nonumber\\
   &&\hspace{1.24cm}=\frac{\sigma_x+\sqrt{3}\sigma_z}{2}\otimes \frac{\sigma_x+\sqrt{3}\sigma_z}{2}\nonumber\\
    &&C_3 \otimes C_3 =(C_1 \otimes C_1)\cdot (C_2 \otimes C_2)=-\sigma_y\otimes \sigma_y
\end{eqnarray}
Hence, from  Eq.~(\ref{rho odd}) we get the two-qubit maximally entangled state \ba \label{rhophi+}\rho_{\phi^+}=\frac{1}{4}\qty(\openone_2\otimes\openone_2+\sigma_x\otimes \sigma_x+\sigma_z\otimes \sigma_z-\sigma_y\otimes \sigma_y)\ea this produces the optimal quantum violation $\qty(\mathcal{G}_3)_Q^{\ opt} =3\csc{\frac{\pi}{6}}$. 
\subsection{Derivation of optimal quantum violation of GBI for \texorpdfstring{$n=4$}{n=4}}\label{gissin4}
\noindent
Substituting $n=4$ in Eq.~(\ref{gisinineq}), we get
\begin{eqnarray}
&&\mathcal{G}_4=A_1\qty(B_1+B_2+B_3+B_4)+A_2\qty(B_1+B_2+B_3-B_4)\nonumber\\
&&+ A_3\qty(B_1+B_2-B_3-B_4)+A_1\qty(B_1-B_2-B_3-B_4)\leq 8 \quad
\end{eqnarray}
Following the SOS approach as outlined in Sec.~\ref{SOSsec}, we get that  the optimal value of $(\mathcal{G}_{4})_Q$ is obtained if $\Tr[ \Gamma_{4}\ \rho_{AB}]=0$ which implies
\begin{equation}\label{g4mu}
   \qty(\mathcal{G}_{4})_Q^{\ opt}= \text{max}\sum\limits\limits_{i=1}^{4} \mu_{4,i}
\end{equation}
where $\mu_{{4},i}=||\sum\limits\limits_{j=1}^{5-i}{B}_j-\sum\limits\limits_{j=6-i}^{4} {B}_j||_{\rho_{AB}},\forall i\in[4]$. Using the inequality in Eq.~(\ref{cnv}), we get
\ba\label{G4q}
(\mathcal{ G}_{4})_Q&\leq&\sqrt{4\sum\limits_{i=1}^4\left(\mu_{4,i}\right)^2}
 \ea
 The equality  holds only  when $\mu_{4,i}=\mu_{4,j},\forall i\neq j\in[4]$. Now putting the value each $\mu_{4,i}$ in Eq.~(\ref{G4q}) and after simplifying we get,
\begin{eqnarray}
    \qty(\mathcal{G}_4)_Q&\leq&\Bigg[4\bigg(16+2\Big[\eac{B_1}{B_2}+\eac{B_2}{B_3}+\eac{B_3}{B_4}\nonumber\\
        &&-\eac{B_1}{B_4}\Big]\bigg) \Bigg]^\half
\end{eqnarray}        

Further, taking, $B_2=\frac{B_1+B_3}{\nu^1_4}$ and $B_4=\frac{B_3-B_1}{\nu^2_4}$,
\begin{eqnarray}    \qty(\mathcal{G}_4)_Q\leq\sqrt{4\Bigg(16+4\bigg(\sqrt{2+\eac{B_1}{B_3}}+\sqrt{2-\eac{B_1}{B_3}}\bigg)\Bigg)}\quad \ \ \
\end{eqnarray}
Maximization requires $\eac{B_1}{B_3}=0$. Thus, the optimal quantum value is given
\begin{eqnarray}
    \qty(\mathcal{G}_4)_Q^{\ opt}=4\sqrt{4+2\sqrt{2}}=4\csc{\frac{\pi}{8}}
\end{eqnarray}
Consequently, one finds the observable relations in the following form. $\eac{B_1}{B_3}=0$ implies $\nu^1_4=\nu^2_4=\sqrt{2}$ which leads to $B_2=\frac{B_1+B_3}{\sqrt{2}}$ and $B_4=\frac{B_3-B_1}{\sqrt{2}}$. From this we can further show that $\eac{B_1}{B_2}=\eac{B_2}{B_3}=\eac{B_3}{B_4}=2\cos{\frac{\pi}{4}}$,  $\eac{B_1}{B_4}=2\cos{\frac{3\pi}{4}}$ and $\mu_{4,i}=\csc{\frac{\pi}{8}},\forall i\in[4]$. In a similar way, Bob's observables follow the relation, i.e.,
\begin{eqnarray}
    \eac{B_j}{B_{j+x}}=2\cos{\frac{\pi \ x}{4}},\quad \forall x\in[4-j]
\end{eqnarray}
Similarly, from the appendix.~(\ref{Bobo}) we get the Alice's observables follows the following property
\begin{eqnarray}\label{g4ai}
    \eac{A_i}{A_{i+x}}=2\cos{\frac{\pi x}{4}},\quad \forall x\in[4-i]
\end{eqnarray}

Using Eq.~(\ref{rho odd}), we obtain the desired state $\rho_{AB}$ by substituting  $n=4$ in Appendix~\ref{stateeven}. This yields
\ba C_1 \otimes C_1 &=& {A}_1 \otimes \mathcal{B}_1, \quad 
C_2 \otimes C_2={A}_3 \otimes \mathcal{B}_3\nonumber\\
C_3 \otimes C_3 &=& \frac{1}{4}\big[A_2, A_4\big]\otimes \big[\mathcal{B}_2, \mathcal{B}_4\big]=(C_1 \otimes C_1)\cdot (C_2 \otimes C_2) \quad\ea 
where, $\mathcal{B}_1 = \frac{B_1+B_2+B_3+B_4}{\mu_{4,1}}, \mathcal{B}_2 = \frac{B_1+B_2+B_3-B_4}{\mu_{4,2}}, \mathcal{B}_3 = \frac{B_1+B_2-B_3-B_4}{\mu_{4,3}}, \mathcal{B}_4=\frac{B_1-B_2-B_3-B_4}{\mu_{4,4}}$ and $\mu_{4,i}=\csc{\frac{\pi}{8}},\forall i\in[4]$.

As an example, in a two-qubit system, we provide a set of observables are the following. 
 \ba
  &&A_i=-\sin{i\frac{\pi}{4}} \ \X+\cos{i\frac{\pi}{4}} \ \Z,\forall i\in[4]\nonumber\\
  &&B_j=-\sin{\frac{(1+2 j)\pi}{8}} \ \X-\cos{\frac{(1+2j)\pi}{8}} \ \Z,\ \forall j\in[4]\nonumber
  \ea
  which again provides
\begin{eqnarray}
    &&C_1 \otimes C_1=\qty(\frac{\sigma_x-\sigma_z}{\sqrt{2}})\otimes \qty(\frac{\sigma_x-\sigma_z}{\sqrt{2}})\nonumber\\
    &&C_2 \otimes C_2=\qty(\frac{\sigma_x+\sigma_z}{\sqrt{2}})\otimes \qty(\frac{\sigma_x+\sigma_z}{\sqrt{2}})\nonumber\\
    &&C_3 \otimes C_3 =(C_1 \otimes C_1)\cdot (C_2 \otimes C_2)=-\sigma_y\otimes \sigma_y
\end{eqnarray}
This, in turn, provides the entangled state $\rho_{\phi^+}$ as in Eq. (\ref{rhophi+}) that produces $(\mathcal{G}_{4})^{opt}_Q=4\csc{\frac{\pi}{8}}$.

In Appendix~(\ref{SOS6,7,11}), we provide a detailed derivation for the optimal quantum value when $n=5, 6, $ and $11$. Upon examination, it becomes apparent that the derivation process exhibits increased complexity when dealing with an odd number of settings, denoted by $n$. Addressing this challenge, we specifically develop a sophisticated solution for $n=11$.
\subsection{Derivation of optimal quantum violation of GBI for arbitrary  \texorpdfstring{$n$}{n}}
With the help of the results for specific values of  $n$, we generalize the results for the $n$ settings per party by rewriting Eq.~(\ref{gnbo}). Using the inequality in Eq.~(\ref{cnv}) as

 \begin{eqnarray}
 \mathcal{
G}_{n}&\leq&\sqrt{n\qty[\sum\limits\limits_{i=1}^n (\mu_{n,i})^2]}\nonumber\\
&=&\Bigg[n\sum\limits\limits_{i=1}^n\Big[n+\sum\limits\limits_{j'> j= 1}^{n-i+1}\langle\{B_{j},B_{j'}\}\rangle+\sum\limits\limits_{j'>j={n-i+2}}^{n}\langle\{B_{j},B_{j'}\}\rangle\nonumber\\
    &&\hspace{1cm}-\sum\limits\limits_{j= 1}^{n-i+1}\sum\limits\limits_{j'= {n-i+2}}^{n}\langle\{B_{j},B_{j'}\}\rangle\Big]\Bigg]^\frac{1}{2}
 \end{eqnarray}
The optimization conditions imply that Bob's observables fulfill the following relations.
\begin{eqnarray}
    &&B_j=\frac{B_{j-1}+B_{j+1}}{\sqrt{2+\langle\{B_{j-1}, B_{j+1}\}\rangle}},\forall \  j\in [n]\label{ai}\\
    &&\big{\langle}\{B_j,B_{j+x}\}\big{\rangle}=2\cos\frac{\pi \ x}{n}, \forall j\in[n], x\in[n-j]\label{aiix}\\
    &&\sum\limits_{j=1}^n (-1)^{j+1} B_j =0,\forall n\in odd 
\end{eqnarray}
For a detailed derivation of Eq.~(\ref{ai}), see Appendix~\ref{Bobo}. Furthermore, based on the definition of $\mathcal{B}_i$, we obtain
\begin{eqnarray}\label{cbi}
   &&\big{\langle}\{\mathcal{B}_i,\mathcal{B}_{i+x}\}\big{\rangle}=2\cos\frac{\pi \ x}{n}, \forall i\in[n], x\in[n-i]
\end{eqnarray}
where $B_{n+1}=-B_1$ and $  B_0=-B_n$. Using Eq. (\ref{aiix}), we get $\mu_{n,i}=\csc{\frac{\pi}{2 n}}, \forall i\in[n]$. Note that computing a general $\mu_{n,i}$ is highly nontrivial and quite involved. For $n = 3,4,5,6$ and for larger values like $11$, the value of $\mu_{n,i}$ coincides exactly with $\csc{\frac{\pi}{2n}}$, and this equality persists for arbitrary $n$.
Thus, we find the optimal quantum value
\ba\label{cnopt}(\mathcal{G}_{n})^{opt}_{Q}=n\csc{\frac{\pi}{2 n}}\ea
Note that deriving $(\mathcal{G}_n)^{opt}_Q$ for even and odd values of $n$ calls for slightly different methods.

From the optimization requirements, we additionally obtain that for all $i,j\in[n]$
\begin{eqnarray}\label{rand0m}
  \bra{\psi}_{AB}A_i\otimes B_j\ket{\psi}_{AB}
    &=&\frac{\sum\limits\limits_{k=1}^{n-i+1}\langle\{B_j,{B}_k\}\rangle-\sum\limits\limits_{k=n-i+2}^n \langle\{B_j,{B}_k\}\rangle}{2\csc{\frac{\pi}{2n}}}\quad \ 
\end{eqnarray}
The detailed proof of Eq. (\ref{rand0m}) is provided in the Appendix.~\ref{Bobo}. 

As shown in the Appendix.~\ref{Bobo}, the optimization condition provides that Alice's observables satisfy the following relations.
\begin{eqnarray}\label{bobi}
   &&A_i=\frac{A_{i-1}+A_{i+1}}{\sqrt{2+\langle\{A_{i-1}, A_{i+1}\}\rangle}},   \  A_0=-A_n ,\ \forall i\in [n]
\end{eqnarray}
and
\begin{eqnarray}
   &&\big{\langle}\{A_i,A_{i+x}\}\big{\rangle}=2\cos\frac{\pi \ x}{n}, \forall i\in[n], x\in[n-i]\label{biix}
\end{eqnarray}
which leads to
\begin{eqnarray}
    &&\mathcal{A}_j\otimes B_j\ket{\psi}_{AB}=\ket{\psi}_{AB}, \  \forall j\in[n]\label{Bobsc1}
\end{eqnarray}
Where $\mathcal{A}_j=\frac{\sum\limits\limits_{i=1}^{n-j+1}{A}_j-\sum\limits\limits_{i=n-j+2}^n {A}_j}{\mu'_{n,j}}$ and $\mu'_{n,j}=||\sum\limits\limits_{i=1}^{n-j+1}{A}_i-\sum\limits\limits_{i=n-j+2}^n {A}_i||_{\rho_{AB}}=\csc{\frac{\pi}{2n}}$. From the construction of $\mathcal{A}_j$, we show that 
\ba \label{Bobscmain}\langle\{\mathcal{A}_j,\mathcal{A}_{j+x}\}\rangle=2\cos\frac{\pi \ x}{n}, \forall j\in[n], x\in[n-j]\ea
Thus, we show that, at the optimal quantum value, the sets of observables of Alice and Bob satisfy a similar set of relations. Interestingly, under the optimal condition, both Eq.~(\ref{selfn}) and (\ref{Bobsc1}) hold simultaneously. Detailed derivations are shown in the Appendix.~(\ref{Bobo}).

The required state $\rho_{AB}$ for odd $n$
\ba\label{C'so}
C_1 \otimes C_1 &=& \mathcal{A}_{\frac{n+1}{2}} \otimes B_{\frac{n+1}{2}}\\\na
C_2 \otimes C_2&=&\frac{1}{\lfloor \frac{n}{2} \rfloor}\sum_{i=1}^{{\lfloor \frac{n}{2}\rfloor}}\frac{(\mathcal{A}_{i}-\mathcal{A}_{n+1-i}) \otimes (B_{i}-B_{n+1-i})}{(2-\langle\{B_{i},B_{n+1-i}\}\rangle)}\ea
and 
\ba C_3 \otimes C_3 &=&\frac{1}{\lfloor \frac{n}{2} \rfloor}\sum_{i=1}^{{\lfloor \frac{n}{2}\rfloor}}\frac{\mathcal{A}_{\frac{n+1}{2}}(\mathcal{A}_{i}-\mathcal{A}_{n+1-i}) \otimes B_{\frac{n+1}{2}}(B_{i}-B_{n+1-i})}{(2-\langle\{B_{i},B_{n+1-i}\}\rangle)}\nonumber\\
&=&(C_1 \otimes C_1)\cdot (C_2 \otimes C_2)\na
\ea
The detailed derivation is placed in  Appendix.~\ref{cseodd},  along with an explicit derivation for $n=5$.\\

Similarly, for even $n$,
\ba\label{C'se}
C_1 \otimes C_1 &=& {A}_1 \otimes \mathcal{B}_1\\\na
C_2 \otimes C_2&=&{A}_{\frac{n}{2}+1} \otimes \mathcal{B}_{\frac{n}{2}+1}\ea 
and \ba 
C_3 \otimes C_3 &=& \frac{1}{4(\frac{n}{2}-1)}\sum\limits_{i=2}^{\frac{n}{2}}\big[A_i, A_{i+\frac{n}{2}}\big]\otimes \big[\mathcal{B}_i,\mathcal{B}_{i+\frac{n}{2}}\big]\nonumber\\
&=&(C_1 \otimes C_1)\cdot (C_2 \otimes C_2)\nonumber
\ea
A full derivation, including the explicit case $n=4$, is provided in Appendix~\ref{stateeven}.
\section{Self-testing of state and observables using Swap-circuit} \label{sec3}

 We now demonstrate that the state and measurements can be uniquely characterized when the optimal quantum violation of the GBI is obtained. The swap circuit in the self-testing protocol is used to mimic a physical experiment in a black-box setting, presuming that the reference experiment functions in a known minimum-dimensional quantum system. In such cases, the action of a local unitary transfers the properties of the physical state to the ancillary system, thus enabling the certification of both the state and the measurements. The optimal quantum violation of the GBI requires the condition in Eq.~(\ref{selfn}) to be satisfied. 
 
 Consider that the required bipartite state is $\ket{\psi}_{AB}$, and the corresponding local observables of Alice(Bob) are $X_m$ and $Z_m$, with $m\in\{A,B\}$ Swap-circuit aims to replicate the same measurement statistics as the physical setup while creating an isometry that acts jointly on the physical system and the reference system. This is mathematically expressed as 
\begin{eqnarray}
\label{stest}
\Phi(\mathcal{O}_t\ket{\psi}_{AB}\otimes\ket{ancilla}_{A'B'} 
 )&=&\ket{\chi}_{AB}\otimes U_t \ket{ent}_{A'B'}
\end{eqnarray}
where $\mathcal{O}_1=\openone_d, \mathcal{O}_2=X_m$ and $\mathcal{O}_3=Z_m, \forall m\in[A, B]$. Also,  $\ket{\psi}_{AB}$ is the physical state, $\ket{ancilla}_{A'B'}$ is the ancillary system, $\ket{\chi}_{AB}$ is the so-called junk' state, $\ket{ent}_{A'B'}$ is the entangled state of an ancillary system and $U_t$ is the unitary operation acting on the ancillary system. For $t=1$, $U_1=\openone_2$. We construct the observables $X_m$ and $Z_m$  in such a way that they satisfy the normalization condition and the self-testing conditions. The construction of observables is different for different input values $n$. For odd $n$, we construct
\begin{eqnarray}
\label{stest1}\na
    Z_A&=&A_\frac{n+1}{2}, 
    X_A =\frac{1}{\lfloor\frac{n}{2}\rfloor}\sum_{i=1}^{{\lfloor \frac{n}{2}\rfloor}}\frac{A_{i}-A_{n+1-i}}{\sqrt{2-\langle\{A_{i},A_{n+1-i}\}\rangle}}\nonumber\\
    Z_B&=&\mathcal{B}_\frac{n+1}{2},
    X_B = \frac{1}{\lfloor\frac{n}{2}\rfloor}\sum_{i=1}^{{\lfloor \frac{n}{2}\rfloor}}\frac{\mathcal{B}_{i}-\mathcal{B}_{n+1-i}}{\sqrt{2-\langle\{\mathcal{B}_{i},\mathcal{B}_{n+1-i}\}\rangle}}\label{n4}
\end{eqnarray}
and for even $n$,
\begin{eqnarray}
    Z_A &=&\frac{2}{n}\sum_{i=1}^{ \frac{n}{2}}\hspace{-0.2cm}\frac{A_{i}+A_{n+1-i}}{\sqrt{2+\langle\{A_{i},A_{n+1-i}\}\rangle}},
    X_A =\frac{2}{n}\sum_{i=1}^{ \frac{n}{2}}\hspace{-0.2cm}\frac{A_{i}-A_{n+1-i}}{\sqrt{2-\langle\{A_{i},A_{n+1-i}\}\rangle}}\nonumber\\ \na 
    Z_B &=&\frac{2}{n}\sum_{i=1}^{ \frac{n}{2}}\hspace{-0.2cm}\frac{\mathcal{B}_{i}+\mathcal{B}_{n+1-i}}{\sqrt{2+\langle\{\mathcal{B}_{i},\mathcal{B}_{n+1+i}\}\rangle}},
    X_B =\frac{2}{n}\sum_{i=1}^{ \frac{n}{2}}\hspace{-0.2cm}\frac{\mathcal{B}_{i}-\mathcal{B}_{n+1-i}}{\sqrt{2-\langle\{\mathcal{B}_{i},\mathcal{B}_{n+1-i}\}\rangle}}\\   
\end{eqnarray}
\begin{figure}
    \centering
    \includegraphics[width=8cm, height=4cm]{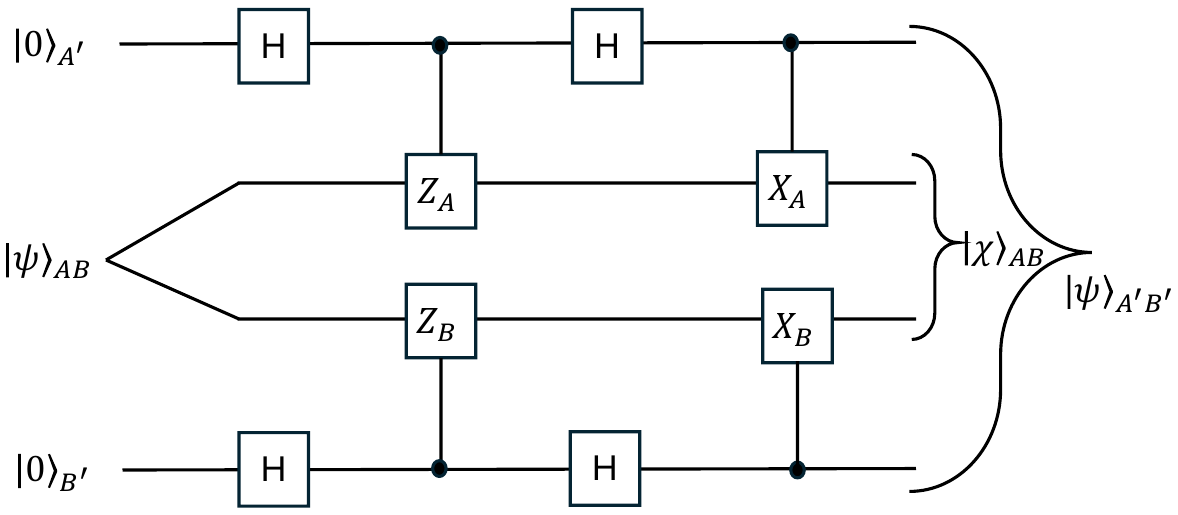}
    \caption{Swap circuit for self-testing of state  and observables for arbitrary $n$}\label{Oddc}
\end{figure}
    It is straightforward to check that
\begin{eqnarray}\label{circuitself}
    Z_A\ket{\psi}_{AB}=Z_B\ket{\psi}_{AB}&,& X_A\ket{\psi}_{AB}=X_B\ket{\psi}_{AB}\nonumber\\
    \{Z_A, X_A\}\ket{\psi}_{AB} &=& \{Z_B, X_A\}\ket{\psi}_{AB}=0.
\end{eqnarray}
 A detailed construction of isometries for even and odd values of $n$ are provided in the Appendix.~\ref{c5} and \ref{c4}.

For experimental authentication, it will be more helpful to find the isometry considering known dimensions. If $\ps_{AB}\in \mathcal{H}_A\otimes \mathcal{H}_B$ is the physical state, which satisfies Eq.~(\ref{circuitself}) and  $A_i\in \mathcal{H}_A$, $B_j\in \mathcal{H}_B$ are the observables that give the optimal value of $\mathcal{G}_n$. Hence, from the swap circuit in Fig.~\ref{Oddc}, we prove that there exist a local isometry $\Phi$ and a local ancilla $\ket{00}_{A'B'}$ so that
\begin{eqnarray}
    \Phi(\mathcal{O}_t\ket{\psi}_{AB}\otimes\ket{00}_{A'B'} 
 )&=&\ket{\chi}_{AB}\otimes U_t \ \ket{\phi^+}_{A'B'}
\end{eqnarray}
where $\ket{\chi}_{AB}=\frac{1+Z_A}{\sqrt{2}}\ket{\psi}_{AB}$ is the junk state. $\mathcal{O}_1=\openone_2, \mathcal{O}_2=X_m$ and $\mathcal{O}_3=Z_m,  \forall m\in\{A, B\}$, and $U_t$s are Pauli observables. Now we can express $A_i$ and $B_j$ in terms of $Z_A, Z_B, X_A, X_B$, therefore, the same relations will hold for $A_i$ and $B_j$, i.e.,
\begin{eqnarray}
\phi(A_i\ket{\psi}_{AB}\otimes\ket{00}_{A'B'} 
 )&=&\ket{\chi}_{AB}\otimes (A'_i\otimes \openone_2)\ket{\phi^+}_{A'B'}\\
\phi(B_j\ket{\psi}_{AB}\otimes\ket{00}_{A'B'} 
 )&=&\ket{\chi}_{AB}\otimes (\openone_2\otimes B'_j)\ket{\phi^+}_{A'B'}\\
 \phi(A_i\otimes B_j\ket{\psi}_{AB}\otimes\ket{00}_{A'B'} 
 )&=&\ket{\chi}_{AB}\otimes (A'_i\otimes B'_j)\ket{\phi^+}_{A'B'}
\end{eqnarray}
The detailed calculation is provided in the appendix.~\ref{c5} and \ref{c4}. Thus, we demonstrate that the self-testing protocol based on the optimal quantum value $(\mathcal{G}_n)_Q^{opt}$ given in Eq.~(\ref{cnopt}) provides the equivalence between the reference and physical experiments.

\section{Robust self-testing using swap circuit} \label{Sec4}
We note that in a real experiment, noise and imperfections in implementation do not allow us to achieve the optimal quantum value. Then it is crucial to analyze the robustness of a self-testing protocol in the presence of noise. Here, we provide a rigorous analysis of the robustness of our protocol in the presence of noise that leads to sub-optimal quantum violation. This analysis does not require the exact root cause of the error; instead, we may assume that it arises solely from the imperfect implementation of observables \cite{Bamps2015, Acin2020}.

Assume that the imperfect version of the observables ($\Tilde{X}_m, \Tilde{Z}_m$) differs from the ideal observables ($X_m, Z_m$) with $m\in\{A,B\}$, so that

\begin{eqnarray}
    ||(\Tilde{X}_m-X_m)\ket{\psi}_{AB}||\leq \alpha_m, \ \ \ ||(\Tilde{Z}_m-Z_m)\ket{\psi}_{AB}||\leq \beta_m
\end{eqnarray}
where $(\alpha_m, \beta_m)\in \mathbb{R}^+$. In an ideal condition, $ \alpha_m=\beta_m=0$. However, due to the presence of noise, the observables are not necessarily unitary. Therefore, in Eq.~(\ref{kni}) $\mathcal{K}_{n,i}$ changes to $0\leq||\Tilde{\mathcal{K}}_{n,i}\ket{\psi}_{AB}||\leq \xi_{n,i}$ instead of $||\mathcal{K}_{n,i}\ket{\psi}_{AB}||=0$. Consequently, 
\begin{eqnarray}
    \Tr[\Tilde{\Gamma}_n\  \rho_{AB}]&=& \frac{1}{2}\sum_{i=1}^{n} \mu_{n,i} \bra{\psi}_{AB}\Tilde{\mathcal{K}}^{\dag}_{n,i} \Tilde{\mathcal{K}}_{n,i}\ket{\psi}_{AB}\nonumber\\
    &=& \frac{1}{2}\sum_{i=1}^{n} \mu_{n,i}  \ \xi_{n,i}^2
\end{eqnarray}
This leads to a sub-optimal quantum value of the Bell functional
\begin{eqnarray}\label{Cn noise}
    \Tilde{(\mathcal{G}_{n})}_{Q}=&n\csc {\frac{\pi}{2n}}-\xi
\end{eqnarray}
where $\xi=\frac{1}{2}\sum_{i=1}^{n} \mu_{n,i}  \ \xi_{n,i}^2\geq 0$. For robust self-testing of state and observables, we show that the trace distance between the output of the perfect and imperfect implementations is
\begin{eqnarray}
    ||\Tilde{\Phi}(\Tilde{\mathcal{O}}_t\ket{\psi}_{AB}\otimes\ket{00}_{A'B'})&-&\Phi(\mathcal{O}_t\ket{\psi}_{AB}\otimes\ket{00}_{A'B'})||\nonumber\\
 &&\hspace{8mm} \leq F_t(\alpha_A,\alpha_B,\beta_A,\beta_B)
\end{eqnarray}
where $\mathcal{O}_1=\openone_d, $ $\mathcal{O}_2=X_m$ , $\mathcal{O}_3=Z_m, \forall m\in\{A,B\}$ and 
\ba \na F_1(\alpha_A,\alpha_B,\beta_A,\beta_B)&=&F_S(\alpha_A,\alpha_B,\beta_A,\beta_B) \\
F_2(\alpha_A,\alpha_B,\beta_A,\beta_B)&=&F_{O_x}(\alpha_A,\alpha_B,\beta_A,\beta_B)\\\na 
F_3(\alpha_A,\alpha_B,\beta_A,\beta_B)&=&F_{O_Z}(\alpha_A,\alpha_B,\beta_A,\beta_B)\ea with 
\begin{eqnarray}
    &&\lim_{\{\alpha_A,\alpha_B,\beta_A,\beta_B\}\to 0} F_t(\alpha_A,\alpha_B,\beta_A,\beta_B)=0.
\end{eqnarray}
We give the detailed derivations in the Appendix.~\ref{robustselftesting}.

Note that errors can arise from imperfections in state preparation, observable implementation, or both. However, this is extremely challenging to demonstrate, and therefore, we restrict our analysis by considering the imperfection that arises from one party's implementation of observables.  
If Alice implements imperfect observables and the error is the same for both implementations, then $\alpha_A=\beta_A=\epsilon\geq 0$. If the error of each Alice observable is $\delta$, then 
\ba ||(\Tilde{A}_i-A_i)\ket{\psi}_{AB}||\leq \delta\implies \Tilde{A}_i\approx A_i+\delta \ \openone_d, \forall i\in [n], \delta \geq 0\na \\
\ea  
which leads to $||(\Tilde{Z}_A-Z_A)\ket{\psi}_{AB}||=||(\Tilde{A}_{\frac{n+1}{2}}-A_{\frac{n+1}{2}})\ket{\psi}_{AB}||\leq \delta$, and therefore we may take $\delta\approx\epsilon$. Then the output isometry differs from the ideal one, as
\begin{eqnarray}
    &&||\Tilde{\Phi}(\ket{\psi}_{AB}\otimes\ket{00}_{A'B'})-\Phi(\ket{\psi}_{AB}\otimes\ket{00}_{A'B'})||\leq 4\epsilon +\epsilon^2\\
    &&||\Tilde{\Phi}(\Tilde{O}\ket{\psi}_{AB}\otimes\ket{00}_{A'B'})-\Phi(O\ket{\psi}_{AB}\otimes\ket{00}_{A'B'})||\nonumber\\
    && \quad \hspace{25mm}\leq \epsilon^3+5\epsilon^2+8\epsilon, \quad \forall O\in\{X_A,Z_A\}
\end{eqnarray}
We derive the relation between $\epsilon$ and the deviation $\xi$ from the optimal value (in Eq.~(\ref{Cn noise})) as $\epsilon =\sqrt{\frac{2\xi }{n \csc \left(\frac{\pi }{2 n}\right)}}$. Detailed derivation is provided in the Appendix~\ref{spodd}. To determine the robust self-testing bounds for both state and observables, we define the relative observed violations $(r)$ as a function of $n$ and $\xi$, where
\begin{eqnarray}
    &&r=\frac{\Tilde{(\mathcal{G}_{n})}_{Q}-(\mathcal{G}_n)_C}{(\mathcal{G}_{n})^{opt}_{Q}-(\mathcal{G}_n)_C}= 1-\frac{\xi}{ n \csc{\frac{\pi}{2n}}-\lfloor\frac{n^2+1}{2}\rfloor}
\end{eqnarray}
Alternatively, $\xi$ can be written in terms of $r$ and $n$ as
\begin{eqnarray}
        &&\xi =(1-r) \left( n \csc{\frac{\pi}{2n}}-\lfloor\frac{n^2+1}{2}\rfloor\right)
\end{eqnarray}
where $\Tilde{(\mathcal{G}_{n})}_{Q}$ and $ (\mathcal{G}_{n})^{opt}_{Q}$ are defined in Eqs.~(\ref{Cn noise}) and (\ref{cnopt}) respectively. Here $(\mathcal{G}_n)_c =\lfloor\frac{n^2+1}{2}\rfloor$ is the classical bound of the GBI.

Hence, the trace distance between the observed and ideal states in terms of the relative observed value $r$ is
\begin{eqnarray}
    &&||\Tilde{\Phi}(\ket{\psi}_{AB}\otimes\ket{00}_{A'B'})-\Phi(\ket{\psi}_{AB}\otimes\ket{00}_{A'B'})||\leq  f_s(r)\quad
\end{eqnarray}
 where  
 \begin{eqnarray}
 f_s(r)&=&\frac{2(1-r)\qty(n\csc{\frac{\pi}{2 n}}-\lfloor\frac{n^2+1}{2}\rfloor)}{n\csc{\frac{\pi}{2 n}}}\nonumber\\
 &&+4\sqrt{2}\sqrt{\frac{(1-r)\qty(n\csc{\frac{\pi}{2 n}}-\lfloor\frac{n^2+1}{2}\rfloor)}{n\csc{\frac{\pi}{2 n}}}}
 \end{eqnarray} 
 \begin{figure}[ht]
\centering
\begin{tikzpicture}
\begin{axis}[
    xlabel={Relative observed violation $(r)$},
    ylabel={Fidelity $F_s(r)$},
    xmin=0.93, xmax=1, 
    ymin=0.5, ymax=1,
    xtick={0.93, 0.94, 0.95, 0.96, 0.97, 0.98, 0.99, 1.00}, 
    xticklabel style={
        font=\small,
        /pgf/number format/fixed, 
        /pgf/number format/precision=2 
    },
    legend style={at={(0.02,0.58)}, anchor=south west, draw=none},
    width=8.5cm,
    height=6.2cm,
    domain=0.93:1, 
    samples=1000, 
    axis lines=box,
    tick label style={font=\small}, 
    label style={font=\small}, 
    legend cell align={left}
]

\addplot[red, thick]
    {((1/6)*x - (2*sqrt(1 - x))/sqrt(3) + 5/6)^2};
\addlegendentry{$n = 3$}

\addplot[blue, thick]
    {((1 - (13/20)*(sqrt(5) - 1))*x - (2*sqrt(10)/5)*sqrt(((13/4)*(sqrt(5) - 1) - 5)*(x - 1)) + (13/20)*(sqrt(5) - 1))^2};
\addlegendentry{$n = 5$}

\addplot[green, thick]
    {((1 - (25/7)*sin(deg(pi/14)))*x - (2*sqrt(14)/7)*sqrt((25*sin(deg(pi/14)) - 7)*(x - 1)) + (25/7)*sin(deg(pi/14)))^2};
\addlegendentry{$n = 7$}

\addplot[violet, thick]
    {((1 - (61/11)*sin(deg(pi/22)))*x - (2*sqrt(22)/11)*sqrt((61*sin(deg(pi/22)) - 11)*(x - 1)) + (61/11)*sin(deg(pi/22)))^2};
\addlegendentry{$n = 11$}

\end{axis}
\end{tikzpicture}
\caption{Trade-off between relative observed violation ($r$) and fidelity $F_s(r)$ for $n=3,5,7$ and $11$. Here, we are taking only that region where fidelity converges to one.}\label{osn1} 
\end{figure}
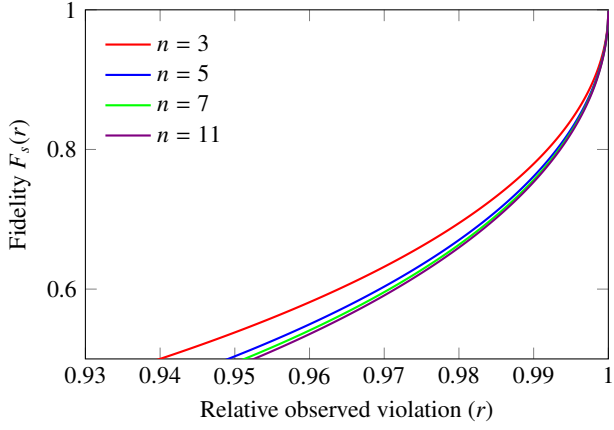
\begin{figure}[ht]
\centering
\begin{tikzpicture}
\begin{axis}[
    xlabel={Relative observed violation $(r)$},
    ylabel={Fidelity $F_o(r)$},
    xmin=0.98, xmax=1, 
    ymin=0.5, ymax=1,
    xtick={0.98, 0.99, 1.00}, 
    xticklabel style={
        font=\small,
        /pgf/number format/fixed, 
        /pgf/number format/precision=2 
    },
    legend style={at={(0.02,0.58)}, anchor=south west, draw=none},
    width=8.5cm,
    height=6.2cm,
    domain=0.98:1, 
    samples=200,
    axis lines=box,
    tick label style={font=\small}, 
    label style={font=\small}, 
    legend cell align={left}]

\addplot[red, thick]
    {((0.5 * ( -pow(1 - x, 1.5)/(3*sqrt(3)) + (5*(x - 1))/3 - (8*sqrt(1 - x))/sqrt(3) )) + 1)^2};
\addlegendentry{$n = 3$}

\addplot[blue, thick]
    {((0.5 * (
  - (1/5)*2*sqrt(2/5)*pow( ((13/4)*(sqrt(5) - 1) - 5)*(x - 1) , 1.5)
  + 2*(5 - (13/4)*(sqrt(5) - 1))*(x - 1)
  - 8*sqrt(2/5)*sqrt( ((13/4)*(sqrt(5) - 1) - 5)*(x - 1) )
)) + 1)^2};
\addlegendentry{$n = 5$}

\addplot[green!60!black, thick]
    {((0.5 * (
  - (1/7)*2*sqrt(2/7)*pow( (x - 1)*(25*sin(deg(pi/14)) - 7) , 1.5) 
  + (10/7)*(x - 1)*(7 - 25*sin(deg(pi/14)))
  - 8*sqrt(2/7)*sqrt((x - 1)*(25*sin(deg(pi/14)) - 7))
)) + 1)^2};
\addlegendentry{$n = 7$}

\addplot[purple, thick]
    {((0.5 * (
  - (1/11)*2*sqrt(2/11)*pow((x - 1)*(61*sin(deg(pi/22)) - 11), 1.5) 
  + (10/11)*(x - 1)*(11 - 61*sin(deg(pi/22)))
  - 8*sqrt(2/11)*sqrt((x - 1)*(61*sin(deg(pi/22)) - 11))
)) + 1)^2};
\addlegendentry{$n = 11$}

\end{axis}
\end{tikzpicture}
\caption{Trade-off between relative observed violation ($r$) and fidelity $F_o(r) \ \forall O\in\{X_B,Z_B\}$ for $n=3,5,7$ and $11$. Here, we are taking only that region where fidelity converges to one.}\label{osn2}

\end{figure}
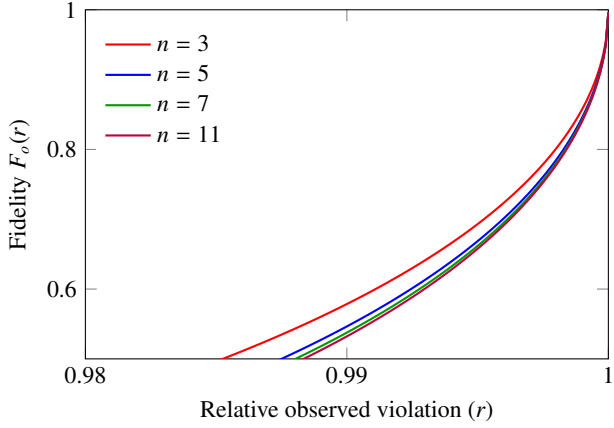
 The detailed derivation is provided in the Appendix.~\ref{spodd}. Now, using the Fuchs-Van de Graaf inequality \cite{Wilde2013}, the approximate relation between the trace distance $f_s(r)$ and robust fidelity $F_s(r)$ is 
 \begin{eqnarray}
     2\qty(1-\sqrt{F_s(r)})\leq f_s(r)\leq 2\sqrt{1-F_s(r)}
 \end{eqnarray}
 This gives the lower bound of fidelity in terms of trace distance, as
 \begin{eqnarray}
     F_k(r)\geq \qty(1-\frac{1}{2}f_k(r))^2\quad \forall k\in\{s,o\}
 \end{eqnarray}
where, for the observables, we define the fidelity $F_o(r)$ in terms of the trace distance $f_o(r)$ where  
\begin{eqnarray}
\nonumber
f_o(r)&=&2\sqrt{2}\qty(\frac{(1-r)\qty(n\csc{\frac{\pi}{2 n}}-\lfloor\frac{n^2+1}{2}\rfloor)}{n\csc{\frac{\pi}{2 n}}})^{\frac{3}{2}}\nonumber\\
&&+8\sqrt{2}\qty(\frac{(1-r)\qty(n\csc{\frac{\pi}{2 n}}-\lfloor\frac{n^2+1}{2}\rfloor)}{n\csc{\frac{\pi}{2 n}}})^\frac{1}{2}\nonumber\\
&&+\frac{10(1-r)\qty(n\csc{\frac{\pi}{2 n}}-\lfloor\frac{n^2+1}{2}\rfloor)}{n\csc{\frac{\pi}{2 n}}}  
\end{eqnarray}
Figures~\ref{osn1} and \ref{osn2} depict the interplay between the lower bound of the fidelity function and the corresponding relative observed violation. These plots indicate that, when the swap circuit is utilized, the ex-tractability of both quantum states and observables remains stable as long as the fidelity is greater than $\frac{1}{2}$. For the case of $n=3$, this critical fidelity threshold is attained at $r = 0.93998 \ (0\leq\epsilon\leq 0.1414)$ for the state and at $r = 0.9852 \ (0\leq\epsilon\leq 0.0701)$ for the observables. 

We note here that increasing the number of measurement settings $n$ can enhance the value of observed violations, but at the same time it makes both the state and the observable extraction procedure more vulnerable to noise and experimental imperfections(see Figs.~\ref{osn1} and \ref{osn2}). A more detailed analysis is presented in Appendix~\ref{spodd}.
 \section{Summary and Discussion}  
\label{Sec5}
To summarize, we have demonstrated the self-testing of quantum states and measurements based on the optimal quantum violation of the GBI  \cite{Gisin1999} featuring  an arbitrary number $n$ of dichotomic measurements per party. We introduced a systematic and analytically transparent SOS method to derive  this optimal violation without assuming the dimension of the quantum system. The SOS method enabled us  to derive the explicit functional relations between local observables and the required entangled state directly from the optimization conditions.  Owing to its generality, the SOS method developed here can be readily applicable to derive the optimal quantum value of a broad class of arbitrary-input Bell functionals.

We further developed a swap-circuit–based self-testing method that certifies a black-box quantum device by constructing a local isometry that exchanges its properties with the known reference system under the assumption of minimal dimension. To connect the theoretical framework with experimental reality, we carried out a detailed robustness analysis of this swap-circuit self-testing protocol. This analysis concentrates on errors in measurement implementation, typically dominant experimental noise, by quantifying how far the certified state can deviate from the ideal target state. We also characterized how this robustness depends on the number of measurement settings per party when only one party (e.g., Alice) employs imperfect observables.  

The DI self-testing framework developed here, which accommodates an arbitrary number of distinct measurement settings, effectively certifies a broad family of observables and thereby paves the way for diverse applications in quantum computational tasks. An additional promising direction for subsequent investigations is the construction and analysis of analogous Bell inequalities that incorporate measurement settings with more than two possible outcomes, both in bipartite and multipartite scenarios.  It could be interesting to examine to what extent such generalized protocols enable robust DI self-testing and certification of genuine randomness. The systematic exploration of these questions constitutes a compelling avenue for future research.

\section*{Acknowledgments}
RP acknowledges the financial support from the Council of Scientific and Industrial Research (CSIR, 09/1001(12429)/2021-EMR-I), Government of India. AKP acknowledges the support
from Research Grant No. SERB/CRG/2021/004258, Government of India.

\vspace{3cm}
\appendix
\begin{widetext}
\section{Relation between the Alice's observables}\label{Bobo}
In the main text, Eq.~(\ref{aiix}) characterizes the relationship between Bob's observables. To obtain these relations, we proceed as follows. The optimization conditions (cf. Eq.~(\ref{selfn}) in the main text) are
\begin{eqnarray}\label{rum1}
    A_{i}\otimes \mathcal{B}_{i}\ket{\psi}_{AB}=\ket{\psi}_{AB}\quad \forall i\in[n]
\end{eqnarray}
where we have $\mathcal{B}_{i}=\frac{\sum\limits\limits_{j=1}^{n-i+1}{B}_j-\sum\limits\limits_{j=n-i+2}^n {B}_j}{\csc{\frac{\pi}{2n}}}$.
Hence,  we can rewrite Eq.~(\ref{rum1}) as follows. 
\begin{eqnarray}
&&A_i\otimes \frac{\sum\limits\limits_{j=1}^{n-i+1}{B}_j-\sum\limits\limits_{j=n-i+2}^n {B}_j}{\csc{\frac{\pi}{2n}}}\ket{\psi}_{AB}=\ket{\psi}_{AB}\nonumber\\
&&A_i\otimes \openone_d\ket{\psi}_{AB}=\openone_d\otimes\frac{\sum\limits\limits_{j=1}^{n-i+1}{B}_j-\sum\limits\limits_{j=n-i+2}^n {B}_j}{\csc{\frac{\pi}{2n}}}\ket{\psi}_{AB}\label{rum2}\\
    &&\bra{\psi}_{AB}A_i\otimes \openone_d=\bra{\psi}_{AB}\openone_d\otimes\frac{\sum\limits\limits_{j=1}^{n-i+1}{B}_j-\sum\limits\limits_{j=n-i+2}^n {B}_j}{\csc{\frac{\pi}{2n}}}\label{rum2t}
\end{eqnarray}
Since each of the observables $A_i$ and $B_k$ is Hermitian, $\bra{\psi}_{AB} A_i\otimes B_k\ket{\psi}_{AB}$ is real for each $i,k\in[n]$. Hence, using Eq.~(\ref{rum2}) and (\ref{rum2t}), we get 
\begin{eqnarray}
    \bra{\psi}_{AB}A_i\otimes B_k\ket{\psi}_{AB}= \begin{cases}
        \bra{\psi}_{AB}\openone_d\otimes B_k \frac{\sum\limits\limits_{j=1}^{n-i+1}{B}_j-\sum\limits\limits_{j=n-i+2}^n {B}_j}{\csc{\frac{\pi}{2n}}}\ket{\psi}_{AB}\nonumber\\
\bra{\psi}_{AB}\openone_d\otimes \frac{\sum\limits\limits_{j=1}^{n-i+1}{B}_j-\sum\limits\limits_{j=n-i+2}^n {B}_j}{\csc{\frac{\pi}{2n}}}B_k \ket{\psi}_{AB}
    \end{cases}&=&\frac{1}{2}\bra{\psi}_{AB}\openone_d\otimes \frac{\sum\limits\limits_{j=1}^{n-i+1}\{B_k,{B}_j\}-\sum\limits\limits_{j=n-i+2}^n \{B_k,{B}_j\}}{\csc{\frac{\pi}{2n}}} \ket{\psi}_{AB}\nonumber\\
    &=&\frac{1}{2\csc{\frac{\pi}{2n}}} \qty(\sum\limits\limits_{j=1}^{n-i+1}\langle\{B_k,{B}_j\}\rangle-\sum\limits\limits_{j=n-i+2}^n \langle\{B_k,{B}_j\}\rangle)
\end{eqnarray}
For notational convenience, we relabel the index $k$ as $j$ and $j$ as $k$, which yields
\begin{eqnarray}
    \bra{\psi}_{AB}A_i\otimes B_j\ket{\psi}_{AB}&=&\frac{1}{2\csc{\frac{\pi}{2n}}} \qty(\sum\limits\limits_{k=1}^{n-i+1}\langle\{B_j,{B}_k\}\rangle-\sum\limits\limits_{k=n-i+2}^n \langle\{B_j,{B}_k\}\rangle)\label{rand0}
\end{eqnarray}
Now our motivation is to prove $\mathcal{A}_j\otimes B_j\ket{\psi}_{AB}=\ket{\psi}_{AB},\forall j\in[n]$, where $\mathcal{A}_j=\frac{\sum\limits\limits_{i=1}^{n-j+1}{A}_i-\sum\limits\limits_{i=n-j+2}^n {A}_i}{\csc{\frac{\pi}{2n}}},\forall j\in[n]$. As each of the $\mathcal{A}_j$ has the $n$ number of $A_j$, hence from Eq.~(\ref{rand0}), it becomes very difficult to prove the desired one. Hence, here we take the value of $n$ as three. Now, putting $n=3$ and $i,j=[3]$ in Eq.~(\ref{rand0}), we get the following.
\begin{eqnarray}\label{s3o}
     &&\bra{\psi}_{AB}A_1\otimes B_1\ket{\psi}_{AB}=\bra{\psi}_{AB}A_1\otimes B_3\ket{\psi}_{AB}=\bra{\psi}_{AB}A_2\otimes B_2\ket{\psi}_{AB}=-\bra{\psi}_{AB}A_2\otimes B_3\ket{\psi}_{AB}=\bra{\psi}_{AB}A_3\otimes B_1\ket{\psi}_{AB}=\frac{1}{2}\nonumber\\
     &&\bra{\psi}_{AB}A_3\otimes B_2\ket{\psi}_{AB}=-\frac{1}{2};\quad \bra{\psi}_{AB}A_1\otimes B_2\ket{\psi}_{AB}=\bra{\psi}_{AB}A_2\otimes B_1\ket{\psi}_{AB}=-\bra{\psi}_{AB}A_3\otimes B_3\ket{\psi}_{AB}=1
\end{eqnarray}
Hence from Eq.~(\ref{s3o}) we can write 
\begin{eqnarray}\label{s31}
    &&\bra{\psi}_{AB}(A_1+A_2+A3)\otimes B_1\ket{\psi}_{AB}=2;\quad \bra{\psi}_{AB}(A_1+A_2-A3)\otimes B_2\ket{\psi}_{AB}=2;\quad \bra{\psi}_{AB}(A_1-A_2-A3)\otimes B_1\ket{\psi}_{AB}=2\nonumber\\
    &&\bra{\psi}_{AB}\frac{(A_1+A_2+A3)}{2}\otimes B_1\ket{\psi}_{AB}=1;\quad \bra{\psi}_{AB}\frac{(A_1+A_2-A3)}{2}\otimes B_2\ket{\psi}_{AB}=1;\quad \bra{\psi}_{AB}\frac{(A_1-A_2-A3)}{2}\otimes B_1\ket{\psi}_{AB}=1
\end{eqnarray}
Hence Eq.~(\ref{s31}) can be generalized as 
\begin{eqnarray}\label{s34}
    \bra{\psi}_{AB}\frac{\sum\limits\limits_{i=1}^{3-j+1}{A}_i-\sum\limits\limits_{i=3-j+2}^3 {A}_i}{\csc{\frac{\pi}{6}}} \otimes B_j \ket{\psi}_{AB}=1\Rightarrow \mathcal{A}_j\otimes B_j \ket{\psi}_{AB}=\ket{\psi}_{AB}, \forall j\in[3], \  \mathcal{A}_j=\frac{\sum\limits\limits_{i=1}^{3-j+1}{A}_i-\sum\limits\limits_{j=3-i+2}^3 {A}_i}{\csc{\frac{\pi}{6}}}
\end{eqnarray}
Again from Eq.~(\ref{s31}) it is straight forward that $\frac{(A_1+A_2+A3)}{2},
\frac{(A_1+A_2-A3)}{2},
\frac{(A_1-A_2-A3)}{2}$ are normalized, which implies the following. 
\begin{eqnarray}\label{s33}
    &&\sqrt{3+\langle\{A_1,A_2\}\rangle+\langle\{A_2,A_3\}\rangle+\langle\{A_1,A_3\}\rangle}=\sqrt{3+\langle\{A_1,A_2\}\rangle-\langle\{A_2,A_3\}\rangle-\langle\{A_1,A_3\}\rangle}=2\nonumber\\
    &&\sqrt{3-\langle\{A_1,A_2\}\rangle+\langle\{A_2,A_3\}\rangle-\langle\{A_1,A_3\}\rangle}=2
\end{eqnarray}
Solving Eq.~(\ref{s33}) we get
\begin{eqnarray}\label{s35}
    \langle\{A_1,A_2\}\rangle=\langle\{A_2,A_3\}\rangle=-\langle\{A_1,A_3\}\rangle=1=2\cos{\frac{\pi}{3}}\Rightarrow \langle\{A_i,A_{i+x}\}\rangle =2\cos{\frac{\pi \ x}{3}},\quad\forall i\in[3],x\in[3-i]
\end{eqnarray}

In Eq.~(\ref{s34}) and (\ref{s35}) we have shown it for $n=3$, but it is appropriate for any arbitrary $n$. Hence, we can write self-testing conditions with respect to Alice, i.e.,
\begin{eqnarray}
    &&\mathcal{A}_j\otimes B_j \ket{\psi}_{AB}=\ket{\psi}_{AB}, \  \mathcal{A}_j=\frac{\sum\limits\limits_{i=1}^{n-j+1}{A}_i-\sum\limits\limits_{i=n-j+2}^n {A}_i}{\csc{\frac{\pi}{2n}}}, \forall j\in[n]\label{ososc}\\
    &&\langle\{A_i,A_{i+x}\}\rangle =2\cos{\frac{\pi \ x}{n}},\quad\forall i\in[n],x\in[n-i]\label{aliceco}
\end{eqnarray}
Hence, putting $x=1$ and $-1$ in Eq.~(\ref{aliceco}), we get 
\begin{eqnarray}\label{aliceai}
    \langle\{A_i,A_{i+1}\}\rangle =2\cos{\frac{\pi}{n}};\quad \langle\{A_i,A_{i-1}\}\rangle =2\cos{\frac{\pi}{n}}\Rightarrow \langle\{A_i,\frac{A_{i+1}+A_{i-1}}{2\cos{\frac{\pi }{n}}}\}\rangle =2
\end{eqnarray}
As $A_i$s are dichotomic normalized observables, hence, Eq.~(\ref{aliceai}) only holds when 
\begin{eqnarray}
    A_i=\frac{A_{i+1}+A_{i-1}}{2\cos{\frac{\pi }{n}}},\forall i\in[n]
\end{eqnarray}
Let us now derive the analogous condition for Bob’s observables. Using the optimization condition provided in the main text, we obtain
\begin{eqnarray}
    \langle\{B_j,B_{j+x}\}\rangle =2\cos{\frac{\pi x}{n}},\quad\forall j\in[n],\,x\in[n-j]\label{bobaco}
\end{eqnarray}
Substituting $x=1$ and $x=-1$ into Eq.~(\ref{bobaco}), we find
\begin{eqnarray}\label{bobai}
    \langle\{B_j,B_{j+1}\}\rangle =2\cos{\frac{\pi}{n}},\quad 
    \langle\{B_j,B_{j-1}\}\rangle =2\cos{\frac{\pi}{n}}
    \Rightarrow 
    \Big\langle\Big\{B_j,\frac{B_{j+1}+B_{j-1}}{2\cos{\frac{\pi}{n}}}\Big\}\Big\rangle =2
\end{eqnarray}
Since the $B_j$ are dichotomic, normalized observables, Eq.~(\ref{bobai}) can only be satisfied if
\begin{eqnarray}
    B_j=\frac{B_{j+1}+B_{j-1}}{2\cos{\frac{\pi }{n}}},\quad\forall j\in[n]
\end{eqnarray}
\section{Explicit derivation of the optimal quantum violation for \texorpdfstring{$n=5,6$ and $11$}{n=3 and n=5}}\label{SOS6,7,11}
 Here, we provide an explicit derivation of the optimal quantum violation $ (\mathcal{G}_n)^{opt}_Q$ for $n=5, 6$ and $11$ along with the state and observables, using our dimension-independent SOS approach.
   
\subsection{Derivation of optimal quantum violation of GBI for \texorpdfstring{$n=5$}{n=5}}\label{sos5}
Consider a bipartite setup with Alice and Bob, where each party measures five dichotomic observables, denoted by $A_i$ for Alice and $B_j$ for Bob, with $i,j \in [5]$. The corresponding GBI is obtained by substituting $n=5$ into Eq.~(\ref{gisinineq}) in the main text, we get
\begin{equation}
\begin{split}
    \mathcal{G}_5&=\sum\limits\limits_{i=1}^5 \qty(\sum\limits\limits_{j=1}^{6-i}{A}_i{B}_j-\sum\limits\limits_{j=7-i}^5 {A}_i{B}_j)\leq 13
\end{split}
\end{equation}
Following the  SOS approach as outlined in Sec.~\ref{SOSsec}, we get that  the optimal value of $(\mathcal{G}_{5})_Q$ is obtained if $\Tr[ \Gamma_5\ \rho_{AB}]=0$ and  thus we get 
\begin{equation}
   \qty(\mathcal{G}_5)_Q^{\ opt}= \text{max}\sum\limits\limits_{i=1}^5 \mu_{5,i}
\end{equation}
where $\mu_{5,i}=||\sum\limits\limits_{j=1}^{6-i}{B}_j-\sum\limits\limits_{j=7-i}^5 {B}_j||_{\rho_{AB}},\forall i\in[5]$.Using the inequality in Eq.~(\ref{cnv}), we get
\ba
(\mathcal{ G}_{5})_Q&\leq&\sqrt{5\sum\limits_{i=1}^5\left[\mu_{5,i}\right]^2}
 \ea
 The equality  holds only  when $\mu_{5,i}=\mu_{5,j},\forall i\neq j\in[5]$. Further, we can write
\begin{eqnarray}\label{g5del}
     (\mathcal{G}_5)_Q &\leq&\Bigg[5\bigg(25+3\Big[\eac{B_1}{B_2}+\eac{B_2}{B_3}+\eac{B_3}{B_4}+\eac{B_4}{B_5}-\eac{B_1}{B_5}\Big] + \eac{B_1}{B_3}-\eac{B_1}{B_4}\nonumber\\
     &&+\eac{B_2}{B_4}-\eac{B_2}{B_5}+\eac{B_3}{B_5}\bigg)\Bigg]^{\half}
\end{eqnarray}
To solve this, we consider a positive operator $\Delta_5 = \left(B_1-B_2+B_3-B_4+B_5\right)$, such that
\begin{eqnarray}
  \Delta_5^2&=&5\I- \bigg(\eac{B_1}{B_2}+\eac{B_2}{B_3}+\eac{B_3}{B_4}+\eac{B_4}{B_5} -\eac{B_1}{B_5}\bigg)+  \eac{B_1}{B_3}-\eac{B_1}{B_4}\nonumber\\
  &&+\eac{B_2}{B_4}-\eac{B_2}{B_5}+\eac{B_3}{B_5}  
\end{eqnarray}
For convenience, we set $\Delta_5^2=\langle (\Delta_5)^2\rangle$, so that Eq.~(\ref{g5del}) can be cast in the following form
\begin{eqnarray}
        (\mathcal{G}_5)_Q & \leq&\Bigg[200+20\bigg(\eac{B_1}{B_3}-\eac{B_1}{B_4}+\eac{B_2}{B_4}-\eac{B_2}{B_5}+\eac{B_3}{B_5}\bigg)-15\Delta_5^2\Bigg]^\half
\end{eqnarray}
In this way, to obtain the maximum of $(\mathcal{G}_5)_Q$, the value of $\Delta_5^2$ must be zero. To optimize, let us consider the following.
\begin{eqnarray}
    \delta_5&=&\eac{B_1}{B_3}-\eac{B_1}{B_4}+\eac{B_2}{B_4} -\eac{B_2}{B_5}+\eac{B_3}{B_5}\nonumber\\
\end{eqnarray}

Taking into account $B_4=\frac{B_2-B_1}{\mu^1_{5}}, B_5=\frac{B_3-B_2}{\mu^2_{5}}$, where $\mu^1_{5}=\sqrt{2-\eac{B_1}{B_2}}$ and $\mu^2_{5}=\sqrt{2-\eac{B_2}{B_3}}$, we arrive at
\begin{equation}
    \delta_5\leq2\sqrt{2\qty(4-\eac{B_1}{B_2}-\eac{B_2}{B_3})}+\eac{B_1}{B_3}.
\end{equation}
The equality only holds when $\mu^1_{5}=\mu^2_{5}$, leads to  $\eac{B_1}{B_2}=\eac{B_2}{B_3}$, again considering $B_2=\frac{B_1+B_3}{\nu^3_5}$, where $\nu^3_5=\sqrt{2+\eac{B_1}{B_3}}$, we get
\begin{equation}
    \delta_5\leq 2\sqrt{2\qty(4-2\sqrt{2+\eac{B_1}{B_3}})}+\eac{B_1}{B_3}.
\end{equation}
Maximal value of $\delta_5=5\qty(\frac{-1+\sqrt{5}}{2})$, is achieved when $\eac{B_1}{B_3}=\frac{\sqrt{5}-1}{2}=2\cos{\frac{2\pi}{5}}$. Subsequently, \[\nu^3_5=\frac{\sqrt{5}+1}{2}\] which, further implies $B_2=\frac{\left(\sqrt{5}-1\right)}{2}(B_1+B_3)$. Hence $\eac{B_1}{B_2}=\eac{B_2}{B_3}=\frac{\sqrt{5}+1}{2}=2\cos{\frac{\pi}{5}}$,from  which we get \[\nu^1_5=\nu^2_5=\sqrt{\frac{1}{2} \left(\sqrt{5}+5\right)}\] 

Hence, $B_4=\sqrt{\frac{2}{\sqrt{5}+5}}(B_2-B_1)$, $B_5=\sqrt{\frac{2}{\sqrt{5}+5}}(B_3-B_2)$, which gives $\eac{B_2}{B_4}=\eac{B_3}{B_5}=\frac{\sqrt{5}-1}{2}=2\cos{\frac{2\pi}{5}}$ and $\eac{B_1}{B_4}=\eac{B_2}{B_5}=\frac{1-\sqrt{5}}{2}=2\cos{\frac{3\pi}{5}}$. In a similar way, we can derive the remaining relations between Bob's observables, which follow the following relation, i.e.,
\begin{eqnarray}
    \eac{B_j}{B_{j+x}}=2\cos{\frac{\pi\ x}{5}},\quad\forall x\in[5-j]
\end{eqnarray}
Consequently, from the normalization
\begin{eqnarray}
    \mu_{5,i} = 1+\sqrt{5}=\csc{\frac{\pi}{10}},\quad \forall i\in [5]
\end{eqnarray}
Therefore, the optimal violation of GBI for $n=5$ is \[\qty(\mathcal{G}_5)_Q^{\ opt} = 5(1+\sqrt{5})=5 \csc{\frac{\pi}{10}}\] 
Similarly, from the appendix.~(\ref{Bobo}) we get the Alice's observables follows the following property
\begin{eqnarray}
    \eac{A_i}{A_{i+x}}=2\cos{\frac{\pi x}{5}},\quad \forall x\in[5-i], i\in[5]
\end{eqnarray}
Following  Eq.~(\ref{rho odd}) of the main text, we can write the required state $\rho_{AB}$ by substituting  $n=5$ in Eq.~(\ref{C'so}). Thus we get 
\ba\label{state5} C_1 \otimes C_1 &=& \mathcal{A}_3 \otimes B_3, \
C_2 \otimes C_2=\dfrac{1}{2}\Bigg[\frac{(\mathcal{A}_1-\mathcal{A}_5) \otimes (B_1-B_5)}{(2-\langle\{B_1,B_5\}\rangle)}+\frac{(\mathcal{A}_2-\mathcal{A}_4) \otimes (B_2-B_4)}{(2-\langle\{B_2,B_4\}\rangle)}\Bigg], \ 
C_3 \otimes C_3 =(C_1 \otimes C_1)\cdot (C_2 \otimes C_2)\nonumber\\
\ea
where, $\mathcal{A}_j=\frac{1}{\csc{\frac{\pi}{10}}}\qty(\sum\limits\limits_{i=1}^{6-j}A_i-\sum\limits\limits_{i=7-j}^{5}A_i),\quad \forall j\in[5]$. The detailed derivation is given in Appendix.~(\ref{Bobo}).\\
As an example, in a two-qubit system, we provide a set of observables as follows. 
 \ba
  &&A_i=-\sin{i\frac{\pi}{5}} \ \X+\cos{i\frac{\pi}{5}} \ \Z, \quad B_j=-\sin{\frac{(1+j)\pi}{5}} \ \X-\cos{\frac{(1+j)\pi}{5}} \ \Z,\quad \forall i,j\in[5]
  \ea
  which again provides
\begin{eqnarray}
    &&C_1 \otimes C_1 = \mathcal{A}_3 \otimes B_3=\qty(\sin{\frac{4\pi}{5}} \ \X+\cos{\frac{4\pi}{5}} \ \Z)\otimes \qty(\sin{\frac{4\pi}{5}} \ \X+\cos{\frac{4\pi}{5}} \ \Z)\nonumber\\
    &&C_2 \otimes C_2=\dfrac{1}{2}\Bigg[\frac{(\mathcal{A}_1-\mathcal{A}_5) \otimes (B_1-B_5)}{(2-\langle\{B_1,B_5\}\rangle)}+\frac{(\mathcal{A}_2-\mathcal{A}_4) \otimes (B_2-B_4)}{(2-\langle\{B_2,B_4\}\rangle)}\Bigg]=\qty(\sin{\frac{4\pi}{5}} \ \Z-\cos{\frac{4\pi}{5}} \ \X)\otimes \qty(\sin{\frac{4\pi}{5}} \ \Z-\cos{\frac{4\pi}{5}} \ \X)\nonumber\\
    &&C_3 \otimes C_3 =(C_1 \otimes C_1)\cdot (C_2 \otimes C_2)=-\sigma_y\otimes \sigma_y\nonumber
\end{eqnarray}
This again leads to the entangled state $\rho_{\phi^+}$ as in Eq. (\ref{rhophi+}), providing $(\mathcal{G}_{5})^{opt}_Q=5\csc{\frac{\pi}{10}}$.\\

\subsection{The derivation of optimal quantum violation of GBI for \texorpdfstring{$n=6$}{n=6}}
We consider a bipartite scenario involving Alice and Bob, where each party has access to six dichotomic observables, labeled $A_i$ for Alice and $B_j$ for Bob, with $i,j \in [6]$. The associated GBI is derived by setting $n=6$ in Eq.~(\ref{gisinineq}) in the main text, yielding
\begin{equation}
    \begin{split}
    \mathcal{G}_6 &= \sum\limits\limits_{i=1}^6 \qty(\sum\limits\limits_{j=1}^{7-i}{A}_i{B}_j-\sum\limits\limits_{j=8-i}^6 A_i{B}_j)\leq 18.
    \end{split}
\end{equation}
Following the SOS approach as outlined in Sec.~\ref{SOSsec}, we get that  the optimal value of $(\mathcal{G}_{6})_Q$ is obtained if $\Tr[ \Gamma_{6}\ \rho_{AB}]=0$ and thus we get 
\begin{equation}\label{g6mu}
   \qty(\mathcal{G}_{6})_Q^{\ opt}= \text{max}\sum\limits\limits_{i=1}^{6} \mu_{6,i}.
\end{equation}
where $\mu_{{6},i}=||\sum\limits\limits_{j=1}^{7-i}{B}_j-\sum\limits\limits_{j=8-i}^{6} {B}_j||_{\rho_{AB}},\forall i\in[6]$. Using the inequality in Eq. (\ref{cnv}) of the main text, we get
\ba\label{G6q}
(\mathcal{ G}_{6})_Q&\leq&\sqrt{6\sum\limits_{i=1}^6\left(\mu_{6,i}\right)^2}
 \ea
 The equality  holds only  when $\mu_{6,i}=\mu_{6,j},\forall i\neq j\in[6]$. Now, putting the value for each $\mu_{6,i}$ in Eq.~(\ref{G6q}) and after simplifying we get,
\begin{eqnarray}
        (\mathcal{G}_6)_Q&\leq&\Bigg[6\bigg(36+4\Big(\eac{B_1}{B_2}+\eac{B_2}{B _3}+\eac{B_3}{B_4} +\eac{B_4}{B_5}+\eac{B_5}{B_6}-\eac{B_1}{B_6}\Big)\nonumber\\
        &&+2\Big(\eac{B_1}{B_3}-\eac{B_1}{B_5}+\eac{B_2}{B_4}-\eac{B_2}{B_6}+ \eac{B_3}{B_5}+\eac{B_4}{B_6}\Big)\bigg)\Bigg]^\half
\end{eqnarray}
To optimize this further, we consider the following observable relations.
\begin{equation}
    \begin{split}\label{g6obs1}
        B_2=\frac{B_1+B_3}{\nu_6^1},\quad  B_4=\frac{B_3+B_5}{\nu_6^2},\quad 
        B_6=\frac{B_5-B_1}{\nu_6^3}.
    \end{split}
\end{equation}
Following a similar method as discussed for the case of $n=5$, we finally reach
\begin{eqnarray}
    (\mathcal{G}_6)_Q&\leq&\Bigg[6\bigg(36+8\sqrt{3\Big(6+\eac{B_1}{B_3}+\eac{B_3}{B_5}-\eac{B_1}{B_5}\Big)}+ 2\Big(\eac{B_1}{B_3}-\eac{B_1}{B_5}+\eac{B_2}{B_4}-\eac{B_2}{B_6}\nonumber\\
    &&+ \eac{B_3}{B_5}+\eac{B_4}{B_6}\Big)\bigg)\Bigg]^\half.
\end{eqnarray}

In addition, we consider
\begin{equation}\label{g6obs2}
    B_5=\frac{B_3-B_1}{\nu_6^4},\quad B_4=\frac{B_2+B_6}{\nu_6^5}.
\end{equation}
Thus, the optimal upper bound for $n=6$ can be written as
\begin{eqnarray}\label{g6q1}
    &&(\mathcal{G}_6)_Q=\sqrt{6 \sum\limits_{i=1}^6 \mu_{6,i}^2}\nonumber\\
    &&=\Bigg[6\bigg(36+8\sqrt{3\Big(6+\eac{B_1}{B_3}+2\sqrt{2-\eac{B_1}{B_3}}\Big)}+2\Big(\eac{B_1}{B_3}+2\sqrt{2-\eac{B_1}{B_3}}+2\sqrt{2+\eac{B_2}{B_6}}-\eac{B_2}{B_6}\Big)\bigg)\Bigg]^\half \ \ \quad
\end{eqnarray}
Again from earlier we know that $B_2=\frac{B_1+B_3}{\sqrt{2+\eac{B_1}{B_3}}}$ and without loss of generality we take $B_6=\frac{a\ B_3-B_1}{\sqrt{1+a^2-a\eac{B_1}{B_3}}},\forall a\in(0,1)$, which leads to $\eac{B_2}{B_6}=\frac{(a-1)(2+\eac{B_1}{B_3})}{\sqrt{(2+\eac{B_1}{B_3})(1+a^2-a\eac{B_1}{B_3})}}$, putting this in Eq.~(\ref{g6q1}) we get
\begin{eqnarray}\label{g6q2}
   (\mathcal{G}_6)_Q&=&\Bigg[6\bigg(36+8\sqrt{3\Big(6+\eac{B_1}{B_3}+2\sqrt{2-\eac{B_1}{B_3}}\Big)}+2\Big(\eac{B_1}{B_3}+2\sqrt{2-\eac{B_1}{B_3}}\nonumber\\
   &&+2\sqrt{2+\frac{(a-1)(2+\eac{B_1}{B_3})}{\sqrt{(2+\eac{B_1}{B_3})(1+a^2-a\eac{B_1}{B_3})}}}-\frac{(a-1)(2+\eac{B_1}{B_3})}{\sqrt{(2+\eac{B_1}{B_3})(1+a^2-a\eac{B_1}{B_3})}}\Big)\bigg)\Bigg]^\half \ \ \quad
\end{eqnarray}

Maximization of $(\mathcal{G}_{6})_Q$ provides the optimal value of $\eac{B_1}{B_3}=2\cos{\frac{2\pi}{6}}$ and $a=\frac{1}{2}$. Thus, the optimal value,
\begin{equation}
    (\mathcal{G}_{6})_Q^{\ opt}=12\sqrt{2+\sqrt{3}}=6\csc{\frac{\pi}{12}}
\end{equation}
From this we can further conclude that
$B_{6}=\frac{B_3-2B_1}{\sqrt{3}}$, which implies that $\eac{B_1}{B_{6}}=2\cos{\frac{5\pi}{6}}$ and $\eac{B_3}{B_{6}}=2\cos{\frac{3\pi}{6}}$, from where $\mu_{6}^{1}=\sqrt{2+2\cos{\frac{2\pi}{6}}}, \mu_{6}^{4}=\sqrt{2-2\cos{\frac{2\pi}{6}}}$. In a similar way, Bob's observables follow the following relations.
\begin{eqnarray}
    \eac{B_j}{B_{j+x}}=2\cos{\frac{\pi x}{6}},\quad \forall x\in[6-j]
\end{eqnarray}
Hence, $\nu_6^2=\sqrt{2+2\cos{\frac{2\pi}{6}}}$, $\nu_6^3=\sqrt{2-2\cos{\frac{4\pi}{6}}}$, $\nu_6^5=\sqrt{2+2\cos{\frac{4\pi}{6}}}$ and $\mu_{6,i}=\mu_{6,j}=\csc{\frac{\pi}{12}},\forall i\neq j\in[6]$ under the optimal conditions. Alice's observables follow similar relations due to the symmetric property of the inequality, i.e.,
\begin{eqnarray}\label{ac6}
    \eac{A_i}{A_{i+x}}=2\cos{\frac{\pi x}{6}},\quad \forall x\in[6-i]
\end{eqnarray}

The following Eq. (\ref{rho odd}) of the main text, we can write the required state $\rho_{AB}$ by substituting  $n=6$ in Eq.~(\ref{C'so}). Thus we get
\ba C_1 \otimes C_1 &=& {A}_1 \otimes \mathcal{B}_1, C_2 \otimes C_2={A}_4 \otimes \mathcal{B}_4,
C_3 \otimes C_3 = \frac{1}{8}\qty(\big[A_2,A_5]\otimes \big[\mathcal{B}_2 , \mathcal{B}_5\big] +\big[A_3, A_6]\otimes \big[\mathcal{B}_3, \mathcal{B}_6\big])=(C_1 \otimes C_1)\cdot (C_2 \otimes C_2) \quad\ea

where, $\mathcal{\mathcal{B}}_i=\frac{\sum\limits_{j=1}^{7-i}B_j-\sum\limits_{j=8-i}^{6}B_j}{\mu_{6,i}}$ and $\mu_{6,i}=\csc{\frac{\pi}{12}},\forall i\in[6]$.

As an example, in a two-qubit system, we provide a set of observables as follows. 
 \ba
  A_i=-\sin{i\frac{\pi}{6}} \ \X+\cos{i\frac{\pi}{6}} \ \Z, \quad B_j=-\sin{\frac{(3+2 j)\pi}{8}} \ \X-\cos{\frac{(3+2j)\pi}{8}} \ \Z,\quad \forall i,j\in[6]
  \ea
  which again provides
\begin{eqnarray}
    &&C_1 \otimes C_1=\qty(\sin{\frac{\pi}{6}} \ \X-\cos{\frac{\pi}{6}} \ \Z)\otimes \qty(\sin{\frac{\pi}{6}} \ \X-\cos{\frac{\pi}{6}} \ \Z);\quad C_2 \otimes C_2=\qty(\sin{\frac{4\pi}{6}} \ \X-\cos{\frac{4\pi}{6}} \ \Z)\otimes \qty(\sin{\frac{4\pi}{6}} \ \X-\cos{\frac{4\pi}{6}} \ \Z)\nonumber\\
    &&C_3 \otimes C_3 =(C_1 \otimes C_1)\cdot (C_2 \otimes C_2)=-\sigma_y\otimes \sigma_y
\end{eqnarray}
This, in turn, provides the entangled state $\rho_{\phi^+}$ as in Eq. (\ref{rhophi+}) that produces $(\mathcal{G}_{6})^{opt}_Q=6\csc{\frac{\pi}{12}}$.

\subsection{The derivation of optimal quantum bound of GBI for n=11}
Similarly, if Alice and Bob measure eleven dichotomic observables denoted by $A_i$ for Alice and $B_j$ for Bob, with $i,j \in [11]$. The corresponding GBI is obtained by substituting $n=11$ into Eq.~(\ref{gisinineq}) in the main text, we get
\begin{equation}
    \mathcal{G}_{11} = \sum\limits_{i=1}^{11} \qty(\sum\limits_{j=1}^{12-i}{A}_i{B}_j-\sum\limits_{j=13-i}^{11}{A}_i{B}_j)\leq 61.
\end{equation}
Following the SOS approach as outlined in Sec.~\ref{SOSsec}, we get that  the optimal value of $(\mathcal{G}_{11})_Q$ is obtained if $\Tr[ \Gamma_{11}\ \rho_{AB}]=0$ and thus we get 
\begin{equation}
   \qty(\mathcal{G}_{11})_Q^{\ opt}= \text{max}\sum\limits\limits_{i=1}^{11} \mu_{11,i}.
\end{equation}
where $\mu_{{11},i}=||\sum\limits\limits_{j=1}^{12-i}{B}_j-\sum\limits\limits_{j=13-i}^{11} {B}_j||_{\rho_{AB}},\forall i\in[11]$. Using the inequality in Eq. (\ref{cnv}) of the main text, we get
\ba\label{G11q}
(\mathcal{ G}_{11})_Q&\leq&\sqrt{11\sum\limits_{i=1}^{11}\left(\mu_{11,i}\right)^2}
 \ea
 The equality  holds only  when $\mu_{11,i}=\mu_{11,j},\forall i\neq j\in[11]$. We put each $\mu_{11,i}$ in Eq.~(\ref{G11q}) and after simplifying, we get 
\begin{equation}\label{g11opt}
\qty(\mathcal{G}_{11})_Q\leq\sqrt{11(121+\delta)}
\end{equation}
Here, $\delta$ is defined as follows.
\begin{eqnarray}\label{del}
\delta&=&9\delta_1+3\delta_2+7\delta_3+\delta_4+5\delta_5
\end{eqnarray}
Here $\delta_i,\forall i\in[5]$ is defined as follows
\begin{eqnarray}
    \delta_1&=&\eac{B_1}{B_2}+\eac{B_2}{B_3}+\eac{B_3}{B_4}+\eac{B_4}{B_5}+\eac{B_5}{B_6}
    +\eac{B_6}{B_7}+\eac{B_7}{B_8}+\eac{B_8}{B_9}\nonumber\\&&+\eac{B_9}{B_{10}}+\eac{B_{10}}{B_{11}}-\eac{B_1}{B_{11}}\\
    \delta_2&=&\eac{B_1}{B_5}+\eac{B_2}{B_6}+\eac{B_3}{B_7}+\eac{B_4}{B_8}+\eac{B_5}{B_9}+\eac{B_6}{B_{10}}+\eac{B_7}{B_{11}}-\eac{B_1}{B_8}\nonumber\\&&-\eac{B_2}{B_9}-\eac{B_3}{B_{10}}-\eac{B_4}{B_{11}}\\
    \delta_3&=&\eac{B_1}{B_3}+\eac{B_2}{B_4}+\eac{B_3}{B_5}+\eac{B_4}{B_6}+\eac{B_5}{B_7}+\eac{B_6}{B_8}+\eac{B_7}{B_9}+\eac{B_8}{B_{10}}\nonumber\\&&+\eac{B_9}{B_{11}}-\eac{B_1}{B_{10}}-\eac{B_2}{B_{11}}\\
    \delta_4&=&\eac{B_1}{B_6}-\eac{B_1}{B_7}+\eac{B_2}{B_7}-\eac{B_2}{B_8}+\eac{B_3}{B_8}-\eac{B_3}{B_9}+\eac{B_4}{B_9}-\eac{B_4}{B_{10}}\nonumber\\&&+\eac{B_5}{B_{10}}-\eac{B_5}{B_{11}}+\eac{B_6}{B_{11}}\\
    \delta_5&=&\eac{B_1}{B_4}+\eac{B_2}{B_5}+\eac{B_3}{B_6}+\eac{B_4}{B_7}\eac{B_5}{B_8}+\eac{B_6}{B_9}+\eac{B_7}{B_{10}}+\eac{B_8}{B_{11}}\nonumber\\
    &&-\eac{B_1}{B_9}-\eac{B_2}{B_{10}}-\eac{B_3}{B_{11}}
\end{eqnarray}
As it is obvious that the optimization gets rigorous for a higher number of settings and hence the approximations. To simplify further, we use $\Bigg(\Delta_{11}=B_1-B_2+B_3-B_4+B_5-B_6+B_7-B_8+B_9-B_{10}+B_{11}\Bigg)$, such that
\begin{equation}\label{delta11}
    \Delta_{11}^2 =11-\delta_1+\delta_2+\delta_3-\delta_4-\delta_5 
\end{equation}
For convenience, we define $\Delta_{11}^2=\langle (\Delta_{11})^2\rangle$. Upon substituting Eq.~\ref{delta11} into Eq.~\ref{g11opt}, we obtain
\begin{eqnarray}\label{gd}
         \qty(\mathcal{G}_{11})_Q&\leq&\Big[11(132+8(\delta_1+\delta_3)+4(\delta_2+\delta_5)-\Delta_{11}^2)\Big]^\frac{1}{2}
\end{eqnarray}
Since $\Delta_{11}$ is a positive quantity, the maximum is obtained only when $\Delta_{11}^2 =0$. Hence, Eq.~(\ref{delta11}) become
\begin{eqnarray}\label{d1}
    11-\delta_1+\delta_2+\delta_3-\delta_4-\delta_5=0
\end{eqnarray}
With $\Delta_{11}^2 =0$ and with the help of Eq.~(\ref{d1}) we replace $\delta_2$ and rewrite the Eq.~(\ref{gd}) as
Eq.\eqref{g11opt} we get
\begin{eqnarray}\label{gd1}
         \qty(\mathcal{G}_{11})_Q&\leq&\Big[11(88+12\delta_1+4\delta_3+4\delta_4+8\delta_5)\Big]^{\frac{1}{2}}
\end{eqnarray}
Thereafter, for further simplification, we consider the four sums separately, to attempt to simplify the first sum, we use the observable relations $B_2=\frac{B_1+B_3}{\nu_{11}^1}, B_4=\frac{B_3+B_5}{\nu_{11}^2}, B_6=\frac{B_5+B_7}{\nu_{11}^3}, B_8=\frac{B_7+B_9}{\nu_{11}^4}, B_{10}=\frac{B_9+B_{11}}{\nu_{11}^5}, B_3=\frac{B_1+B_5}{\nu_{11}^6}, B_7=\frac{B_5+B_9}{\nu_{11}^7}, B_5=\frac{B_1+B_9}{\nu_{11}^8}$.$\mu_{11}^i$s are defined as $\nu_{11}^k=\sqrt{2\pm\{B_j,B_{j'}\}}\ \forall j\neq j'$. Using the observable relation and applying SOS, the sum reduces to
\begin{eqnarray}
       \delta_1&=&\eac{B_1}{B_2}+\eac{B_2}{B_3}+\eac{B_3}{B_4}+\eac{B_4}{B_5}+\eac{B_5}{B_6}
    +\eac{B_6}{B_7}+\eac{B_7}{B_8}+\eac{B_8}{B_9}+\eac{B_9}{B_{10}}+\eac{B_{10}}{B_{11}}\nonumber\\&&-\eac{B_1}{B_{11}}\nonumber\\ 
    &=& \eac{B_2}{B_1+B_3}+\eac{B_4}{B_3+B_5}+\eac{B_6}{B_5+B_7}+\eac{B_8}{B_7+B_9}+\eac{B_{10}}{B_9+B_{11}}-\eac{B_1}{B_{11}}\nonumber\\
    &=&2\Bigg[\sqrt{2+\eac{B_1}{B_3}}+\sqrt{2+\eac{B_3}{B_5}}+\sqrt{2+\eac{B_5}{B_7}}+\sqrt{2+\eac{B_7}{B_9}}+\sqrt{2+\eac{B_9}{B_{11}}}\Bigg]-\eac{B_1}{B_{11}}\nonumber\\
    &\leq&2\Bigg[5\qty(10+\eac{B_3}{B_1+B_5}+\eac{B_7}{B_5+B_9}+\eac{B_9}{B_{11}})\Bigg]^{\frac{1}{2}}-\eac{B_1}{B_{11}}\nonumber\\
    &=&2\Bigg[5\qty(10+2\sqrt{2+\eac{B_1}{B_5}}+2\sqrt{2+\eac{B_5}{B_9}}+\eac{B_9}{B_{11}})\Bigg]^{\frac{1}{2}}-\eac{B_1}{B_{11}}\nonumber\\
     &\leq&2\Bigg[5\qty(10+2\sqrt{2\qty(4+\eac{B_5}{B_1+B_9})}+\eac{B_9}{B_{11}})\Bigg]^{\frac{1}{2}}-\eac{B_1}{B_{11}}\nonumber\\
     &=& 2\Bigg[5\qty(10+2\sqrt{2\qty(4+2\sqrt{2+\eac{B_1}{B_9}})}+\eac{B_9}{B_{11}})\Bigg]^{\frac{1}{2}}-\eac{B_1}{B_{11}}
\end{eqnarray}
Similarly, for the second sum, we consider the relations $B_3=\frac{B_1+B_5}{\nu_{11}^6}, B_2 =\frac{B_4-B_{11}}{\mu_{11}^{9}}, B_{10} =\frac{B_8-B_1}{\nu_{11}^{10}}, B_6 =\frac{B_4+B_8}{\nu_{11}^{11}}, B_7 =\frac{B_5+B_9}{\mu_{11}^{7}}, B_1 =\frac{B_5-B_8}{\nu_{11}^{12}}, B_4 =\frac{B_8-B_{11}}{\nu_{11}^{13}}, B_8 =\frac{B_5+B_{11}}{\nu_{11}^{14}}, B_5 =\frac{B_1+B_9}{\mu_{11}^{8}}$. Therefore, the sum can be written as

\begin{eqnarray}
    \delta_3&=&\eac{B_1}{B_3}+\eac{B_2}{B_4}+\eac{B_3}{B_5}+\eac{B_4}{B_6}+\eac{B_5}{B_7}+\eac{B_6}{B_8}+\eac{B_7}{B_9}+\eac{B_8}{B_{10}}+\eac{B_9}{B_{11}}-\eac{B_1}{B_{10}}\nonumber\\&&-\eac{B_2}{B_{11}}\nonumber\\
    &=& \eac{B_3}{B_1+B_5}+\eac{B_2}{B_4-B_{11}}+\eac{B_{10}}{B_8-B_1}+\eac{B_6}{B_4+B_8}+\eac{B_7}{B_5+B_9}+\eac{B_9}{B_{11}}\nonumber\\
    &=&2\Bigg[\sqrt{2+\eac{B_1}{B_5}}+\sqrt{2-\eac{B_4}{B_{11}}}+\sqrt{2-\eac{B_1}{B_8}}+\sqrt{2+\eac{B_4}{B_8}}+\sqrt{2+\eac{B_5}{B_9}}\Bigg]+\eac{B_9}{B_{11}}\nonumber\\
    &\leq&2\Bigg[5\qty(10+\eac{B_1}{B_5-B_8}+\eac{B_4}{B_8-B_{11}}+\eac{B_5}{B_9})\Bigg]^{\frac{1}{2}}+\eac{B_9}{B_{11}}\nonumber\\
    &=&2\Bigg[5\Bigg(10+2\sqrt{2-\eac{B_5}{B_8}}+2\sqrt{2-\eac{B_8}{B_{11}}}+\eac{B_5}{B_9}\Bigg)\Bigg]^{\frac{1}{2}}+\eac{B_9}{B_{11}}\nonumber\\
     &\leq&2\Bigg[5\qty(10+2\sqrt{2\qty(4-\eac{B_8}{B_5+B_{11}})}+\eac{B_5}{B_9})\Bigg]^{\frac{1}{2}}+\eac{B_9}{B_{11}}\nonumber\\
     &=& 2\Bigg[5\qty(10+2\sqrt{2\qty(4-2\sqrt{2+\eac{B_5}{B_{11}}})}+\eac{B_5}{B_9})\Bigg]^{\frac{1}{2}}+\eac{B_9}{B_{11}}\quad \text{$\qty[As, \ B_5=\frac{B_1+B_9}{\sqrt{2+\eac{B_1}{B_9}}}]$}\nonumber\\
      &=& 2\Bigg[5\Bigg(10+2\sqrt{2\qty(4-2\sqrt{2+\frac{\eac{B_1}{B_{11}}+\eac{B_9}{B_{11}}}{\sqrt{2+\eac{B_1}{B_9}}}})}+\sqrt{2+\eac{B_1}{B_9}}\Bigg)\Bigg]^{\frac{1}{2}}+\eac{B_9}{B_{11}}
\end{eqnarray}
Similarly, for the third sum, we consider the relations $B_6=\frac{B_1+B_{11}}{\nu_{11}^{15}}, B_7 =\frac{B_2-B_1}{\nu_{11}^{16}}, B_8 =\frac{B_3-B_2}{\nu_{11}^{17}}, B_9 =\frac{B_4-B_3}{\nu_{11}^{18}}, B_{10} =\frac{B_5-B_4}{\nu_{11}^{19}}, B_2 =\frac{B_1+B_3}{\mu_{11}^{1}}, B_4 =\frac{B_3+B_5}{\mu_{11}^{2}}, B_3 =\frac{B_1+B_5}{\mu_{11}^{6}}, B_5 =\frac{B_1+B_9}{\mu_{11}^{8}}$. Hence, the third term.

\begin{eqnarray}
    \delta_4&=&\eac{B_1}{B_6}-\eac{B_1}{B_7}+\eac{B_2}{B_7}-\eac{B_2}{B_8}+\eac{B_3}{B_8}-\eac{B_3}{B_9}+\eac{B_4}{B_9}-\eac{B_4}{B_{10}}+\eac{B_5}{B_{10}}-\eac{B_5}{B_{11}}\nonumber\\&&+\eac{B_6}{B_{11}}\nonumber\\
    &=& \eac{B_6}{B_1+B_{11}}+\eac{B_7}{B_2-B_1}+\eac{B_8}{B_3-B_2}+\eac{B_9}{B_4-B_3}+\eac{B_{10}}{B_5-B_4}-\eac{B_5}{B_{11}}\nonumber\\
    &=&2\Bigg[\sqrt{2+\eac{B_1}{B_{11}}}+\sqrt{2-\eac{B_1}{B_2}}+\sqrt{2-\eac{B_2}{B_3}}+\sqrt{2-\eac{B_3}{B_4}}+\sqrt{2-\eac{B_4}{B_5}}\Bigg]-\eac{B_5}{B_{11}}\nonumber\\
    &\leq&2\Bigg[5\qty(10+\eac{B_1}{B_{11}}-\eac{B_2}{B_1+B_3}-\eac{B_4}{B_3+B_5})\Bigg]^{\frac{1}{2}}-\eac{B_5}{B_{11}}\nonumber\\
    &=&2\Bigg[5\Bigg(10-2\sqrt{2+\eac{B_1}{B_3}}-2\sqrt{2+\eac{B_3}{B_5}}+\eac{B_1}{B_{11}}\Bigg)\Bigg]^{\frac{1}{2}}-\eac{B_5}{B_{11}}\nonumber\\
     &\leq&2\Bigg[5\qty(10-2\sqrt{2\qty(4+\eac{B_3}{B_1+B_5})}+\eac{B_1}{B_{11}})\Bigg]^{\frac{1}{2}}-\eac{B_5}{B_{11}}\nonumber\\
      &=&2\Bigg[5\qty(10-2\sqrt{2\qty(4+2\sqrt{2+\eac{B_1}{B_5}})}+\eac{B_1}{B_{11}})\Bigg]^{\frac{1}{2}}-\eac{B_5}{B_{11}}\quad\text{$\qty[As, \ B_5=\frac{B_1+B_9}{\sqrt{2+\eac{B_1}{B_9}}}]$}\nonumber\\
    &=&2\Bigg[5\qty(10-2\sqrt{2\qty(4+2\sqrt{2+\frac{2+\eac{B_1}{B_9}}{\sqrt{2+\eac{B_1}{B_9}}}})}+\eac{B_1}{B_{11}})\Bigg]^{\frac{1}{2}}-\frac{\eac{B_1}{B_{11}}+\eac{B_9}{B_{11}}}{\sqrt{2+\eac{B_1}{B_9}}}
\end{eqnarray}
Similarly, for the fourth sum, we consider the relations $B_1=\frac{B_4-B_9}{\nu_{11}^{20}}, B_2 =\frac{B_5-B_{10}}{\nu_{11}^{21}}, B_3 =\frac{B_6-B_{11}}{\nu_{11}^{22}}, B_7 =\frac{B_4+B_{10}}{\nu_{11}^{23}}, B_8 =\frac{B_5+B_{11}}{\nu_{11}^{14}}, B_4 =\frac{B_9-B_{10}}{\nu_{11}^{24}}, B_5 =\frac{B_{10}-B_{11}}{\nu_{11}^{25}}, B_{10} =\frac{B_9+B_{11}}{\nu_{11}^{26}}, B_6 =\frac{B_1+B_{11}}{\nu_{11}^{15}}$. And the fourth term can be written as
\begin{eqnarray}
    \delta_5&=&\eac{B_1}{B_4}+\eac{B_2}{B_5}+\eac{B_3}{B_6}+\eac{B_4}{B_7}+\eac{B_5}{B_8}+\eac{B_6}{B_9}+\eac{B_7}{B_{10}}+\eac{B_8}{B_{11}}-\eac{B_1}{B_9}-\eac{B_2}{B_{10}}\nonumber\\
    &&-\eac{B_3}{B_{11}}\nonumber\\
    &=& \eac{B_1}{B_4-B_9}+\eac{B_2}{B_5-B_{10}}+\eac{B_3}{B_6-B_{11}}+\eac{B_7}{B_4+B_{10}}+\eac{B_8}{B_5+B_{11}}+\eac{B_6}{B_9}\nonumber\\
    &=&2\Bigg[\sqrt{2-\eac{B_4}{B_9}}+\sqrt{2-\eac{B_5}{B_{10}}}+\sqrt{2-\eac{B_6}{B_{11}}}+\sqrt{2+\eac{B_4}{B_{10}}}+\sqrt{2+\eac{B_5}{B_{11}}}\Bigg]+\eac{B_6}{B_9}\nonumber\\
    &\leq&2\Bigg[5\qty(10-\eac{B_4}{B_9-B_{10}}-\eac{B_5}{B_{10}-B_{11}}-\eac{B_6}{B_{11}})\Bigg]^{\frac{1}{2}}+\eac{B_6}{B_9}\nonumber\\
    &=&2\Bigg[5\Bigg(10-2\sqrt{2-\eac{B_9}{B_{10}}}-2\sqrt{2-\eac{B_{10}}{B_{11}}}-\eac{B_6}{B_{11}}\Bigg)\Bigg]^{\frac{1}{2}}+\eac{B_6}{B_9}\nonumber\\
     &\leq&2\Bigg[5\qty(10-2\sqrt{2\qty(4-\eac{B_{10}}{B_9+B_{11}})}-\eac{B_6}{B_{11}})\Bigg]^{\frac{1}{2}}+\eac{B_6}{B_9}\nonumber\\
      &=&2\Bigg[5\qty(10-2\sqrt{2\qty(4-2\sqrt{2+\eac{B_9}{B_{11}}})}-\eac{B_6}{B_{11}})\Bigg]^{\frac{1}{2}}+\eac{B_6}{B_9}\quad \text{$\qty[As, \ B_6=\frac{B_1+B_{11}}{\sqrt{2+\eac{B_1}{B_{11}}}}]$}\nonumber\\
    &=&2\Bigg[5\Big(10-2\sqrt{2\qty(4-2\sqrt{2+\eac{B_9}{B_{11}}})}-\sqrt{2+\eac{B_1}{B_{11}}}\Big)\Bigg]^{\frac{1}{2}}+\frac{\eac{B_1}{B_9}+\eac{B_9}{B_{11}}}{\sqrt{2+\eac{B_1}{B_{11}}}}
\end{eqnarray}
Therefore, finally, optimization requires the maximization of the term.
\begin{eqnarray}
        \qty(\mathcal{G}_{11})_Q &\leq&\Big[11(88+12\delta_1+4\delta_3+4\delta_4+8\delta_5)\Big]^{\frac{1}{2}}\nonumber\\
        &=&\Bigg[11\Bigg(88+12\Bigg(2\Bigg[5\Bigg(10+2\sqrt{2\qty(4+2\sqrt{2+\eac{B_1}{B_9}})}+\eac{B_9}{B_{11}}\Bigg)\Big]^{\frac{1}{2}}-\eac{B_1}{B_{11}}\Bigg)\nonumber\\&&+4\Bigg(2\Bigg[5\Bigg(10+2\sqrt{2\Bigg(4-2\sqrt{2+\frac{\eac{B_1}{B_{11}}+\eac{B_9}{B_{11}}}{\sqrt{2+\eac{B_1}{B_9}}}}\Bigg)}+\sqrt{2+\eac{B_1}{B_9}}\Bigg)\Bigg]^{\frac{1}{2}}+\eac{B_9}{B_{11}}\Bigg)\nonumber\\&&+4\Bigg(2\Bigg[5\Bigg(10-2\sqrt{2\qty(4+2\sqrt{2+\frac{2+\eac{B_1}{B_9}}{\sqrt{2+\eac{B_1}{B_9}}}})}+\eac{B_1}{B_{11}}\Bigg)\Bigg]^{\frac{1}{2}}-\frac{\eac{B_1}{B_{11}}+\eac{B_9}{B_{11}}}{\sqrt{2+\eac{B_1}{B_9}}}\Bigg)\nonumber\\&&+8\Bigg(2\Bigg[5\Bigg(10-2\sqrt{2\qty(4-2\sqrt{2+\eac{B_9}{B_{11}}})}-\sqrt{2+\eac{B_1}{B_{11}}}\Bigg)\Bigg]^{\frac{1}{2}}+\frac{\eac{B_1}{B_9}+\eac{B_9}{B_{11}}}{\sqrt{2+\eac{B_1}{B_{11}}}}\Bigg)\Bigg)\Bigg]^{\frac{1}{2}}\nonumber\\
\end{eqnarray}
Without loss of generality, consider $B_{11}=\frac{a B_9-B_1}{\sqrt{1+a^2-a\eac{B_1}{B_9}}}$ and $a\in(0,1)$.
Maximization of $(\mathcal{G}_{11})_Q$ provides the optimal value of $\eac{B_1}{B_9}=-2\sin{\frac{5\pi}{22}}$ and $a=\cos{\frac{\pi}{3}}$. Thus, the optimal value,
\begin{equation}
    (\mathcal{G}_{11})_Q^{\ opt}= 11\csc{\frac{\pi}{22}}
\end{equation}

From this we can further conclude that
$B_{11}=\frac{\cos{\frac{\pi}{3}} B_9-B_1}{\sqrt{1+\cos^2{\frac{\pi}{3}}+2\sin{\frac{5\pi}{22}}\cos{\frac{\pi}{3}}}}$, which implies that $\eac{B_9}{B_{11}}=2\cos{\frac{2\pi}{11}}$ and $\eac{B_1}{B_{11}}=2\cos{\frac{10\pi}{11}}$, from where $\nu_{11}^{20}=\sqrt{2-2\sin{\frac{5\pi}{22}}}, \nu_{11}^{15}=\sqrt{2+2\cos{\frac{10\pi}{11}}}, \mu_{11}^{28}=\sqrt{2+2\cos{\frac{2\pi}{11}}}$. In a similar way, Bob's observables follow the following relations.
\begin{eqnarray}\label{b7}
    \eac{B_j}{B_{j+x}}=2\cos{\frac{\pi x}{11}},\quad \forall x\in[11-j]
\end{eqnarray}
Hence, $\nu_{11}^1=\nu_{11}^2=\nu_{11}^3=\nu_{11}^4=\nu_{11}^5=\nu_{11}^{26}=\sqrt{2+2\cos{\frac{2\pi}{11}}}$, $\nu_{11}^6=\mu_{11}^{7}=\nu_{11}^{11}=\sqrt{2+2\cos{\frac{4\pi}{11}}}$, $\nu_{11}^8=\sqrt{2+2\cos{\frac{8\pi}{11}}}$, $\nu_{11}^{14}=\nu_{11}^{23}=\sqrt{2+2\cos{\frac{6\pi}{11}}}$, $\nu_{11}^{15}=\sqrt{2+2\cos{\frac{10\pi}{11}}}$, $\nu_{11}^9=\nu_{11}^{10}=\sqrt{2-2\cos{\frac{7\pi}{11}}}$,$\nu_{11}^{12}=\nu_{11}^{13}=\sqrt{2-2\cos{\frac{3\pi}{11}}}$, $\nu_{11}^{16}=\nu_{11}^{17}=\nu_{11}^{18}=\nu_{11}^{19}=\nu_{11}^{24}=\nu_{11}^{25}=\sqrt{2-2\cos{\frac{\pi}{11}}}$, $\nu_{11}^{20}=\nu_{11}^{21}=\nu_{11}^{22}=\sqrt{2-2\cos{\frac{5\pi}{11}}}$ and $\mu_{11,i}=\mu_{11,j}=\csc{\frac{\pi}{22}},\forall i\neq j\in[11]$ under the optimal conditions. Alice's observables follow similar relations due to the symmetric property of the inequality, i.e.,
\begin{eqnarray}\label{a8}
    \eac{A_i}{A_{i+x}}=2\cos{\frac{\pi x}{11}},\quad \forall x\in[11-i]
\end{eqnarray}

The following Eq. (\ref{rho odd}) of the main text, we can write the required state $\rho_{AB}$ by substituting  $n=11$ in Eq.~(\ref{C'so}). Thus, we get
\begin{eqnarray}
   C_1 \otimes C_1 &=& \mathcal{A}_6 \otimes B_6 \nonumber\\
    C_2 \otimes C_2&=&\frac{1}{5}\Bigg[\frac{(\mathcal{A}_1 \otimes B_1 - \mathcal{A}_1 \otimes B_{11} - \mathcal{A}_{11} \otimes B_1 + \mathcal{A}_{11} \otimes B_{11})}{2-\{B_1,B_{11}\}}+\frac{(\mathcal{A}_2 \otimes B_2 - \mathcal{A}_2 \otimes B_{10} - \mathcal{A}_{10} \otimes B_2 + \mathcal{A}_{10} \otimes B_{10})}{2-\{B_2,B_{10}\}}\nonumber\\&& \quad+\frac{(\mathcal{A}_3 \otimes B_3 - \mathcal{A}_3 \otimes B_9 - \mathcal{A}_9 \otimes B_3 + \mathcal{A}_9 \otimes B_9)}{2-\{B_3,B_{9}\}}+\frac{(\mathcal{A}_4 \otimes B_4 - \mathcal{A}_4 \otimes B_8 - \mathcal{A}_8 \otimes B_4 + \mathcal{A}_8 \otimes B_8)}{2-\{B_4,B_8\}}\\&&\quad+\frac{(\mathcal{A}_5 \otimes B_5 - \mathcal{A}_5 \otimes B_7 - \mathcal{A}_7 \otimes B_5 + \mathcal{A}_7 \otimes B_7)}{2-\{B_5,B_7\}}\Bigg]\nonumber
\end{eqnarray}
and $C_3 \otimes C_3 =(C_1 \otimes C_1)\cdot (C_2 \otimes C_2)$
where $\mathcal{A}_j=\frac{\sum\limits\limits_{i=1}^{12-j}A_i-\sum\limits_{i=13-j}^{11}A_i}{\csc{\frac{\pi}{22}}}$ .

As an example, in a two-qubit system, we provide a set of observables that are the following. 
 \ba
  A_i=-\sin{i\frac{\pi}{11}} \ \X+\cos{i\frac{\pi}{11}} \ \Z, \quad B_j=-\sin{\frac{(4+j)\pi}{11}} \ \X-\cos{\frac{(4+j)\pi}{11}} \ \Z,\quad \forall i,j\in [11] \ea
  which again provides
\begin{eqnarray}
    &&C_1 \otimes C_1=\qty(\sin{\frac{10\pi}{11}}\X+\cos{\frac{10\pi}{11}}\Z)\otimes \qty(\sin{\frac{10\pi}{11}}\X+\cos{\frac{10\pi}{11}}\Z);\quad C_2 \otimes C_2=\qty(\sin{\frac{10\pi}{11}}\X-\cos{\frac{10\pi}{11}}\Z)\otimes \qty(\sin{\frac{10\pi}{11}}\X-\cos{\frac{10\pi}{11}}\Z)\nonumber\\
    &&C_3 \otimes C_3 =(C_1 \otimes C_1)\cdot (C_2 \otimes C_2)=-\sigma_y\otimes \sigma_y
\end{eqnarray}
This, in turn, provides the entangled state $\rho_{\phi^+}$ as in Eq. (\ref{rhophi+}) that produces $(\mathcal{G}_{11})^{opt}_Q=11\csc{\frac{\pi}{22}}$.
\section{Derivation of the  required  entangled state for optimal quantum violation}\label{cse}
Deriving the explicit form of the required quantum state from the sum-of-squares (SOS) conditions, in order to attain the optimal quantum value for an arbitrary local dimension \(d\) with an arbitrary number \(n\) of measurement settings, is a highly nontrivial task. This construction problem is mathematically demanding yet also conceptually rich, making it a particularly compelling subject for further theoretical investigation. In the next section, we define the state for both odd and even settings.
\subsection{Derivation of the  required entangled state for odd \texorpdfstring{$n$}{n}}\label{cseodd}
At the optimal quantum value $(\mathcal{G}_{n})^{opt}_Q$ given in the main text Eq.~(\ref{Bobsc1}), the following condition is obtained
    \begin{eqnarray}\label{L1}
        \mathcal{A}_j\otimes B_j\ket{\psi}_{AB}&=&\ket{\psi}_{AB}, \ \forall j\in[n].
    \end{eqnarray}
    
    For $j=1$ and $j=n$, we have the relations 
     \begin{eqnarray}\label{i1}
        \mathcal{A}_1\otimes B_1\ket{\psi}_{AB}&=&\ket{\psi}_{AB}\\\label{in}
        \mathcal{A}_n\otimes B_n\ket{\psi}_{AB}&=&\ket{\psi}_{AB}
    \end{eqnarray}
We begin by noting that $C_1\otimes C_1=\mathcal{A}_{\frac{n+1}{2}}\otimes B_{\frac{n+1}{2}}$, which leads to $\Tr[C_1\otimes C_1\,\rho_{AB}]=1$. Next, we left-multiply Eq.~(\ref{i1}) and Eq.~(\ref{in}) by $\openone_d\otimes B_n B_1$ and $\openone_d\otimes B_1 B_n$, respectively, to obtain
    \begin{eqnarray}
        \mathcal{A}_1\otimes B_n\ket{\psi}_{AB}&=&\openone_d\otimes B_n B_1\ket{\psi}_{AB}\label{A1B5}\\
        \mathcal{A}_n\otimes B_1\ket{\psi}_{AB}&=&\openone_d\otimes B_1 B_n\ket{\psi}_{AB}\label{A5B1}
    \end{eqnarray}
    Adding Eq.~(\ref{i1}),(\ref{in}) and subtracting Eq.~(\ref{A1B5}),(\ref{A5B1}) we get,
    \begin{eqnarray}
        &&\mathcal{A}_1\otimes B_1+\mathcal{A}_n\otimes B_n-\mathcal{A}_1\otimes B_n-\mathcal{A}_n\otimes B_1\ket{\psi}_{AB}= (2\openone_d\otimes\openone_d-\openone_d\otimes\{B_1,B_n\})\ket{\psi}_{AB}\nonumber\\
        &&\bra{\psi}_{AB}\mathcal{A}_1\otimes B_1+\mathcal{A}_n\otimes B_n-\mathcal{A}_1\otimes B_n-\mathcal{A}_n\otimes B_1\ket{\psi}_{AB}= (2-\langle\{B_1,B_n\}\rangle)\nonumber\\
        &&\bra{\psi}_{AB}\frac{\mathcal{A}_1\otimes B_1+\mathcal{A}_n\otimes B_n-\mathcal{A}_1\otimes B_n-\mathcal{A}_n\otimes B_1}{(2-\langle\{B_1,B_n\}\rangle)}\ket{\psi}_{AB}=1\nonumber\\
    &&\frac{\mathcal{A}_1\otimes B_1+\mathcal{A}_n\otimes B_n-\mathcal{A}_1\otimes B_n-\mathcal{A}_n\otimes B_1}{(2-\langle\{B_1,B_n\}\rangle)}\ket{\psi}_{AB}= \ket{\psi}_{AB}\quad\bigg[\text{$(2-\langle\{B_1,B_n\}\rangle)$ is normalization constant}\bigg]\label{A1AnB1Bn}
    \end{eqnarray}
Following similar steps, we also get for any pair $\mathcal{A}_j,\mathcal{A}_{n+1-j}, \  j\in[n]$ we get
\ba\label{AiBi}
\frac{(\mathcal{A}_{j} \otimes B_{j} + \mathcal{A}_{n+1-j} \otimes B_{n+1-j} - \mathcal{A}_{j} \otimes B_{n+1-j} - \mathcal{A}_{n+1-j} \otimes B_{j})}{(2-\langle\{B_{j},B_{n+1-j}\}\rangle)}\ket{\psi}_{AB}&=& \ket{\psi}_{AB}
\ea 
Summing all these for $j \in \bigg[\lfloor\frac{n}{2}\rfloor\bigg]$ and then simplifying, we obtain
\begin{eqnarray}
    \frac{1}{\lfloor \frac{n}{2} \rfloor}\sum_{j=1}^{{\lfloor \frac{n}{2}\rfloor}}\frac{(\mathcal{A}_{j}-\mathcal{A}_{n+1-j}) \otimes (B_{j}-B_{n+1-j})}{(2-\langle\{B_{j},B_{n+1-j}\}\rangle)}\ket{\psi}_{AB}&=& \ket{\psi}_{AB}\label{c2c2}
\end{eqnarray}
Hence, we can consider
\ba C_2\otimes C_2 = \frac{1}{\lfloor \frac{n}{2} \rfloor}\sum_{i=1}^{{\lfloor \frac{n}{2}\rfloor}}\frac{(\mathcal{A}_{i}-\mathcal{A}_{n+1-i}) \otimes (B_{i}-B_{n+1-i})}{(2-\langle\{B_{i},B_{n+1-i}\}\rangle)}\label{c2c2n}
\ea 
From Eq.~(\ref{c2c2n}), it follows directly that $\Tr[C_2\otimes C_2 \ \rho_{AB}]=1$. Moreover, by pre-multiplying Eq.~(\ref{c2c2}) with $\mathcal{A}_{\frac{n+1}{2}}\otimes B_{\frac{n+1}{2}}$, we obtain
\begin{eqnarray}
     \frac{1}{\lfloor \frac{n}{2} \rfloor}\sum_{j=1}^{{\lfloor \frac{n}{2}\rfloor}}\frac{\mathcal{A}_{\frac{n+1}{2}}(\mathcal{A}_{j}-\mathcal{A}_{n+1-j}) \otimes B_{\frac{n+1}{2}}(B_{j}-B_{n+1-j})}{(2-\langle\{B_{j},B_{n+1-j}\}\rangle)}\ket{\psi}_{AB}&=& \ket{\psi}_{AB}\nonumber
\end{eqnarray}
Similarly, We can consider  
\ba C_3\otimes C_3= \frac{1}{\lfloor \frac{n}{2} \rfloor}\sum_{i=1}^{{\lfloor \frac{n}{2}\rfloor}}\frac{\mathcal{A}_{\frac{n+1}{2}}(\mathcal{A}_{i}-\mathcal{A}_{n+1-i}) \otimes B_{\frac{n+1}{2}}(B_{i}-B_{n+1-i})}{(2-\langle\{B_{i},B_{n+1-i}\}\rangle)}\ea 
Consequently, $\Tr[C_3\otimes C_3 \ \rho_{AB}]=1$. It is straightforward to verify that $C_3\otimes C_3 = (C_1\otimes C_1)(C_2\otimes C_2)$ and that the condition $\langle[C_{f}\otimes C_{f}, C_{f'}\otimes C_{f'}]_{f\neq {f'}}\rangle_{\rho_{AB}}=0,  \ \forall {f}\neq {f'}\in [d^2-1]$ together imply $\Tr[C_{f}\otimes C_{f} \ \rho_{AB}]=1$ for all ${f}\in[d^2-1]$. Moreover, any arbitrary bipartite state in $d$ dimensions can be expressed as a function of the operators $C_{f}\otimes C_{f}$ as 
\begin{eqnarray}\label{state 7}
    \rho_{AB} = \frac{1}{d^2} \qty[\openone_d\otimes\openone_d + C_1 \otimes C_1+C_2 \otimes C_2+C_3 \otimes C_3 + \sum_{f=4}^{d^2-1} C_{f} \otimes C_{f}]
\end{eqnarray}
Again, altogether, we get the following.
\begin{eqnarray}
    C_1\otimes C_1\ket{\psi}_{AB}=C_2\otimes C_2\ket{\psi}_{AB}=C_3\otimes C_3\ket{\psi}_{AB}=\ket{\psi}_{AB}
\end{eqnarray} 
Hence, we can directly show that $\langle [C_{f}\otimes C_{f},C_{f'}\otimes C_{f'}]\rangle_{\rho_{AB}}=0,\forall f\neq f'\in\{1,2,3\}$.

The state $\rho_{AB}$ achieves the optimal quantum violation when $\Tr[(\mathcal{A}_j\otimes B_j)\,\rho_{AB}]=1$ for all $j\in[n]$. As $n$ increases, the corresponding derivation becomes significantly more involved and lengthy. For clarity, we explicitly present the case $n=5$. Using the above argument from Eq.~(\ref{state5}), we obtain
\ba C_1 \otimes C_1 &=& \mathcal{A}_3 \otimes B_3\nonumber\\
C_2 \otimes C_2&=&\dfrac{1}{2}\Bigg[\frac{(\mathcal{A}_1-\mathcal{A}_5) \otimes (B_1-B_5)}{(2-\langle\{B_1,B_5\}\rangle)}+\frac{(\mathcal{A}_2-\mathcal{A}_4) \otimes (B_2-B_4)}{(2-\langle\{B_2,B_4\}\rangle)}\Bigg]\nonumber\\
C_3 \otimes C_3 &=&\dfrac{1}{2}\Bigg[\frac{\mathcal{A}_3(\mathcal{A}_1-\mathcal{A}_5) \otimes B_3(B_1-B_5)}{(2-\langle\{B_1,B_5\}\rangle)}+\frac{\mathcal{A}_3(\mathcal{A}_2-\mathcal{A}_4) \otimes B_3(B_2-B_4)}{(2-\langle\{B_2,B_4\}\rangle)}\Bigg]
\ea 

From the derivation of the optimal quantum bound, we have $\langle\{\mathcal{A}_j,\mathcal{A}_{j+x}\}\rangle=2 \cos{\frac{\pi x}{5}},\forall x\in[n-j]$ which can be further generalized to $\{\mathcal{A}_j,\mathcal{A}_{j+x}\}=\{B_j, B_{j+x}\}=2\openone_d \cos{\frac{\pi x}{5}}$  only for $\rho_{AB}$, which directly implies
\ba \label{ix} \Tr[\mathcal{A}_j \mathcal{A}_{j+x}]=\Tr[B_j B_{j+x}]=d\cos{\frac{\pi x}{5}}\ea
To prove $\Tr[\mathcal{A}_1\otimes B_1 \ \rho_{AB}]=1$, we use the Eq.~(\ref{ix}) and calculate the following.
\begin{eqnarray}\label{A1C1B1C1}
    \Tr[(\mathcal{A}_1\otimes B_1) (C_1 \otimes  C_1)]&=& \Tr[\mathcal{A}_1 \mathcal{A}_3 \otimes B_1 B_3]= d^2 \cos^2{\frac{2\pi}{5}}\\
\label{A1C2B1C2}
    \Tr[(\mathcal{A}_1\otimes B_1)( C_2 \otimes C_2)]&=& \frac{1}{2}\Bigg[\Tr\bigg[\frac{\openone_d \otimes \openone_d-\openone_d \otimes B_1 B_5 - \mathcal{A}_1 \mathcal{A}_5\otimes\openone_d+\mathcal{A}_1 \mathcal{A}_5\otimes B_1 B_5}{4}\bigg]\nonumber\\
    &&\phantom{\frac{1}{2}\Bigg[}+\Tr\bigg[\frac{\mathcal{A}_1 \mathcal{A}_2\otimes B_1 B_2-\mathcal{A}_1 \mathcal{A}_2\otimes B_1 B_4 - \mathcal{A}_1 \mathcal{A}_4\otimes B_1 B_2+\mathcal{A}_1 \mathcal{A}_4\otimes B_1 B_4}{2}\bigg]\Bigg]\nonumber\\
      &=& \frac{d^2}{4}\Bigg[1-\cos{\frac{4\pi}{5}}+\frac{(\cos{\frac{\pi}{5}}-\cos{\frac{3\pi}{5}})^2}{1-\cos{\frac{2\pi}{5}}}\Bigg] \\
     \label{A1C3B1C3}\Tr[(\mathcal{A}_1\otimes B_1)( C_3 \otimes C_3)]&=&\frac{1}{2}\Bigg[\Tr\bigg[\frac{(\mathcal{A}_1\mathcal{A}_3\mathcal{A}_1 \otimes B_1 B_3 B_1 - \mathcal{A}_1 \mathcal{A}_3\mathcal{A}_1 \otimes B_1 B_3 B_5 - \mathcal{A}_1 \mathcal{A}_3\mathcal{A}_5 \otimes B_1 B_3 B_1 +\mathcal{A}_1 \mathcal{A}_3\mathcal{A}_5 \otimes B_1 B_3 B_5)}{2-2\cos{\frac{4\pi}{5}}}\bigg]\nonumber\\
    &&\phantom{\frac{1}{2}\Bigg[}+\Tr\bigg[\frac{(\mathcal{A}_1\mathcal{A}_3\mathcal{A}_2 \otimes B_1 B_3 B_2 - \mathcal{A}_1 \mathcal{A}_3\mathcal{A}_2 \otimes B_1 B_3 B_4 - \mathcal{A}_1 \mathcal{A}_3\mathcal{A}_4 \otimes B_1 B_3 B_2 -\mathcal{A}_1 \mathcal{A}_3 \mathcal{A}_4 \otimes B_1 B_3 B_4)}{2-2\cos{\frac{2\pi}{5}}}\bigg]\Bigg]\nonumber\\
\end{eqnarray}
From the optimization in the section.~\ref{sos5}, we have found that Bob's observables have the following relations
\ba \label{b72}
B_4=\frac{B_2-B_1}{\nu^1_5}, B_5=\frac{B_3-B_2}{\nu^2_5},  B_2=\frac{B_1+B_3}{\nu^3_5} \ea 
  where $\nu^1_5=\nu^2_5=2\cos{\frac{2\pi}{5}}, \nu^3_5=2\cos{\frac{\pi}{5}}$. Since each observable is dichotomic, we have $(B_j)^2= \openone_d$ and $\Tr[B_j]=0,\forall j\in[5]$. 
Now, using the cyclic property of the trace operator, we get \ba \label{tr1}\Tr [B_i B_j B_i]=\Tr[B_j]=0, \ \forall i\neq j\in[5]\ea  Hence, using Eq. (\ref{b72}) and (\ref{tr1}),  we calculate the following. 
    \ba \Tr[B_1 B_3 B_5]&=&\Tr[B_1 B_3 \frac{B_3-B_2}{\nu^2_5}]=\frac{1}{\nu^2_5}\Tr[B_1 B_3 \qty(B_3-\frac{B_1+B_3}{\nu^3_5})]=\frac{1}{\nu^2_5}\Tr[\qty(B_1-\frac{B_1 B_3 B_1+B_1}{\nu^3_5})]=0\ea 
We can also calculate 
\begin{eqnarray}
    &&\Tr[B_1 B_3 B_2]=\Tr[B_1 B_3 \frac{B_1+B_3}{\nu^3_5}]=\frac{1}{\nu^3_5}\Tr[B_1+B_3]=0\\
    &&\Tr[B_1 B_3 B_4]=\Tr[B_1B_3\frac{B_2-B_1}{\nu^1_5}]=\frac{1}{\nu^1_5}\Tr[B_1B_3\qty(\frac{B_1+B_3}{\nu^3_5}-B_1)]=\frac{1}{\nu^1_5}\Tr[\qty(\frac{B_1B_3B_1+B_1}{\nu^3_5}-B_1B_3B_1)]=0\label{tr6}
\end{eqnarray}
Using  Eq.~(\ref{tr1})-(\ref{tr6}) in Eq.~(\ref{A1C3B1C3}), we get
\begin{eqnarray}
    \Tr[(\mathcal{A}_1\otimes B_1)( C_3 \otimes C_3)]=0
\end{eqnarray}
Note that the observables $\mathcal{A}_j,\forall j\in[n]$ satisfy relations analogous to those of $B_j$. Hence, we see that \(\frac{1}{d^2}\Tr[(\mathcal{A}_1 \otimes B_1)(C_1 \otimes C_1 + C_2 \otimes C_2)] = 1\), which gives $\Tr[(\mathcal{A}_1 \otimes B_1)\rho_{AB}]=1$, and therefore, the other terms $(C_{f} \otimes C_{f}), \forall {f}\in[d^2-1]\setminus [2]$ do not contribute. The same conclusion applies to the other correlation terms \((\mathcal{A}_j \otimes B_j)\). Thus, for all \(j\in [5]\), the only contributing terms for \(\Tr[(\mathcal{A}_j \otimes B_j) \rho_{AB}] = 1\) are \(C_1 \otimes C_1\) and \(C_2 \otimes C_2\). For all \(j\in [5]\), this results in the condition \(\Tr\left[\sum_{{f}=3}^{d^2-1} (\mathcal{A}_j \otimes B_j)(C_{f} \otimes C_{f})\right] = 0\).
Since we have $\Tr[(\mathcal{A}_1 \otimes B_1)\openone_d]=0$, we obtain 
\begin{eqnarray}
    \Tr[\mathcal{A}_1\otimes B_1 \ \rho_{AB}]&=&\frac{1}{d^2}\qty[  d^2 \cos^2{\frac{2\pi}{5}}+\frac{d^2}{4}\bigg(1-\cos{\frac{4\pi}{5}}+\frac{(\cos{\frac{\pi}{5}}-\cos{\frac{3\pi}{5}})^2}{1-\cos{\frac{2\pi}{5}}}\bigg)]=1
\end{eqnarray}
Following a similar procedure, we can show that $\Tr[\mathcal{A}_j\otimes B_j \ \rho_{AB}]=1, \forall j\in[5]$.  This shows that the state defined in Eq. (\ref{state 7}) is required to obtain the optimal quantum violation $(\mathcal{G}_5)^{opt}_Q$.
\subsection{Derivation of the  required  state for even  \texorpdfstring{$n$}{n}}\label{stateeven}
The optimization condition used to derive $(\mathcal{G}_n)^{opt}_Q$ is taken from Eq.~(\ref{selfn}) in the main text.
  \begin{eqnarray}
    A_i\otimes \mathcal{B}_i \ps_{AB}&=&\ps_{AB},\quad \forall i\in[n]
\end{eqnarray}
as derived in  Eq.~(\ref{selfn}), where $\mathcal{B}_i=\frac{1}{\mu_{n,i}}\qty(\sum\limits_{j=1}^{n-i+1}{B}_j-\sum\limits_{j=n-i+2}^n {B}_j)$ and $B_0=-B_n$. For $i=1$ and $i=\frac{n}{2}+1$, we have the relations 
 \ba A_{1}\otimes \mathcal{B}_1 \ket{\psi}_{AB} &=& \ket{\psi}_{AB}, \ 
A_{\frac{n}{2}+1}\otimes  \mathcal{B}_{\frac{n}{2}+1} \ket{\psi}_{AB} = \ket{\psi}_{AB}
 \ea 
We consider $C_1\otimes C_1=A_1\otimes \mathcal{B}_{1}$ and $C_2\otimes C_2=A_{\frac{n}{2}+1}\otimes \mathcal{B}_{\frac{n}{2}+1}$ which  implies  that \ba \Tr[C_1\otimes C_1 \ \rho_{AB}]=\Tr[C_2\otimes C_2 \ \rho_{AB}]=1\ea 
For $i=2$ and $i=\frac{n}{2}+2$, we have the relations 
 \ba \label{even2}A_{2}\otimes \mathcal{B}_2 \ket{\psi}_{AB} &=& \ket{\psi}_{AB}, \\ \label{evenn/2}
A_{\frac{n}{2}+2}\otimes  \mathcal{B}_{\frac{n}{2}+2} \ket{\psi}_{AB} &=& \ket{\psi}_{AB}
 \ea 
From the optimization, we have $\langle\{A_i,A_{i+x}\}\rangle=2\cos\frac{\pi \ x}{n},  \forall i\in[n], x\in[n-i]$ implying that \ba \label{antin/2}\langle\{A_i,A_{i+\frac{n}{2}}\}\rangle=0\ea 
Pre-multiplying $A_{\frac{n}{2}+2}\otimes  \mathcal{B}_{\frac{n}{2}+2}$ and $A_{2}\otimes \mathcal{B}_2$  in  Eq.~(\ref{even2}) and Eq.~(\ref{evenn/2}) respectively, and using Eq. (\ref{antin/2}),  \textcolor{red}{}we have
 \begin{eqnarray}
    A_{\frac{n}{2}+2} A_2\otimes \mathcal{B}_{\frac{n}{2}+2}\mathcal{B}_2 \ket{\psi}_{AB}&=&\ket{\psi}_{AB}\label{es9}\\
   A_2 A_{\frac{n}{2}+2} \otimes \mathcal{B}_{\frac{n}{2}+2}\mathcal{B}_2 \ket{\psi}_{AB}&=&-\ket{\psi}_{AB}\label{es10}\\
   A_2 A_{\frac{n}{2}+2}\otimes \mathcal{B}_2 \mathcal{B}_{\frac{n}{2}+2} \ket{\psi}_{AB}&=&\ket{\psi}_{AB}\label{es11}\\
   A_{\frac{n}{2}+2} A_2 \otimes \mathcal{B}_2 \mathcal{B}_{\frac{n}{2}+2} \ket{\psi}_{AB}&=&-\ket{\psi}_{AB}\label{es12}\\
\end{eqnarray}
Now adding Eq.~(\ref{es9}), Eq.~(\ref{es11}) and subtracting Eq.~(\ref{es10}), Eq.~(\ref{es12}) we get
\begin{eqnarray}
    \frac{1}{4}\bigg(A_{\frac{n}{2}+2} A_2\otimes \mathcal{B}_{\frac{n}{2}+2}\mathcal{B}_2+A_2 A_{\frac{n}{2}+2}\otimes \mathcal{B}_2 \mathcal{B}_{\frac{n}{2}+2}-A_2 A_{\frac{n}{2}+2} \otimes \mathcal{B}_{\frac{n}{2}+2}\mathcal{B}_2-A_{\frac{n}{2}+2} A_2 \otimes \mathcal{B}_2 \mathcal{B}_{\frac{n}{2}+2}\bigg)\ket{\psi}_{AB}&=&\ket{\psi}_{AB}
\end{eqnarray}

Similarly, for each pair $\mathcal{B}_{i}, \mathcal{B}_{\frac{n}{2}+i}$, we find the relations
\ba 
 \frac{1}{4}\bigg(A_{\frac{n}{2}+i} A_i\otimes \mathcal{B}_{\frac{n}{2}+i}\mathcal{B}_i+A_i A_{\frac{n}{2}+i}\otimes \mathcal{B}_i \mathcal{B}_{\frac{n}{2}+i}-A_i A_{\frac{n}{2}+i} \otimes \mathcal{B}_{\frac{n}{2}+i}\mathcal{B}_i-A_{\frac{n}{2}+i} A_i \otimes \mathcal{B}_i \mathcal{B}_{\frac{n}{2}+i}\bigg)\ket{\psi}_{AB}&=&\ket{\psi}_{AB}\label{es13}
\ea 

Adding all these for $i\in[\frac{n}{2}]\setminus \{1\}$ and   simplifying,  we get
\begin{eqnarray}
    &&\frac{1}{4(\frac{n}{2}-1)}\sum\limits_{i=2}^{\frac{n}{2}}\Big(A_i A_{i+\frac{n}{2}}\otimes \mathcal{B}_i \mathcal{B}_{i+\frac{n}{2}} + A_{i+\frac{n}{2}} A_i\otimes \mathcal{B}_{i+\frac{n}{2}} \mathcal{B}_i - A_i A_{i+\frac{n}{2}}\otimes \mathcal{B}_{i+\frac{n}{2}} \mathcal{B}_i - A_{i+\frac{n}{2}} A_i\otimes \mathcal{B}_i \mathcal{B}_{i+\frac{n}{2}}\Big)\ket{\psi}_{AB}=\ket{\psi}_{AB}\nonumber\\
    &&\frac{1}{4(\frac{n}{2}-1)}\sum\limits_{i=2}^{\frac{n}{2}}\big[A_i, A_{i+\frac{n}{2}}\big]\otimes \big[\mathcal{B}_i,\mathcal{B}_{i+\frac{n}{2}}\big]\ket{\psi}_{AB}=\ket{\psi}_{AB}
\end{eqnarray}
Consider
\begin{eqnarray} C_3 \otimes C_3 &=& \frac{1}{4(\frac{n}{2}-1)}\sum\limits_{i=2}^{\frac{n}{2}}\Big(A_i A_{i+\frac{n}{2}}\otimes \mathcal{B}_i \mathcal{B}_{i+\frac{n}{2}} + A_{i+\frac{n}{2}} A_i\otimes \mathcal{B}_{i+\frac{n}{2}} \mathcal{B}_i - A_i A_{i+\frac{n}{2}}\otimes \mathcal{B}_{i+\frac{n}{2}} \mathcal{B}_i - A_{i+\frac{n}{2}} A_i\otimes \mathcal{B}_i \mathcal{B}_{i+\frac{n}{2}}\Big)\nonumber\\
&=&\frac{1}{4(\frac{n}{2}-1)}\sum\limits_{i=2}^{\frac{n}{2}}\big[A_i, A_{i+\frac{n}{2}}\big]\otimes \big[\mathcal{B}_i,\mathcal{B}_{i+\frac{n}{2}}\big]
\end{eqnarray}
With some further simplification, we can write 
\begin{eqnarray}
    C_3 \otimes C_3 &=& (C_1 \otimes C_1)(C_2\otimes C_2)
\end{eqnarray}
  From the optimization conditions of the SOS approach, it can be shown that $\langle[C_{f}\otimes C_{f}, C_{f'}\otimes C_{f'}]\rangle_{\rho_{AB}}=0, \ \forall f\neq f'\in [d^2-1]$ along with $\Tr[C_{f}\otimes C_{f} \ \rho_{AB}]=1,\forall f\in [d^2-1]$.  Now,  any arbitrary $d$-dimensional bipartite state can be written as a function of $C_{f}\otimes C_{f}$ as follows.
\begin{eqnarray}\label{state 4}
    \rho_{AB} = \frac{1}{d^2} \qty[\openone_d\otimes\openone_d + C_1 \otimes C_1+C_2 \otimes C_2+C_3 \otimes C_3 + \sum_{f=4}^{d^2-1} C_{f} \otimes C_{f}]
\end{eqnarray}
Again, altogether, we get the following.
\begin{eqnarray}
    C_1\otimes C_1\ket{\psi}_{AB}=C_2\otimes C_2\ket{\psi}_{AB}=C_3\otimes C_3\ket{\psi}_{AB}=\ket{\psi}_{AB}
\end{eqnarray} 
Hence, we can directly show that $\langle [C_{f}\otimes C_{f},C_{f'}\otimes C_{f'}]\rangle_{\rho_{AB}}=0,\forall f\neq f'\in\{1,2,3\}$.

The state $\rho_{AB}$ achieves the optimal quantum violation whenever $\Tr\!\big[({A}_i\otimes \mathcal{B}_i)\,\rho_{AB}\big]=1$ for all $i\in[n]$. As $n$ increases, the resulting derivation becomes highly intricate. For clarity, we explicitly work out the case $n=4$. Applying the preceding reasoning together with the optimization conditions in (\ref{g4ai}), we obtain
\begin{eqnarray}
C_1 \otimes C_1 &=& {A}_1 \otimes \mathcal{B}_1, \ 
C_2 \otimes C_2={A}_3 \otimes \mathcal{B}_3, \ 
C_3 \otimes C_3 =\frac{1}{4}\big[A_2, A_4\big]\otimes \big[\mathcal{B}_2,\mathcal{B}_{4}\big]
\end{eqnarray}
Now we can further reduce $C_3 \otimes C_3$ as follows.
\begin{eqnarray}\label{c3c3n=4}
    C_3 \otimes C_3 \ket{\psi}_{AB}&=&\frac{1}{4}\big[A_2, A_{4}\big]\otimes \big[\mathcal{B}_2,\mathcal{B}_{4}\big]\ket{\psi}_{AB}\nonumber\\
    &=&\frac{1}{4}(A_2 A_4\otimes \mathcal{B}_2 \mathcal{B}_4 + A_4 A_2\otimes \mathcal{B}_4 \mathcal{B}_2 - A_2 A_4\otimes \mathcal{B}_4 \mathcal{B}_2 - A_4 A_2\otimes \mathcal{B}_2 \mathcal{B}_4)\ket{\psi}_{AB}\nonumber\\
    &=& \frac{1}{4}(A_2 A_4-A_4 A_2)\otimes (\mathcal{B}_2 \mathcal{B}_4-\mathcal{B}_4 \mathcal{B}_2)\ket{\psi}_{AB}\nonumber\\
    &=&( A_2 A_4\otimes \mathcal{B}_2 \mathcal{B}_4)\ket{\psi}_{AB} \quad \bigg[\text{As $\langle\{A_2,A_4\}\rangle=\langle\{\mathcal{B}_2, \mathcal{B}_4 \}\rangle=0$}\bigg]
\end{eqnarray}
Again, from the SOS conditions,  from Eq.~(\ref{bobi}), (\ref{ai}) and (\ref{cbi}) we know that $A_2=\frac{A_1+A_3}{\sqrt{2}}, A_4=\frac{A_3-A_1}{\sqrt{2}}, \mathcal{B}_2=\frac{\mathcal{B}_1+\mathcal{B}_3}{\sqrt{2}}$ and $\mathcal{B}_4=\frac{\mathcal{B}_3-\mathcal{B}_1}{\sqrt{2}}$. Now, using these conditions in Eq.~(\ref{c3c3n=4}) we get
\begin{eqnarray}
    C_3 \otimes C_3 &=& (A_1\otimes \mathcal{B}_1) (A_3\otimes \mathcal{B}_3)= (C_1 \otimes C_1)(C_2\otimes C_2)
\end{eqnarray}
Now, in order to prove the relation $\Tr[(A_i\otimes \mathcal{B}_i) \rho_{AB}]=1$, we need some additional conditions of $A_i$ and $\mathcal{B}_i$. From the optimization for $n=4$, we have $\{A_i,A_{i+x}\}=\{\mathcal{B}_i, \mathcal{B}_{i+x}\}=2\openone_d \cos{\frac{\pi x}{4}}$ only for $\rho_{AB}$, which directly implies that \ba \label{4anti}\Tr[A_i A_{i+x}]=\Tr[\mathcal{B}_i\mathcal{B}_{i+x}]=d^2\cos^2{\frac{\pi x}{4}}\ea and $\Tr[\sum_{{f}=3}^{d^2-1} A_jC_{f} \otimes \mathcal{B}_jC_{f}] =0, \forall j\in[4]$. 
For proving $\Tr[A_1\otimes \mathcal{B}_1 \ \rho_{AB}]=1$,  we will use Eq. (\ref{4anti}) and calculate the following:
\begin{eqnarray}\label{A1beta1c1}
    \Tr[(A_1\otimes  \mathcal{B}_1)(C_1 \otimes C_1)]&=&d^2,\ 
    \Tr[(A_1\otimes  \mathcal{B}_1)(C_2 \otimes C_2)]= \Tr[A_1 A_3 \otimes \mathcal{B}_1 \mathcal{B}_3]
    = 0
\end{eqnarray}
Also, we have 
\begin{eqnarray}\label{A1beta1c3}
    \Tr[(A_1\otimes \mathcal{B}_1)( C_3 \otimes C_3)]&=& \frac{1}{4}\Tr[A_1A_2 A_4\otimes \mathcal{B}_1\mathcal{B}_2 \mathcal{B}_4 + A_1 A_4 A_2\otimes \mathcal{B}_1 \mathcal{B}_4 \mathcal{B}_2 - A_1 A_2 A_4\otimes \mathcal{B}_1 \mathcal{B}_4 \mathcal{B}_2 - A_1 A_4 A_2\otimes \mathcal{B}_1\mathcal{B}_2 \mathcal{B}_4]
\end{eqnarray} Since each observable is dichotomic, we have $(A_i)^2= \openone_d$ and $\Tr[A_i]=\Tr[\mathcal{B}_i]=0,\forall i\in[4]$. 
Now, using the cyclic property of the trace operator, we get $\Tr [   A_iA_jA_i]=\Tr[A_j]=0, \ \forall i\neq j\in[4]$, which gives 
\begin{eqnarray}\label{4c3_1}
    \Tr[A_1 A_2 A_4]&=& \Tr[A_1\frac{(A_3+A_1)}{\sqrt{2}}\frac{(A_3-A_1)}{\sqrt{2}}]=0, \quad 
    \Tr[A_1 A_4 A_2]= \Tr[A_1\frac{(A_3-A_1)}{\sqrt{2}}\frac{(A_3+A_1)}{\sqrt{2}}]=0
    \end{eqnarray}
Using  Eq.~(\ref{4c3_1})  in Eq.~(\ref{A1beta1c3}) we get
\begin{eqnarray}
    \Tr[A_1 C_3 \otimes \mathcal{B}_1 C_3]&=&0
\end{eqnarray}which finally implies that
$\Tr[\mathcal{A}_1\otimes \mathcal{B}_1 \ \rho_{AB}]=\frac{1}{d^2}\qty[d^2]=1$. 
Following a similar procedure, we can show that $\Tr[A_i\otimes \mathcal{B}_i \ \rho_{AB}]=1, \   \forall i\in[4]$. This shows that the state defined in Eq. (\ref{state 4}) is required to obtain the optimal quantum violation $(\mathcal{G}_4)^{opt}_Q$.
\section{Self-testing of quantum states and observables from the optimal quantum value}
\subsection{For odd number of measurement settings  \texorpdfstring{$n$}{n}}\label{c5}
At the optimal quantum value $(\mathcal{G}_{n})^{opt}_Q$ given in the main text Eq.~(\ref{selfn}), the following condition is obtained
    \begin{eqnarray}\label{L1o}
        A_i\otimes \mathcal{B}_i\ket{\psi}_{AB}&=&\ket{\psi}_{AB}, \ \forall i\in[n].
    \end{eqnarray}
    
    For $i=1$ and $i=n$, we have the relations 
     \begin{eqnarray}\label{i1o}
        A_1\otimes \mathcal{B}_1\ket{\psi}_{AB}&=&\ket{\psi}_{AB}\\\label{ino}
        A_n\otimes \mathcal{B}_n\ket{\psi}_{AB}&=&\ket{\psi}_{AB}
    \end{eqnarray}
Pre-multiplying $ A_n A_1\otimes \openone_d$ and $A_1 A_n\otimes \openone_d $ in  the Eq.~(\ref{i1o}) and Eq.~(\ref{ino}) respectively, we get,
    \begin{eqnarray}
        A_n\otimes \mathcal{B}_1\ket{\psi}_{AB}&=&A_n A_1\otimes\openone_d\ket{\psi}_{AB}\label{A1B5o}\\
        A_1\otimes \mathcal{B}_n\ket{\psi}_{AB}&=&A_1A_n\otimes\openone_d\ket{\psi}_{AB}\label{A5B1o}
    \end{eqnarray}
    Adding Eq.~(\ref{i1o}),(\ref{ino}) and subtracting Eq.~(\ref{A1B5o}),(\ref{A5B1o}) we get,
    \begin{eqnarray}
        &&\qty(A_1\otimes \mathcal{B}_1+A_n\otimes \mathcal{B}_n- A_n\otimes \mathcal{B}_1-A_1\otimes \mathcal{B}_n)\ket{\psi}_{AB}= (2\openone_d\otimes\openone_d-\{A_1,A_n\}\otimes\openone_d)\ket{\psi}_{AB}\nonumber\\
        &&\bra{\psi}_{AB}\qty(A_1\otimes \mathcal{B}_1+A_n\otimes \mathcal{B}_n- A_n\otimes \mathcal{B}_1-A_1\otimes \mathcal{B}_n)\ket{\psi}_{AB}= (2-\langle\{A_1,A_n\}\rangle)\nonumber\\
        &&\bra{\psi}_{AB}\frac{\qty(A_1\otimes \mathcal{B}_1+A_n\otimes \mathcal{B}_n- A_n\otimes \mathcal{B}_1-A_1\otimes \mathcal{B}_n)}{(2-\langle\{A_1,A_n\}\rangle)}\ket{\psi}_{AB}=1\nonumber\\
    &&\frac{\qty(A_1\otimes \mathcal{B}_1+A_n\otimes \mathcal{B}_n- A_n\otimes \mathcal{B}_1-A_1\otimes \mathcal{B}_n)}{(2-\langle\{A_1,A_n\}\rangle)}\ket{\psi}_{AB}= \ket{\psi}_{AB}\quad\bigg[\text{$(2-\langle\{A_1,A_n\}\rangle)$ is normalization constant}\bigg]\label{A1AnB1Bno}
    \end{eqnarray}
Following similar steps, we also get for any pair $A_i,A_{n+1-i}, \  i\in[n]$ we get
\begin{eqnarray}\label{AiBio}
    \frac{\qty(A_i\otimes \mathcal{B}_i+A_{n+1-i}\otimes \mathcal{B}_{n+1-i}- A_i\otimes \mathcal{B}_{n+1-i}-A_{n+1-i}\otimes \mathcal{B}_i)}{(2-\langle\{A_i,A_{n+1-i}\}\rangle)}\ket{\psi}_{AB}= \ket{\psi}_{AB}
\end{eqnarray}
From the optimization conditions, we clearly have $\langle\{A_i,A_{n+1-i}\}\rangle=\langle\{\mathcal{B}_i,\mathcal{B}_{n+1-i}\}\rangle$ for each $i\in[n]$.  Using it in Eq.~(\ref{AiBio}),  we get
\begin{eqnarray}
    &&\frac{\qty(A_i\otimes \mathcal{B}_i+A_{n+1-i}\otimes \mathcal{B}_{n+1-i}- A_i\otimes \mathcal{B}_{n+1-i}-A_{n+1-i}\otimes \mathcal{B}_i)}{(2-\langle\{A_i,A_{n+1-i}\}\rangle)}\ket{\psi}_{AB}= \ket{\psi}_{AB}\\
    &&\left(\frac{A_i-A_{n+1-i}}{\sqrt{2-\langle\{A_1,A_{n+1-i}\}\rangle}}\otimes \frac{\mathcal{B}_i-\mathcal{B}_{n+1-i}}{\sqrt{2-\langle\{\mathcal{B}_i,\mathcal{B}_{n+1-i}\}\rangle}}\right)\ket{\psi}_{AB}=\ket{\psi}_{AB} \nonumber\\\label{selfti}
         &&\left(\frac{A_i-A_{n+1-i}}{\sqrt{2-\langle\{A_i,A_{n+1-i}\}\rangle}}\otimes \openone_d\right)\ket{\psi}_{AB}= \left(\openone_d\otimes \frac{\mathcal{B}_i-\mathcal{B}_{n+1-i}}{\sqrt{2-\langle\{\mathcal{B}_i,\mathcal{B}_{n+1-i}\}\rangle}}\right)\ket{\psi}_{AB}
\end{eqnarray}
 Adding  for each  $i\in\bigg[\lfloor\frac{n}{2}\rfloor\bigg]$ and   simplifying it,  we get  
 \begin{eqnarray}
     \left(\frac{1}{\lfloor\frac{n}{2}\rfloor}\sum_{i=1}^{{\lfloor \frac{n}{2}\rfloor}}\frac{A_{i}-A_{n+1-i}}{\sqrt{2-\langle\{A_{i},A_{n+1-i}\}\rangle}}\otimes \openone_d\right)\ket{\psi}_{AB}&=&\left(\openone_d\otimes \frac{1}{\lfloor\frac{n}{2}\rfloor}\sum_{i=1}^{{\lfloor \frac{n}{2}\rfloor}}\frac{\mathcal{B}_{i}-\mathcal{B}_{n+1-i}}{\sqrt{2-\langle\{\mathcal{B}_{i},\mathcal{B}_{n+1-i}\}\rangle}}\right)\ket{\psi}_{AB} \label{ct}
 \end{eqnarray}
 which implies 
 \begin{eqnarray}
     X_A\ket{\psi}_{AB}=X_B\ket{\psi}_{AB}
 \end{eqnarray}
 where $X_A$ and $X_B$ are given by 
\begin{eqnarray}
    X_A &=&\frac{1}{\lfloor\frac{n}{2}\rfloor}\sum_{i=1}^{{\lfloor \frac{n}{2}\rfloor}}\frac{A_{i}-A_{n+1-i}}{\sqrt{2-\langle\{A_{i},A_{n+1-i}\}\rangle}}, \quad 
    X_B = \frac{1}{\lfloor\frac{n}{2}\rfloor}\sum_{i=1}^{{\lfloor \frac{n}{2}\rfloor}}\frac{\mathcal{B}_{i}-\mathcal{B}_{n+1-i}}{\sqrt{2-\langle\{\mathcal{B}_{i},\mathcal{B}_{n+1-i}\}\rangle}}
\end{eqnarray}
Also, from the self-testing condition $(A_{\frac{n+1}{2}}\otimes  \mathcal{B}_{\frac{n+1}{2}}) \ket{\psi}_{AB}=\ket{\psi}_{AB}$, we can write 
\begin{eqnarray}
    (A_{\frac{n+1}{2}}\otimes \openone_d) \ket{\psi}_{AB}&=& (\openone_d \otimes \mathcal{B}_{\frac{n+1}{2}}) \ket{\psi}_{AB}
\end{eqnarray}
 which satisfies the following condition
  \begin{eqnarray}
     Z_A\ket{\psi}_{AB}=Z_B\ket{\psi}_{AB}
 \end{eqnarray}
where $Z_A = A_{\frac{n+1}{2}}$ and $Z_B=\mathcal{B}_{\frac{n+1}{2}}$. Moreover, one can verify that $\{Z_A,X_A\}\ket{\psi}_{AB}=\{Z_B,X_B\}\ket{\psi}_{AB}=0$. If $\ket{\psi}_{AB}\in\mathcal{H}_A\otimes\mathcal{H}_B$ is the appropriate state and $A_i\in \mathcal{H}_A,\ B_j\in\mathcal{H}_B,\ (\forall i,j\in[n])$ are the associated observables that yield the optimal quantum violation, then, using Fig.~\ref{Oddc} from the main text, it follows that there exist a local unitary operation $\Phi$ and an ancillary state $\ket{00}_{A'B'}$ such that
\begin{eqnarray}\label{phic}
    \Phi(\ket{\psi}_{AB}\otimes\ket{00}_{A'B'} 
 )&=&\frac{1}{4}\Bigg[(1+Z_A)(1+Z_B)\ket{\psi}_{AB}\ket{00}+X_B(1+Z_A)(1-Z_B)\ket{\psi}_{AB}\ket{01}\nonumber\\
 &&\phantom{\frac{1}{4}\Bigg[}+X_A(1-Z_A)(1+Z_B)\ket{\psi}_{AB}\ket{10}+X_AX_B(1-Z_A)(1-Z_B)\ket{\psi}_{AB}\ket{11}]
\end{eqnarray}
Using the self-testing properties in  Eq.~(\ref{circuitself}) of the main text,  we can rewrite Eq.~(\ref{phic}) as follows.
\begin{eqnarray}
    \Phi(\ket{\psi}_{AB}\otimes\ket{00}_{A'B'} 
 )&=&\ket{\chi}_{AB}\otimes\ket{\phi^+}_{A'B'}
\end{eqnarray}
where $\ket{\chi}_{AB}=\frac{1+Z_A}{\sqrt{2}}\ket{\psi}_{AB}$. This directly depicts the self-testing of a two-qubit maximally entangled state using the optimal quantum value of the Bell functional $\mathscr{C}_{n}$. Note that we can consider 
 \begin{eqnarray}\label{ABxz}
&&A_i=\cos\qty(\frac{(i-1)\pi}{n})Z_A+\sin\qty(\frac{(i-1)\pi}{n})X_A,\quad B_j=-\cos{\qty(\frac{(1-n-2j)\pi}{2n})}Z_A -\sin{\qty(\frac{(1-n-2j)\pi}{2n})}X_A,\quad\forall i,j\in [n]
  \end{eqnarray}

Clearly, here $A_i$ and $B_j$ can be written in terms of $Z_A, Z_B, X_A$ and $ X_B$. It is sufficient to demonstrate the function of local isometry for $Z_A, Z_B, X_A$ and $ X_B$. Hence, from the self-testing circuit, we get the following.  
\begin{eqnarray}\label{D1}
    \Phi(X_A\ket{\psi}_{AB}\otimes\ket{00}_{A'B'} 
 )&=&\frac{1}{4}\Bigg[(1+Z_A)X_A(1+Z_B)\ket{\psi}_{AB}\ket{00}+X_B(1+Z_A)X_A(1-Z_B)\ket{\psi}_{AB}\ket{01}\nonumber\\
 &&+X_A(1-Z_A)X_A(1+Z_B)\ket{\psi}_{AB}\ket{10}+X_AX_B(1-Z_A)X_A(1-Z_B)\ket{\psi}_{AB}\ket{11}\Bigg]
\end{eqnarray}
Using $X_A\ket{\psi}_{AB}=X_B\ket{\psi}_{AB}$, $Z_A\ket{\psi}_{AB}=Z_B\ket{\psi}_{AB}$ and $\{Z_A,X_A\}\ket{\psi}_{AB}=\{Z_B,X_B\}\ket{\psi}_{AB}=0$  in Eq.~(\ref{D1}) we get
\begin{eqnarray}\label{D2}
    \Phi(X_A\ket{\psi}_{AB}\otimes\ket{00}_{A'B'})=\frac{1+Z_A}{\sqrt{2}}\ket{\psi}_{AB}\otimes\frac{\ket{10}+\ket{01}}{\sqrt{2}}=\ket{\chi}_{AB}\otimes (\sigma_x\otimes\openone_d)\ket{\phi^+}_{A'B'}
\end{eqnarray}
In a similar way, we can further show that
\begin{eqnarray}\label{D3} 
\Phi(X_B\ket{\psi}_{AB}\otimes\ket{00}_{A'B'})&=&\frac{1+Z_A}{\sqrt{2}}\ket{\psi}_{AB}\otimes\frac{\ket{01}+\ket{10}}{\sqrt{2}}=\ket{\chi}_{AB}\otimes (\openone_d\otimes\sigma_x)\ket{\phi^+}_{A'B'}\\\Phi(Z_A\ket{\psi}_{AB}\otimes\ket{00}_{A'B'})&=&\frac{1+Z_A}{\sqrt{2}}\ket{\psi}_{AB}\otimes\frac{\ket{00}-\ket{11}}{\sqrt{2}}=\ket{\chi}_{AB}\otimes (\sigma_z\otimes\openone_d)\ket{\phi^+}_{A'B'}\label{D4}\\ \Phi(Z_B\ket{\psi}_{AB}\otimes\ket{00}_{A'B'})&=&\frac{1+Z_A}{\sqrt{2}}\ket{\psi}_{AB}\otimes\frac{\ket{00}-\ket{11}}{\sqrt{2}}=\ket{\chi}_{AB}\otimes (\openone_d\otimes\sigma_z)\ket{\phi^+}_{A'B'}\label{D5}
\end{eqnarray}
As we can express $A_i$ and $B_j$ in terms of $Z_A, Z_B, X_A,$ and $X_B$, similar relations hold for $A_i$ and $B_j$, for each $i,j \in [n]$. Hence, using   Eq.~(\ref{D2}),(\ref{D3}),(\ref{D4}) and (\ref{D5}), we can show that
\begin{eqnarray}
\Phi(A_i\ket{\psi}_{AB}\otimes\ket{00}_{A'B'} 
 )&=&\ket{\chi}_{AB}\otimes (A'_i\otimes \openone_d)\ket{\phi^+}_{A'B'}\\
\Phi(B_j\ket{\psi}_{AB}\otimes\ket{00}_{A'B'} 
 )&=&\ket{\chi}_{AB}\otimes (\openone_d\otimes B'_j)\ket{\phi^+}_{A'B'}\\
 \Phi(A_i\otimes B_j\ket{\psi}_{AB}\otimes\ket{00}_{A'B'} 
 )&=&\ket{\chi}_{AB}\otimes (A'_i\otimes B'_j)\ket{\phi^+}_{A'B'}
\end{eqnarray}
The above equations self-test the set of $n$ measurement settings $A_i$ and $B_j$ (for each $i,j\in[n])$ for Alice and Bob, respectively.
\subsection{For even number of measurement settings \texorpdfstring{$n$}{n}}\label{c4}
Using the optimization condition  in as Eq.~(\ref{selfn}) of the main text, for $i=1$ and $n$, we can write 
    \begin{eqnarray}\label{Le1}
        A_1\otimes \mathcal{B}_1\ket{\psi}_{AB}&=&\ket{\psi}_{AB}\\
   \label{Len}
        A_n\otimes \mathcal{B}_n\ket{\psi}_{AB}&=&\ket{\psi}_{AB}
    \end{eqnarray}
    Multiplying $ A_n A_1\otimes\openone_d$ and $ A_1 A_n\otimes\openone_d$ from the left side of the Eq.~(\ref{Le1}) and Eq.~(\ref{Len}) respectively, we get,
    \begin{eqnarray}\label{EA1B4}
        A_n\otimes \mathcal{B}_1\ket{\psi}_{AB}&=&A_n A_1\otimes\openone_d\ket{\psi}_{AB}\\
        \label{EA4B1}
        A_1\otimes \mathcal{B}_n\ket{\psi}_{AB}&=&A_1A_n\otimes \openone_d\ket{\psi}_{AB}
    \end{eqnarray}
    Since the optimization condition provides $\langle\{A_i,A_{n+1-i}\}\rangle=\langle\{\mathcal{B}_i,\mathcal{B}_{n+1-i}\}\rangle$ for each $i\in[n]$,  adding Eq.~(\ref{Le1}),(\ref{Len}); and then subtracting Eq.~(\ref{EA1B4}), (\ref{EA4B1}) we get the following.
    \ba
        &&A_1\otimes \mathcal{B}_1+A_n\otimes \mathcal{B}_n-A_n\otimes \mathcal{B}_1-A_1\otimes \mathcal{B}_n\ket{\psi}_{AB}= (2\openone_d\otimes\openone_d-\{A_1,A_n\}\otimes\openone_d)\ket{\psi}_{AB}\nonumber\\\na 
        &&\bra{\psi}_{AB}A_1\otimes \mathcal{B}_1+A_n\otimes \mathcal{B}_n-A_n\otimes \mathcal{B}_1-A_1\otimes \mathcal{B}_n\ket{\psi}_{AB}= (2-\langle\{A_1,A_n\}\rangle)\nonumber\\\na
        &&\bra{\psi}_{AB}\frac{A_1\otimes \mathcal{B}_1+A_n\otimes \mathcal{B}_n-A_n\otimes \mathcal{B}_1-A_1\otimes \mathcal{B}_n}{(2-\langle\{A_1,A_n\}\rangle)}\ket{\psi}_{AB}= 1\nonumber\\
        &&\frac{A_1-A_n}{\sqrt{2-\langle\{A_1,A_n\}\rangle}}\otimes \frac{ \mathcal{B}_1- \mathcal{B}_n}{\sqrt{2-\langle\{\mathcal{B}_1, \mathcal{B}_n\}\rangle}}\ket{\psi}_{AB}=\ket{\psi}_{AB} \quad \nonumber\ea 
    which implies that \ba     \left(\frac{A_1-A_n}{\sqrt{2-\langle\{A_1,A_n\}\rangle}}\otimes \openone_d\right)\ket{\psi}_{AB}&=&\left(\openone_d\otimes \frac{\mathcal{B}_1-\mathcal{B}_n}{\sqrt{2-\langle\{\mathcal{B}_1,\mathcal{B}_n\}\rangle}}\right)\ket{\psi}_{AB}
    \ea
Following the similar steps, we also get for any pair $A_i,A_{n+1-i}, \  i\in[n]$ we get
\ba
    \left(\frac{A_i-A_{n+1-i}}{\sqrt{2-\langle\{A_i,A_{n+1-i}\}\rangle}}\otimes \openone_d\right)\ket{\psi}_{AB}&=&\left(\openone_d\otimes \frac{\mathcal{B}_i-\mathcal{B}_{n+1-i}}{\sqrt{2-\langle\{\mathcal{B}_i,\mathcal{B}_{n+1-i}\}\rangle}}\right)\ket{\psi}_{AB}
\ea 
Adding  for each $i\in[\frac{n}{2}]$ and simplifying it we get
 \begin{eqnarray}
     \frac{1}{\frac{n}{2}}\left(\sum_{i=1}^{ \frac{n}{2}}\frac{A_{i}-A_{n+1-i}}{\sqrt{2-\langle\{A_{i},A_{n+1-i}\}\rangle}}\otimes \openone_d\right)\ket{\psi}_{AB}&=&\left(\openone_d\otimes  \frac{1}{\frac{n}{2}}\sum_{i=1}^{ \frac{n}{2}}\frac{\mathcal{B}_{i}-\mathcal{B}_{n+1-i}}{\sqrt{2-\langle\{\mathcal{B}_{i},\mathcal{B}_{n+1-i}\}\rangle}}\right)\ket{\psi}_{AB}\label{ect4}
 \end{eqnarray}
 Hence Eq.~(\ref{ect4}) satisfies  the following condition
 \begin{eqnarray}
     X_A\ket{\psi}_{AB}=X_B\ket{\psi}_{AB}
 \end{eqnarray}
 where $X_A$ and $X_B$ are given by 
\begin{eqnarray}
    X_A &=&\frac{2}{n}\sum_{i=1}^{ \frac{n}{2}}\frac{A_{i}-A_{n+1-i}}{\sqrt{2-\langle\{A_{i},A_{n+1-i}\}\rangle}}, \ 
    X_B = \frac{2}{n}\sum_{i=1}^{ \frac{n}{2}}\frac{\mathcal{B}_{i}-\mathcal{B}_{n+1-i}}{\sqrt{2-\langle\{\mathcal{B}_{i},\mathcal{B}_{n+1-i}\}\rangle}}
\end{eqnarray}
Now adding Eq.~(\ref{Le1})-(\ref{EA4B1}) and following the similar process, we get
    \begin{eqnarray}\left(\frac{A_1+A_n}{\sqrt{2+\langle\{A_1,A_n\}\rangle}}\otimes \openone_d\right)\ket{\psi}_{AB}&=&\left(\openone_d\otimes \frac{\mathcal{B}_1+\mathcal{B}_n}{\sqrt{2+\langle\{\mathcal{B}_1,\mathcal{B}_n\}\rangle}}\right)\ket{\psi}_{AB}\label{E+A1A4B1B4}
    \end{eqnarray}
Similarly for any pair $A_i,A_{n+1-i}, \  i\in[n]$ we get
\ba
    \left(\frac{A_i+A_{n+1-i}}{\sqrt{2+\langle\{A_i,A_{n+1-i}\}\rangle}}\otimes \openone_d\right)\ket{\psi}_{AB}&=&\left(\openone_d\otimes \frac{\mathcal{B}_i+\mathcal{B}_{n+1-i}}{\sqrt{2+\langle\{\mathcal{B}_i,\mathcal{B}_{n+1-i}\}\rangle}}\right)\ket{\psi}_{AB}
\ea 
Adding  for each $i\in[\frac{n}{2}]$ and simplifying it we get
 \begin{eqnarray} \label{e+ct4}
     \frac{1}{\frac{n}{2}}\left(\sum_{i=1}^{ \frac{n}{2}}\frac{A_{i}+A_{n+1-i}}{\sqrt{2+\langle\{A_{i},A_{n+1-i}\}\rangle}}\otimes \openone_d\right)\ket{\psi}_{AB}&=&\left(\openone_d\otimes  \frac{1}{\frac{n}{2}}\sum_{i=1}^{ \frac{n}{2}}\frac{\mathcal{B}_{i}+\mathcal{B}_{n+1-i}}{\sqrt{2+\langle\{\mathcal{B}_{i},\mathcal{B}_{n+1-i}\}\rangle}}\right)\ket{\psi}_{AB}
 \end{eqnarray}
 Hence Eq.~(\ref{e+ct4}) satisfies the following condition.
 \begin{eqnarray}
     Z_A\ket{\psi}_{AB}=Z_B\ket{\psi}_{AB}
 \end{eqnarray}
 where $Z_A$ and $Z_B$ are given by 
\begin{eqnarray}
    Z_A &=&\frac{2}{n}\sum_{i=1}^{ \frac{n}{2}}\frac{A_{i}+A_{n+1-i}}{\sqrt{2+\langle\{A_{i},A_{n+1-i}\}\rangle}}, \ 
    Z_B = \frac{2}{n}\sum_{i=1}^{ \frac{n}{2}}\frac{\mathcal{B}_{i}+\mathcal{B}_{n+1-i}}{\sqrt{2+\langle\{\mathcal{B}_{i},\mathcal{B}_{n+1-i}\}\rangle}}
\end{eqnarray}
We can further show that $\{X_A,Z_A\}\ket{\psi}_{AB}=\{X_B,Z_B\}\ket{\psi}_{AB}=0$. \textcolor{red}{}Now the state and the observables can be self-tested by defining a local unitary operation $\Phi$ and an ancillary state $\ket{00}_{A'B'}$ such that
\begin{eqnarray}\label{phi'c}
    \Phi(\ket{\psi}_{AB}\otimes\ket{00}_{A'B'} 
 )&=&\frac{1}{4}\Bigg[(1+Z_A)(1+Z_B)\ket{\psi}_{AB}\ket{00}+X_B(1+Z_A)(1-Z_B)\ket{\psi}_{AB}\ket{01}\nonumber\\
 &&\phantom{\frac{1}{4}\Bigg[}+X_A(1-Z_A)(1+Z_B)\ket{\psi}_{AB}\ket{10}+X_AX_B(1-Z_A)(1-Z_B)\ket{\psi}_{AB}\ket{11}\Bigg]
\end{eqnarray}
Using the self-testing properties in  Eq.~(\ref{circuitself}) of the main text,  we can rewrite Eq.~(\ref{phi'c}) as follows.
\begin{eqnarray}
    \Phi(\ket{\psi}_{AB}\otimes\ket{00}_{A'B'} 
 )&=&\ket{\chi}_{AB}\otimes\ket{\phi^+}_{A'B'}
\end{eqnarray}
where $\ket{\chi}_{AB}=\frac{1+Z_A}{\sqrt{2}}\ket{\psi}_{AB}$. Proceeding as in Eq. (\ref{ABxz}) and applying the same method to the observables, we obtain 
\begin{eqnarray}
&&\Phi(A_i\ket{\psi}_{AB}\otimes\ket{00}_{A'B'} 
 )=\ket{\chi}_{AB}\otimes (A'_i\otimes \openone_d)\ket{\phi^+}_{A'B'}\\
&&\Phi(B_j\ket{\psi}_{AB}\otimes\ket{00}_{A'B'} 
 )=\ket{\chi}_{AB}\otimes (\openone_d\otimes B'_j)\ket{\phi^+}_{A'B'}\\
 &&\Phi(A_i\otimes B_j\ket{\psi}_{AB}\otimes\ket{00}_{A'B'} 
 )=\ket{\chi}_{AB}\otimes (A'_i\otimes B'_j)\ket{\phi^+}_{A'B'}
\end{eqnarray}
This set of equations self-tests the set of $n$ measurements for Alice and Bob.
\section{Comprehensive analysis for robust self-testing of quantum states and observables}\label{robustselftesting}
We assume that the observables $\Tilde{X}_m$ and $ \Tilde{Z}_m$ deviate from the real observables $X_m$ and  $ Z_m$  with the margins $\epsilon_m$ and $\alpha_m$, respectively. Then mathematically, we can write 
\begin{eqnarray}\label{rob}
    ||(\Tilde{X}_m-X_m)\ket{\psi}_{AB}||\leq \alpha_m\Rightarrow  \Tilde{X}_m\approx \alpha_m \openone_d + X_m \ ; \ \ \ ||(\Tilde{Z}_m-Z_m)\ket{\psi}_{AB}||\leq \beta_m \Rightarrow  \Tilde{Z}_m\approx \beta_m \openone_d + Z_m \ ;\ \ \forall m\in\{A, B\}
\end{eqnarray}
 where $||X_m\ket{\psi}_{AB}||=||Z_m\ket{\psi}_{AB}||=1$. 
We derive the following relations using Eq.~(\ref{rob}). 
\ba\label{robzz}
    ||(\Tilde{Z}_m \Tilde{Z}_p-Z_m Z_p)\ket{\psi}_{AB}||&\approx& ||(\Tilde{Z}_m (\beta_p \openone_d + Z_p)-Z_m Z_p)\ket{\psi}_{AB}||\nonumber\\
    &\leq& \beta_p ||\Tilde{Z}_m\ket{\psi}_{AB}||+||(\Tilde{Z}_m-Z_m)Z_p\ket{\psi}_{AB}||\quad (\text{Using Tri-angular inequality})\nonumber\\
    &\leq& \beta_p ||\Tilde{Z}_m\ket{\psi}_{AB}||+\beta_m\ea 
Similarly, we can also derive  \ba \label{robxz}
    ||(\Tilde{X}_m \Tilde{Z}_p-X_m Z_p)\ket{\psi}_{AB}||&\approx& ||(\Tilde{X}_m (\beta_p \openone_d + Z_p))-X_m Z_p)\ket{\psi}_{AB}||
    \leq \beta_p ||\Tilde{X}_m\ket{\psi}_{AB}||+\alpha_m
\\ 
    ||(\Tilde{X}_m \Tilde{X}_p-X_m X_p)\ket{\psi}_{AB}||&\approx& ||(\Tilde{X}_m (\alpha_p \openone_d + X_p))-X_m X_p)\ket{\psi}_{AB}||
    \leq \alpha_p ||\Tilde{X}_m\ket{\psi}_{AB}||+\alpha_m\label{robxx}\ea 
Now using Eq.~(\ref{robxz}), we calculate the following. 
\ba \label{robxzz}
    ||(\Tilde{X}_m \Tilde{Z}_p\Tilde{Z}_k-X_m Z_p Z_k)\ket{\psi}_{AB}||&\approx& ||(\Tilde{X}_m \Tilde{Z}_p(\beta_k \openone_d + Z_k))-X_m Z_p Z_k)\ket{\psi}_{AB}||\nonumber\\
    &\leq& \beta_k ||\Tilde{X}_m\Tilde{Z}_p\ket{\psi}_{AB}||+||(\Tilde{X}_m\Tilde{Z}_p-X_mZ_p)Z_k\ket{\psi}_{AB}||\quad (\text{Using Tri-angular inequality})\nonumber\\
    &\leq& \beta_k ||\Tilde{X}_m\Tilde{Z}_p\ket{\psi}_{AB}||+\beta_p ||\Tilde{X}_m\ket{\psi}_{AB}||+\alpha_m
\end{eqnarray}
Similarly, using Eq.~(\ref{robxx}), we get 
\begin{eqnarray}\label{robxxz}
    ||(\Tilde{X}_m \Tilde{X}_p\Tilde{Z}_k-X_m X_p Z_k)\ket{\psi}_{AB}||&\approx& ||(\Tilde{X}_m \Tilde{X}_p(\beta_k \openone_d + Z_k))-X_m X_p Z_k)\ket{\psi}_{AB}||\nonumber\\
    &\leq& \beta_k ||\Tilde{X}_m \Tilde{X}_p\ket{\psi}_{AB}||+||(\Tilde{X}_m \Tilde{X}_p-X_m X_p) Z_k\ket{\psi}_{AB}||\quad (\text{Using Tri-angular inequality})\nonumber\\
    &\leq& \beta_k ||\Tilde{X}_m \Tilde{X}_p\ket{\psi}_{AB}||+\alpha_p ||\Tilde{X}_m\ket{\psi}_{AB}||+\alpha_m 
\end{eqnarray}  Using Eq.~(\ref{robxxz}), we get 
\begin{eqnarray}\label{robxxzz}
    ||(\Tilde{X}_m \Tilde{X}_p\Tilde{Z}_k\Tilde{Z}_l-X_m X_p Z_k Z_l)\ket{\psi}_{AB}||&\approx& ||(\Tilde{X}_m \Tilde{X}_p\Tilde{Z}_k(\beta_l \openone_d + Z_l))-X_m X_p Z_k Z_l)\ket{\psi}_{AB}||\nonumber\\
    &\leq& \beta_l ||\Tilde{X}_m \Tilde{X}_p\Tilde{Z}_k\ket{\psi}_{AB}||+||(\Tilde{X}_m \Tilde{X}_p\Tilde{Z}_k-X_m X_p Z_k) Z_l\ket{\psi}_{AB}||\quad (\text{Using Tri-angular inequality})\nonumber\\
    &\leq& \beta_l ||\Tilde{X}_m \Tilde{X}_p \Tilde{Z}_k\ket{\psi}_{AB}||+\beta_k ||\Tilde{X}_m \Tilde{X}_p\ket{\psi}_{AB}||+\alpha_p ||\Tilde{X}_m\ket{\psi}_{AB}||+\alpha_m
\end{eqnarray}
\footnote{Tri-angular inequality: $||M\pm N||\leq||M||+||N||$}
\subsection{Robust self-testing of the state}
For the perfect  implementation of the isometry $\Phi$, the output state  can be written as 
\ba\label{robs}
\Phi(\ket{\psi}_{AB}\otimes\ket{00}_{A'B'})&=&\frac{1}{4}\sum_{a,b\in\{0,1\}} (X_A)^a (X_B)^b  \Big(1+(-1)^a Z_A\Big) \Big(1+(-1)^b Z_B\Big) \ket{\psi}_{AB} \ket{ab}_{A'B'}\ea 
Similarly, for the imperfect implementation of the isometry, the output state  can be written as  
\ba \Tilde{\Phi}(\ket{\psi}_{AB}\otimes\ket{00}_{A'B'})&=&\frac{1}{4}\sum_{a,b\in\{0,1\}} (\Tilde{X}_A)^a (\Tilde{X}_B)^b \Big(1+(-1)^a \Tilde{Z}_A\Big) \Big(1+(-1)^b \Tilde{Z}_B\Big) \ket{\psi}_{AB} \ket{ab}_{A'B'}
\ea
This, in turn, provides that 
\begin{eqnarray}\label{robs1}
    ||\Tilde{\Phi}(\ket{\psi}_{AB}\otimes\ket{00}_{A'B'})-\Phi(\ket{\psi}_{AB}\otimes\ket{00}_{A'B'})||&=&\frac{1}{4}\sum_{a,b\in\{0,1\}} \Bigg|\Bigg|\Big[(\Tilde{X}_A)^a (\Tilde{X}_B)^b \Big(1+(-1)^a \Tilde{Z}_A\Big) \Big(1+(-1)^b \Tilde{Z}_B\Big)\nonumber\\
    &&\phantom{\frac{1}{4}\sum_{a,b\in\{0,1\}} \Bigg|\Bigg|\Big[}-(X_A)^a (X_B)^b \Big(1+(-1)^a Z_A\Big) \Big(1+(-1)^b Z_B\Big) \Big] \ket{\psi}_{AB} \ket{ab}_{A'B'}\Bigg|\Bigg|
\end{eqnarray}
Putting $a=0, b=0$ in Eq. (\ref{robs1}) and using Eq.~(\ref{rob}),(\ref{robzz}),  we have 
\begin{eqnarray}\label{robs2}
   \Bigg|\Bigg|\Big[(1+ \Tilde{Z}_A) (1+ \Tilde{Z}_B)
    -(1+ Z_A) (1+ Z_B) \Big] \ket{\psi}_{AB} \ket{00}_{A'B'}\Bigg|\Bigg|\nonumber
    &\leq& \Big[||(\Tilde{Z}_A-Z_A)\ket{\psi}_{AB}||+||(\Tilde{Z}_B-Z_B)\ket{\psi}_{AB}||+||(\Tilde{Z}_A\Tilde{Z}_B-Z_A Z_B)\ket{\psi}_{AB}||\Big]\ket{00}_{A'B'}\nonumber\\
    &\leq&\Big[2\beta_A+\beta_B+\beta_B||\Tilde{Z}_A\ket{\psi}_{AB}||\Big]\ket{00}_{A'B'}\nonumber\\
    &\leq& f_1(\beta_A,\beta_B) \ket{00}_{A'B'}
\end{eqnarray}
Similarly, putting $a=0, b=1$ in Eq. (\ref{robs1}) and using Eq.~(\ref{rob}), (\ref{robxz}) and (\ref{robxzz}),  we have 
\begin{eqnarray}\label{robs3}
    \Bigg|\Bigg|\Big[\Tilde{X}_B(1+ \Tilde{Z}_A) (1-\Tilde{Z}_B)
    -X_B(1+ Z_A) (1-Z_B)\Big] \ket{\psi}_{AB} \ket{01}_{A'B'}\Bigg|\Bigg|
    &\leq&\Big[\beta_B\qty(||\Tilde{X}_B\ket{\psi}_{AB}||+||\Tilde{X}_B\Tilde{Z}_A\ket{\psi}_{AB}||)+4 \alpha_B+2\beta_A ||\Tilde{X}_B\ket{\psi}_{AB}||\Big]\ket{01}_{A'B'}\nonumber\\
    &=& f_2(\alpha_B,\beta_A,\beta_B) \ket{01}_{A'B'}
\end{eqnarray}
Putting $a=1, b=0$ in Eq. (\ref{robs1}) and using Eq.~(\ref{rob}),(\ref{robxz}) and(\ref{robxzz}),  we have 
\begin{eqnarray}\label{robs4}
\Bigg|\Bigg|\Big[\Tilde{X}_A(1-\Tilde{Z}_A) (1+\Tilde{Z}_B)
    -X_A(1-Z_A) (1+Z_B)\Big] \ket{\psi}_{AB} \ket{10}_{A'B'}\Bigg|\Bigg|
    &\leq&\Big[\beta_A\qty(||\Tilde{X}_A\ket{\psi}_{AB}||+||\Tilde{X}_A\Tilde{Z}_B\ket{\psi}_{AB}||)+4 \alpha_A+2\beta_B ||\Tilde{X}_A\ket{\psi}_{AB}||\Big]\ket{10}_{A'B'}\nonumber\\
    &=& f_3(\alpha_A,\beta_A,\beta_B) \ket{10}_{A'B'}
\end{eqnarray}
Similarly, putting  $a=1, b=1$ in Eq. (\ref{robs1}) and using Eq.~(\ref{rob}), (\ref{robxz}), (\ref{robxx}), (\ref{robxzz}) and (\ref{robxxzz}), we have
\begin{eqnarray}\label{robs5}
   \Bigg|\Bigg|\Big[\Tilde{X}_A\Tilde{X}_B (1-\Tilde{Z}_A) (1-\Tilde{Z}_B)
    -X_A X_B (1-Z_A) (1-Z_B)\Big] \ket{\psi}_{AB} \ket{11}_{A'B'}\Bigg|\Bigg|\nonumber
    &=&\Big[\beta_B\qty(||\Tilde{X}_A\Tilde{X}_B\ket{\psi}_{AB}||+||\Tilde{X}_A\Tilde{X}_B\Tilde{Z}_A\ket{\psi}_{AB}||)+2\beta_A||\Tilde{X}_A\Tilde{X}_B\ket{\psi}_{AB}||\nonumber\\
    && +4\alpha_B||\Tilde{X}_A\ket{\psi}_{AB}||+4\alpha_A\Big]\ket{11}_{A'B'}\nonumber\\
    &\leq& f_4(\alpha_A,\alpha_B,\beta_A,\beta_B) \ket{11}_{A'B'}
\end{eqnarray}
 Substituting Eq.~(\ref{robs2})-(\ref{robs5})  in Eq.~(\ref{robs1}) and neglecting the ancillary part, we get the following.
\begin{eqnarray}\label{robsF}
    ||\Tilde{\Phi}(\ket{\psi}_{AB}\otimes\ket{00}_{A'B'})-\Phi(\ket{\psi}_{AB}\otimes\ket{00}_{A'B'})||&\leq&\frac{1}{4}\Big[f_1(\beta_A,\beta_B)+f_2(\alpha_B,\beta_A,\beta_B)+f_3(\alpha_A,\beta_A,\beta_B)+f_4(\alpha_A,\alpha_B,\beta_A,\beta_B)\Big]
    = F_S(\alpha_A,\alpha_B,\beta_A,\beta_B)\nonumber
\end{eqnarray}
Clearly here, we have $\lim\limits_{\{\alpha_A,\alpha_B,\beta_A,\beta_B\}\to 0} F_S(\alpha_A,\alpha_B,\beta_A,\beta_B)=0$,    
which in turn provides \ba||\Tilde{\Phi}(\ket{\psi}_{AB}\otimes\ket{00}_{A'B'})||\approx ||\Phi(\ket{\psi}_{AB}\otimes\ket{00}_{A'B'})||\ea 
\subsection{Robust self-testing of observables}
In order to find the robustness of the observables, we follow a procedure similar to the one stated above. Thus we use Eq.~(\ref{robs1}) and calculate the robustness with observable  $X_m$ ($\forall m\in\{A, B\}$) as follows.
\begin{eqnarray}\label{robob1}
    &&||\Tilde{\Phi}(\Tilde{X}_m\ket{\psi}_{AB}\otimes\ket{00}_{A'B'})-\Phi(X_m\ket{\psi}_{AB}\otimes\ket{00}_{A'B'})||\nonumber\\
    &=&\frac{1}{4}\sum_{a,b\in\{0,1\}} \Bigg|\Bigg|\Big[(\Tilde{X}_A)^a (\Tilde{X}_B)^b \Big(1+(-1)^a \Tilde{Z}_A\Big) \Big(1+(-1)^b \Tilde{Z}_B\Big) \Tilde{X}_m\ket{\psi}_{AB} -(X_A)^a (X_B)^b  \Big(1+(-1)^a Z_A\Big) \Big(1+(-1)^b Z_B\Big) X_m\ket{\psi}_{AB}\Big]  \ket{ab}_{A'B'}\Bigg|\Bigg|\nonumber\\
    &\approx&\frac{1}{4}\sum_{a,b\in\{0,1\}} \Bigg|\Bigg|\Big[(\Tilde{X}_A)^a (\Tilde{X}_B)^b \Big(1+(-1)^a \Tilde{Z}_A\Big) \Big(1+(-1)^b \Tilde{Z}_B\Big) \Big(\alpha_m\openone_d + X_m\Big)\ket{\psi}_{AB} -(X_A)^a (X_B)^b  \Big(1+(-1)^a Z_A\Big) \Big(1+(-1)^b Z_B\Big) X_m\ket{\psi}_{AB}\Big]  \ket{ab}_{A'B'}\Bigg|\Bigg|\nonumber\\
     &\leq&\frac{\alpha_m}{4}\sum_{a,b\in\{0,1\}} \Bigg|\Bigg|(\Tilde{X}_A)^a (\Tilde{X}_B)^b \Big(1+(-1)^a \Tilde{Z}_A\Big) \Big(1+(-1)^b \Tilde{Z}_B\Big)\ket{\psi}_{AB}\ket{ab}_{A'B'}\Bigg|\Bigg|\nonumber\\
     &&+ \frac{1}{4}\sum_{a,b\in\{0,1\}} \Bigg|\Bigg|\Big[(\Tilde{X}_A)^a (\Tilde{X}_B)^b \Big(1+(-1)^a \Tilde{Z}_A\Big) \Big(1+(-1)^b \Tilde{Z}_B\Big)-(X_A)^a (X_B)^b  \Big(1+(-1)^a Z_A\Big) \Big(1+(-1)^b Z_B\Big)\Big]X_m \ket{\psi}_{AB} \ket{ab}_{A'B'}\Bigg|\Bigg|\\
     &\leq&\frac{\alpha_m}{4}\sum_{a,b\in\{0,1\}} \Bigg|\Bigg|(\Tilde{X}_A)^a (\Tilde{X}_B)^b \Big(1+(-1)^a \Tilde{Z}_A\Big) \Big(1+(-1)^b \Tilde{Z}_B\Big)\ket{\psi}_{AB}\ket{ab}_{A'B'}\Bigg|\Bigg|\na \\
     &&+ \frac{1}{4}\sum_{a,b\in\{0,1\}} \Bigg|\Bigg|\Big[(\Tilde{X}_A)^a (\Tilde{X}_B)^b \Big(1+(-1)^a \Tilde{Z}_A\Big) \Big(1+(-1)^b \Tilde{Z}_B\Big)-(X_A)^a (X_B)^b  \Big(1+(-1)^a Z_A\Big) \Big(1+(-1)^b Z_B\Big)\Big]\ket{\psi}_{AB} \ket{ab}_{A'B'}\Bigg|\Bigg|\nonumber\\
    &\leq&\frac{\alpha_m}{4}\sum_{a,b\in\{0,1\}} \Bigg|\Bigg|(\Tilde{X}_A)^a (\Tilde{X}_B)^b \Big(1+(-1)^a \Tilde{Z}_A\Big) \Big(1+(-1)^b \Tilde{Z}_B\Big)\ket{\psi}_{AB}\ket{ab}_{A'B'}\Bigg|\Bigg|+\Big|\Big|\Tilde{\Phi}(\ket{\psi}_{AB}\otimes\ket{00}_{A'B'})-\Phi(\ket{\psi}_{AB}\otimes\ket{00}_{A'B'})\Big|\Big|,\quad (m\in\{A,B\})\nonumber\\\na 
    &=&F_{O_X}(\alpha_A,\alpha_B,\beta_A,\beta_B)
\end{eqnarray}
where $F_{O_X}(\alpha_A,\alpha_B,\beta_A,\beta_B)=F(\alpha_m) + F_S(\alpha_A,\alpha_B,\beta_A,\beta_B),\forall m\in\{A,B\}$.  Note that here, we have \ba 
    \lim\limits _{\{\alpha_A,\alpha_B,\beta_A,\beta_B\}\to 0}  F_{O_X}(\alpha_A,\alpha_B,\beta_A,\beta_B)=0\ea  which in turn provides 
\ba ||\Tilde{\Phi}(\Tilde{X}_m\ket{\psi}_{AB}\otimes\ket{00}_{A'B'})||\approx ||\Phi(X_m\ket{\psi}_{AB}\otimes\ket{00}_{A'B'})||\ea \\
Similarly, we again use Eq.~(\ref{robs1}) and calculate the robustness for the other observable $Z_m$ as follows. 
 \begin{eqnarray}\label{robob2}
    &&||\Tilde{\Phi}(\Tilde{Z}_m\ket{\psi}_{AB}\otimes\ket{00}_{A'B'})-\Phi(Z_m\ket{\psi}_{AB}\otimes\ket{00}_{A'B'})||\nonumber\\
    &=&\frac{1}{4}\sum_{a,b\in\{0,1\}} \Bigg|\Bigg|\Big[(\Tilde{X}_A)^a (\Tilde{X}_B)^b \Big(1+(-1)^a \Tilde{Z}_A\Big) \Big(1+(-1)^b \Tilde{Z}_B\Big) \Tilde{Z}_m\ket{\psi}_{AB} -(X_A)^a (X_B)^b  \Big(1+(-1)^a Z_A\Big) \Big(1+(-1)^b Z_B\Big) Z_m\ket{\psi}_{AB}\Big]  \ket{ab}_{A'B'}\Bigg|\Bigg|\nonumber\\
    &\approx&\frac{1}{4}\sum_{a,b\in\{0,1\}} \Bigg|\Bigg|\Big[(\Tilde{X}_A)^a (\Tilde{X}_B)^b \Big(1+(-1)^a \Tilde{Z}_A\Big) \Big(1+(-1)^b \Tilde{Z}_B\Big) (\beta_m\openone_d + Z_m))\ket{\psi}_{AB} -(X_A)^a (X_B)^b  \Big(1+(-1)^a Z_A\Big) \Big(1+(-1)^b Z_B\Big) Z_m\ket{\psi}_{AB}\Big]  \ket{ab}_{A'B'}\Bigg|\Bigg|\nonumber\\
     &\leq&\frac{\beta_m}{4}\sum_{a,b\in\{0,1\}} \Bigg|\Bigg|(\Tilde{X}_A)^a (\Tilde{X}_B)^b \Big(1+(-1)^a \Tilde{Z}_A\Big) \Big(1+(-1)^b \Tilde{Z}_B\Big)\ket{\psi}_{AB}\ket{ab}_{A'B'}\Bigg|\Bigg|\nonumber\\
     &&+ \frac{1}{4}\sum_{a,b\in\{0,1\}} \Bigg|\Bigg|\Big[(\Tilde{X}_A)^a (\Tilde{X}_B)^b (1+(-1)^a \Tilde{Z}_A) (1+(-1)^b \Tilde{Z}_B)-(X_A)^a (X_B)^b  (1+(-1)^a Z_A) (1+(-1)^b Z_B)\Big]Z_m \ket{\psi}_{AB} \ket{ab}_{A'B'}\Bigg|\Bigg|\nonumber\\
     &\leq&\frac{\beta_m}{4}\sum_{a,b\in\{0,1\}} \Bigg|\Bigg|(\Tilde{X}_A)^a (\Tilde{X}_B)^b \Big(1+(-1)^a \Tilde{Z}_A\Big) \Big(1+(-1)^b \Tilde{Z}_B\Big)\ket{\psi}_{AB}\ket{ab}_{A'B'}\Bigg|\Bigg|\\
     &&+ \frac{1}{4}\sum_{a,b\in\{0,1\}} \Bigg|\Bigg|\Big[(\Tilde{X}_A)^a (\Tilde{X}_B)^b \Big(1+(-1)^a \Tilde{Z}_A\Big) \Big(1+(-1)^b \Tilde{Z}_B\Big)-(X_A)^a (X_B)^b  (1+(-1)^a Z_A) (1+(-1)^b Z_B)\Big]\ket{\psi}_{AB} \ket{ab}_{A'B'}\Bigg|\Bigg|\nonumber\\
    &\leq&\frac{\beta_m}{4}\sum_{a,b\in\{0,1\}} \Bigg|\Bigg|(\Tilde{X}_A)^a (\Tilde{X}_B)^b \Big(1+(-1)^a \Tilde{Z}_A\Big) \Big(1+(-1)^b \Tilde{Z}_B\Big)\ket{\psi}_{AB}\ket{ab}_{A'B'}\Bigg|\Bigg|+\Big|\Big|\Tilde{\Phi}(\ket{\psi}_{AB}\otimes\ket{00}_{A'B'})-\Phi(\ket{\psi}_{AB}\otimes\ket{00}_{A'B'})\Big|\Big|,(m\in\{A,B\}) \nonumber\\
    &=&  F_{O_Z}(\alpha_A,\alpha_B,\beta_A,\beta_B) \
\end{eqnarray}
where $F_{O_Z}(\alpha_A,\alpha_B,\beta_A,\beta_B)=F(\beta_m) + F_S(\alpha_A,\alpha_B,\beta_A,\beta_B),\forall m\in\{A,B\}$. Note that here we have \ba  \lim\limits_{\{\alpha_A,\alpha_B,\beta_A,\beta_B\}\to 0} F_{O_Z}(\alpha_A,\alpha_B,\beta_A,\beta_B)=0\ea  which in turn provides  
\begin{eqnarray}
||\Tilde{\Phi}(\Tilde{Z}_m\ket{\psi}_{AB}\otimes\ket{00}_{A'B'})||\approx ||\Phi(Z_m\ket{\psi}_{AB}\otimes\ket{00}_{A'B'})||
\end{eqnarray}
Thus, we have provided robust self-testing of the state and the set of observables of Alice and Bob.
\subsection{Special case: For odd \textit{n} only first party (Alice) implements imperfect observables}\label{spodd}

If we consider only first-party (Alice) implements the imperfect observables, and  the error in both is the same, i.e.,  $\alpha_A=\beta_A=\epsilon\geq 0$ then \ba  ||(\Tilde{X}_A-X_A)\ket{\psi}_{AB}||\leq \epsilon,\ ||(\Tilde{Z}_A-Z_A)\ket{\psi}_{AB}||\leq \epsilon\ea Again, if the error of each observable of the first party is $\delta$ then \ba ||(\Tilde{A}_i-A_i)\ket{\psi}_{AB}||\leq \delta\implies \Tilde{A}_i\approx A_i+\delta \ \openone_d, \ \forall i\in[n], \ \delta \geq 0\ea \tcr{}According to \cite{Bamps2015}, in the presence of noise, the observable is  $\Tilde{Z}_A=\Tilde{A}_{\frac{n+1}{2}}$, which implies that 
\ba ||(\Tilde{Z}_A-Z_A)\ket{\psi}_{AB}||=||(\Tilde{A}_{\frac{n+1}{2}}-A_{\frac{n+1}{2}})\ket{\psi}_{AB}||\leq \delta \Rightarrow  \delta\approx\epsilon\ea  Hence, the operator $\mathcal{K}_{n,i}$ of Eq. (\ref{kni}) from the main text becomes $\Tilde{\mathcal{K}}_{n,i}$, and thus we get 

\begin{eqnarray}\label{noisygamman}
    \Tr[\Tilde{\Gamma}_n\  \rho_{AB}]&=& \frac{1}{2}\sum_{i=1}^{n} \mu_{n,i} \bra{\psi}_{AB}\Tilde{\mathcal{K}}^{\dag}_{n,i} \Tilde{\mathcal{K}}_{n,i}\ket{\psi}_{AB}= \frac{1}{2}\sum_{i=1}^{n} \mu_{n,i}  \ \xi_{n,i}^2=\xi
\end{eqnarray}
where $\Tilde{\mathcal{K}}_{n,i}$ defined as 
\ba\Tilde{\mathcal{K}}_{n,i}=\openone_d\otimes\mathcal{B}_i -\tilde{A}_i \otimes \openone_d, \forall i\in[n]\quad \qty(\text{where $\mathcal{B}_i=\frac{1}{\mu_{n,i}}\qty(\sum\limits_{j=1}^{n-i+1}{B}_j-\sum\limits_{j=n-i+2}^n {B}_j)$})\ea 
which further implies that 
\begin{eqnarray}
    ||\Tilde{\mathcal{K}}_{n,i}-\mathcal{K}_{n,i}\ket{\psi}_{AB}||\leq||\openone_d\otimes(\Tilde{A}_i-A_i)\ket{\psi}_{AB}||\leq \epsilon \implies\Tilde{\mathcal{K}}_{n,i} \approx \mathcal{K}_{n,i}+\openone_d\otimes\epsilon \openone_d
\end{eqnarray}
Hence, we get  
\begin{eqnarray}\label{noisyLij} \Tr[\Tilde{\mathcal{K}}^{\dag}_{n,i} \Tilde{\mathcal{K}}_{n,i}\rho_{AB}]&=&\langle (\mathcal{K}^\dagger_{n,i}+\openone_d \otimes \epsilon \openone_d)(\mathcal{K}_{n,i}+\openone_d \otimes \epsilon \openone_d)\rangle_{\rho_{AB}}\nonumber\\
    &=& \langle \mathcal{K}^\dagger_{n,i} \mathcal{K}_{n,i} \rangle_{\rho_{AB}}+\epsilon \langle \mathcal{K}^\dagger_{n,i} \rangle_{\rho_{AB}}+\epsilon \langle \mathcal{K}_{n,i} \rangle_{\rho_{AB}}+\epsilon^2
\end{eqnarray}
Since in the ideal scenario $\langle \mathcal{K}^\dagger_{n,i} \mathcal{K}_{n,i} \rangle_{\rho_{AB}}=\langle \mathcal{K}^\dagger_{n,i}\rangle_{\rho_{AB}}=\langle \mathcal{K}_{n,i} \rangle_{\rho_{AB}}=0$, we have the following.
\ba \Tr[\Tilde{\mathcal{K}}^{\dag}_{n,i} \Tilde{\mathcal{K}}_{n,i}\rho_{AB}]&=&\epsilon^2 
\ea
The optimization condition to derive $(\mathcal{
G}_{n})_{Q}^{opt}$ gives $\mu_{n,i}=\csc{\frac{\pi}{2n}}$. Using  $\mu_{n,i}$ and Eq.~(\ref{noisyLij}) in Eq.~(\ref{noisygamman}),  we get 
\begin{eqnarray}\label{Efun}
    \frac{1}{2}\sum_{i=1}^{n} \mu_{n,i} \bra{\psi}_{AB}\Tilde{\mathcal{K}}^{\dag}_{n,i} \Tilde{\mathcal{K}}_{n,i}\ket{\psi}_{AB}=\xi\Rightarrow \frac{1}{2}n \ \epsilon^2 \ \csc{\frac{\pi}{2n}}=\xi\Rightarrow \epsilon =\sqrt{\frac{2\xi }{n \csc \left(\frac{\pi }{2 n}\right)}}
\end{eqnarray}
In this case, robust self-testing of the state from Eq.~(\ref{robs1}) provides 
\begin{eqnarray}\label{stateep}
    &&||\Tilde{\Phi}(\ket{\psi}_{AB}\otimes\ket{00}_{A'B'})-\Phi(\ket{\psi}_{AB}\otimes\ket{00}_{A'B'})||\nonumber\\
    &&=\frac{1}{4}\sum_{a,b\in\{0,1\}} \Bigg|\Bigg|\Big[(\Tilde{X}_A)^a (X_B)^b \Big(1+(-1)^a \Tilde{Z}_A\Big) \Big(1+(-1)^b Z_B\Big) -(X_A)^a (X_B)^b \Big(1+(-1)^a Z_A\Big) \Big(1+(-1)^b Z_B\Big) \Big] \ket{\psi}_{AB} \ket{ab}_{A'B'}\Bigg|\Bigg|\nonumber\\
    &&\leq\frac{1}{4}\sum_{a\in\{0,1\}} \Bigg|\Bigg|\Big[\qty( (\Tilde{X}_A)^a  \Big(1+(-1)^a \Tilde{Z}_A\Big) - (X_A)^a  \Big(1+(-1)^a Z_A\Big))\sum_{b\in\{0,1\}}\Big(1+(-1)^b {Z}_B\Big)\Big] \ket{\psi}_{AB} \ket{ab}_{A'B'}\Bigg|\Bigg|\nonumber\\
    &&\leq\frac{1}{4}\Bigg[\sum_{a\in\{0,1\}} \Bigg|\Bigg|\qty( (\Tilde{X}_A)^a  \Big(1+(-1)^a \Tilde{Z}_A\Big) - (X_A)^a  \Big(1+(-1)^a Z_A\Big))\sum_{b\in\{0,1\}} \ket{\psi}_{AB} \ket{ab}_{A'B'}\Bigg|\Bigg|\nonumber\\
    &&\phantom{\frac{1}{4}\Bigg[}+\sum_{a\in\{0,1\}} \Bigg|\Bigg|\qty( (\Tilde{X}_A)^a  \Big(1+(-1)^a \Tilde{Z}_A\Big) - (X_A)^a  \Big(1+(-1)^a Z_A\Big))\sum_{b\in\{0,1\}}(-1)^b {Z}_B \ket{\psi}_{AB} \ket{ab}_{A'B'}\Bigg|\Bigg|\Bigg]\nonumber\\
    &&\leq\frac{1}{4}\Bigg[\sum_{a\in\{0,1\}} 2\Bigg|\Bigg|\qty( (\Tilde{X}_A)^a  \Big(1+(-1)^a \Tilde{Z}_A\Big) - (X_A)^a  \Big(1+(-1)^a Z_A\Big))\sum_{b\in\{0,1\}} \ket{\psi}_{AB} \ket{ab}_{A'B'}\Bigg|\Bigg|\Bigg]\nonumber\\
     &&\leq\sum_{a\in\{0,1\}} \Bigg|\Bigg|\Big[ (\Tilde{X}_A)^a  \Big(1+(-1)^a \Tilde{Z}_A\Big) - (X_A)^a  \Big(1+(-1)^a Z_A\Big)\Big] \ket{\psi}_{AB} \Bigg|\Bigg|\ \ \ \ \qty( \text{Neglecting the ancillary part  $\ket{ab}_{A'B'}$} )\nonumber\\
     &&\leq\sum_{a\in\{0,1\}} \Bigg|\Bigg|\Big[\qty(( \Tilde{X}_A)^a-(X_a)^a) + (-1)^a \qty((\Tilde{X}_A)^a \Tilde{Z}_A-(X_a)^a Z_A)\Big] \ket{\psi}_{AB} \Bigg|\Bigg|\ \ \ \nonumber\\
     &&\leq ||(\Tilde{X}_A-X_A)\ket{\psi}_{AB}||+||(\Tilde{Z}_A- Z_A)\ket{\psi}_{AB}||+
     ||(\Tilde{X}_A \Tilde{Z}_A-X_A Z_A)\ket{\psi}_{AB}||
\end{eqnarray}
Again, we get 
\begin{eqnarray}\label{xbzbsp}
    ||(\Tilde{X}_A \Tilde{Z}_A-X_A Z_A)\ket{\psi}_{AB}||\leq||(\Tilde{X}_A (Z_A+\epsilon \openone_d)_A-X_A Z_A)\ket{\psi}_{AB}||
    &\leq& ||(\Tilde{X}_A-X_A)\ket{\psi}_{AB}||+\epsilon||\Tilde{X}_A\ket{\psi}_{AB}||\nonumber\\
    &\leq& \epsilon+\epsilon||(\epsilon\openone_d+X_A)\ket{\psi}_{AB}||\nonumber\\
    &\leq&2\epsilon +\epsilon^2
\end{eqnarray}
Substituting Eq.~(\ref{xbzbsp})  in Eq.~(\ref{stateep}) and using $ ||(\Tilde{X}_A-X_A)\ket{\psi}_{AB}||\leq \epsilon,||(\Tilde{Z}_A-Z_A)\ket{\psi}_{AB}||\leq \epsilon$, we get 
\begin{eqnarray}\label{staterobep}
    ||\Tilde{\Phi}(\ket{\psi}_{AB}\otimes\ket{00}_{A'B'})-\Phi(\ket{\psi}_{AB}\otimes\ket{00}_{A'B'})||\leq 4\epsilon +\epsilon^2
\end{eqnarray}
Similarly, the robust self-testing of the observable from Eq.~(\ref{robob1}) provides 

\begin{eqnarray}
&&||\Tilde{\Phi}(\Tilde{X}_A\ket{\psi}_{AB}\otimes\ket{00}_{A'B'})-\Phi(X_A\ket{\psi}_{AB}\otimes\ket{00}_{A'B'})||\nonumber\\
    &&\leq\frac{\epsilon}{4}\sum_{a,b\in\{0,1\}} \Bigg|\Bigg|(\Tilde{X}_A)^a (X_B)^b \Big(1+(-1)^a \Tilde{Z}_A\Big) \Big(1+(-1)^b Z_B\Big)\ket{\psi}_{AB}\ket{ab}_{A'B'}\Bigg|\Bigg|+\Big|\Big|\Tilde{\Phi}(\ket{\psi}_{AB}\otimes\ket{00}_{A'B'})-\Phi(\ket{\psi}_{AB}\otimes\ket{00}_{A'B'})\Big|\Big|\nonumber\\
    &&\leq\frac{\epsilon}{4}\sum_{a\in\{0,1\}} \Bigg|\Bigg| (\Tilde{X}_A)^a  \Big(1+(-1)^a \Tilde{Z}_A\Big)\sum_{b\in\{0,1\}} (X_B)^b \Big(1+(-1)^b Z_B\Big)\ket{\psi}_{AB}\ket{ab}_{A'B'}\Bigg|\Bigg|+4\epsilon+\epsilon^2\nonumber\\
    &&\leq\frac{\epsilon}{4}\sum_{a\in\{0,1\}} \Bigg|\Bigg| (\Tilde{X}_A)^a  \Big(1+(-1)^a \Tilde{Z}_A\Big) (1+Z_B+X_B-X_B Z_B)\ket{\psi}_{AB}\ket{ab}_{A'B'}\Bigg|\Bigg|+4\epsilon+\epsilon^2\nonumber\\
    &&\leq\epsilon\sum_{a\in\{0,1\}} \Bigg|\Bigg| (\Tilde{X}_A)^a  \Big(1+(-1)^a \Tilde{Z}_A\Big) \ket{\psi}_{AB}\Bigg|\Bigg|+4\epsilon+\epsilon^2\nonumber\ \ \ \ \ \qty( \text{Neglecting the ancillary part  $\ket{ab}_{A'B'}$} )\\
    &&\leq \epsilon \Bigg|\Bigg| \qty(1+\Tilde{Z}_A +\Tilde{X}_A-\Tilde{X}_A\Tilde{Z}_A) \ket{\psi}_{AB}\Bigg|\Bigg|+4\epsilon+\epsilon^2\nonumber\\
     &&\leq \epsilon \qty(1+||\Tilde{X}_A \ket{\psi}_{AB}||+||\Tilde{Z}_A \ket{\psi}_{AB}||+||\Tilde{X}_A\Tilde{Z}_A \ket{\psi}_{AB}||)+4\epsilon+\epsilon^2\nonumber\\
     &&\leq \epsilon \qty(1+||(X_A+\epsilon \openone_d) \ket{\psi}_{AB}||+||(Z_A+\epsilon \openone_d)\ket{\psi}_{AB}||+||(X_A+\epsilon \openone_d)(Z_A+\epsilon \openone_d) \ket{\psi}_{AB}||)+4\epsilon+\epsilon^2\nonumber\\
      &&\leq \epsilon^3+5\epsilon^2+8\epsilon
\end{eqnarray}
Similarly, 
\begin{eqnarray}
      &&||\Tilde{\Phi}(\Tilde{Z}_A\ket{\psi}_{AB}\otimes\ket{00}_{A'B'})-\Phi(Z_A\ket{\psi}_{AB}\otimes\ket{00}_{A'B'})||\leq \epsilon^3+5\epsilon^2+8\epsilon
\end{eqnarray}
Recalling all the results.
\begin{eqnarray}
    &&||\Tilde{\Phi}(\ket{\psi}_{AB}\otimes\ket{00}_{A'B'})-\Phi(\ket{\psi}_{AB}\otimes\ket{00}_{A'B'})||\leq 4\epsilon +\epsilon^2\\
    &&||\Tilde{\Phi}(\Tilde{O}\ket{\psi}_{AB}\otimes\ket{00}_{A'B'})-\Phi(O\ket{\psi}_{AB}\otimes\ket{00}_{A'B'})||\leq \epsilon^3+5\epsilon^2+8\epsilon\quad \forall O\in\{X_A,Z_A\}
\end{eqnarray}
and the relation between $\epsilon$ and the observed deviation $\xi$, i.e. $\epsilon =\sqrt{\frac{2\xi }{n \csc \left(\frac{\pi }{2 n}\right)}}$, we determine the robust self-testing bounds for both the state and the observables.  We define the relative observed violations $r$ as a function of the number of inputs $n$ and the observed deviation $\xi$ as follows.
\begin{eqnarray}
    &&r=\frac{\Tilde{(\mathcal{G}_{n})}_{Q}-(\mathcal{G}_n)_C}{(\mathcal{G}_{n})^{opt}_{Q}-(\mathcal{G}_n)_C}= 1-\frac{\xi}{ n \csc{\frac{\pi}{2n}}-\lfloor\frac{n^2+1}{2}\rfloor}
\end{eqnarray}
Alternatively, $\xi$ can be written in terms of $r$ and $n$ as
\begin{eqnarray}
        &&\xi =(1-r) \left( n \csc{\frac{\pi}{2n}}-\lfloor\frac{n^2+1}{2}\rfloor\right)
\end{eqnarray}
where $\Tilde{(\mathcal{G}_{n})}_{Q}$ and $ (\mathcal{G}_{n})^{opt}_{Q}$ are defined in Eqs.~(\ref{Cn noise}) and (\ref{cnopt}) respectively. Here $(\mathcal{G}_n)_c =\lfloor\frac{n^2+1}{2}\rfloor$ is the classical bound of the GBI. 
Hence, the trace distance between the observed and ideal state in terms of the relative operator $r$ is
\begin{eqnarray}
  && ||\Tilde{\Phi}(\ket{\psi}_{AB}\otimes\ket{00}_{A'B'})-\Phi(\ket{\psi}_{AB}\otimes\ket{00}_{A'B'})||\leq f_s(r) \\ 
  &&f_s(r)=\frac{2(1-r)\qty(n\csc{\frac{\pi}{2 n}}-\lfloor\frac{n^2+1}{2}\rfloor)}{n\csc{\frac{\pi}{2 n}}}+4\sqrt{2}\sqrt{\frac{(1-r)\qty(n\csc{\frac{\pi}{2 n}}-\lfloor\frac{n^2+1}{2}\rfloor)}{n\csc{\frac{\pi}{2 n}}}}
\end{eqnarray}
 Now, according to Fuchs-Van de Graaf, the approximate relation between the trace distance $f_s(r)$ and the robust fidelity $F_s(r)$ is 
 \begin{eqnarray}
     2\qty(1-\sqrt{F_s(r)})\leq f_s(r)\leq 2\sqrt{1-F_s(r)}
 \end{eqnarray}
 This gives the lower bound of fidelity in terms of trace distance, as 
 \begin{eqnarray}
     F_s(r)\geq \qty(1-\frac{1}{2}f_s(r))^2
 \end{eqnarray}
Similarly, for the observables, we can define the robust fidelity $F_o(r)$ in terms of the trace distance $f_o(r)$, i.e., 
  \begin{eqnarray}
     F_o(r)\geq \qty(1-\frac{1}{2}f_o(r))^2
 \end{eqnarray}
 where \ba f_o(r)=2\sqrt{2}\qty(\frac{(1-r)\qty(n\csc{\frac{\pi}{2 n}}-\lfloor\frac{n^2+1}{2}\rfloor)}{n\csc{\frac{\pi}{2 n}}})^{\frac{3}{2}}+8\sqrt{2}\qty(\frac{(1-r)\qty(n\csc{\frac{\pi}{2 n}}-\lfloor\frac{n^2+1}{2}\rfloor)}{n\csc{\frac{\pi}{2 n}}})^\frac{1}{2}+\frac{10(1-r)\qty(n\csc{\frac{\pi}{2 n}}-\lfloor\frac{n^2+1}{2}\rfloor)}{n\csc{\frac{\pi}{2 n}}}\ea

\end{widetext}

\vspace{3cm}

\bibliography{references}

\end{document}